\documentclass[a4paper,11pt]{article}
\usepackage{amsmath,amssymb}
\usepackage[dvipdfmx]{graphicx}
\usepackage{graphicx}
\usepackage{color}
\usepackage{mathrsfs}
\usepackage{bm}
\usepackage{multirow}
\usepackage{booktabs}
\usepackage[left]{lineno}
\usepackage{amsmath, amsthm, amssymb}
\usepackage[ruled]{algorithm}
\usepackage{algpseudocode}
\usepackage{algpascal}
\usepackage{algc}

\newcommand\ASTART{\bigskip\noindent\begin{minipage}[b]{0.5\linewidth}}

\newcommand\AENDSKIP{\end{minipage}\bigskip}
\newcommand\AEND{\end{minipage}}

\newcommand{\Part}[3]{ \frac{ \partial^{#3} #1 }{ \partial #2^{#3} } }
\newcommand{\V}[1]{\bm{#1} } 
\newcommand{\Tr}[1]{ \mathop{\rm Tr}_{ #1 } }

\newcommand{\Ave}[1]{\left\langle {#1} \right\rangle} 
\newcommand{\dAve}[1]{\left\langle \left\langle {#1} \right\rangle \right\rangle} 

\newcommand{\Extr}[1]{ \mathop{\rm Extr}_{ #1 } }

\newcommand{\mR}{\mathbb{R}}

\newcommand{\lb}{\left(}
\newcommand{\rb}{\right)}
\newcommand{\lbb}{\left\{}
\newcommand{\rbb}{\right\}}
\newcommand{\lsb}{ \left[ }
\newcommand{\rsb}{ \right] }

\newcommand{\Blbb}{ \Biggl\{ }
\newcommand{\Brbb}{ \Biggr\} }

\newcommand{\Req}[1]{eq.\ (\ref{eq:#1})}
\newcommand{\BReq}[1]{Eq.\ (\ref{eq:#1})}
\newcommand{\NReq}[1]{(\ref{eq:#1})}
\newcommand{\Reqs}[2]{eqs.\ (\ref{eq:#1},\ref{eq:#2})}

\newcommand{\Rfig}[1]{Fig.\ \ref{fig:#1}}

\newcommand{\Lfig}[1]{\label{fig:#1}}
\newcommand{\Leq}[1]{\label{eq:#1}}
\newcommand{\Rsec}[1]{sec.\ \ref{sec:#1}}
\newcommand{\Lsec}[1]{\label{sec:#1}}
\newcommand{\be}{\begin{eqnarray}}
\newcommand{\ee}{\end{eqnarray}}
\newcommand{\ba}{\begin{array}}
\newcommand{\ea}{\end{array}}
\newcommand{\no}{\nonumber}

\newcommand{\subbe}{\begin{subequations}}
\newcommand{\subee}{\end{subequations}}

\newcommand{\mc}[1]{\mathcal{#1}}
\DeclareMathOperator*{\argmin}{arg\,min}

\newcommand{\supp}[1]{{\rm supp}({#1})}
\newcommand{\nnzero}{K}
\newcommand{\MSEx}{\epsilon_{x}}
\newcommand{\MSEy}{\epsilon_{y}}
\newcommand{\MSEp}{\epsilon_{y}}
\newcommand{\POWx}{\sigma_x^2}

\newcommand{\lA}{\leftarrow}

\newcommand{\Lcode}[1]{\label{code:#1}}
\newcommand{\Rcode}[1]{Alg.\ \ref{code:#1}}

\title{Statistical mechanical analysis of sparse linear regression \\ as a variable selection problem}
\author{Tomoyuki Obuchi\thanks{Department of Mathematical and Computing Science, Tokyo Institute of Technology, 2-12-1, Ookayama, Meguro-ku, Tokyo, Japan}, 
Yoshinori Nakanishi-Ohno\thanks{Graduate School of Arts and Sciences, The University of Tokyo, Komaba
3-8-1, Meguro, Tokyo, 153-8902, Japan}
\thanks{Precursory Research for Embryonic Science and Technology, Japan Science and Technology Agency, Honcho 4-1-8, Kawaguchi, Saitama, 332-0012, Japan}, 
Masato Okada\thanks{Graduate School of Frontier Sciences, The University of Tokyo, Kashiwa, Chiba 277-8561, Japan}, and 
Yoshiyuki Kabashima$^{\ast}$}

\begin{document}
\date{}
\maketitle

\begin{abstract}
An algorithmic limit of compressed sensing or related variable-selection problems is analytically evaluated when a design matrix is given by an overcomplete random matrix. The replica method from statistical mechanics is employed to derive the result. The analysis is conducted through evaluation of the entropy, an exponential rate of the number of combinations of variables giving a specific value of fit error to given data which is assumed to be generated from a linear process using the design matrix. This yields the typical achievable limit of the fit error when solving a representative $\ell_0$ problem and includes the presence of unfavourable phase transitions preventing local search algorithms from reaching the minimum-error configuration. The associated phase diagrams are presented. A noteworthy outcome of the phase diagrams is that there exists a wide parameter region where any phase transition is absent from the high temperature to the lowest temperature at which the minimum-error configuration or the ground state is reached. This implies that certain local search algorithms can find the ground state with moderate computational costs in that region. Another noteworthy result is the presence of the random first-order transition in the strong noise case. The theoretical evaluation of the entropy is confirmed by extensive numerical methods using the exchange Monte Carlo and the multi-histogram methods. Another numerical test based on a metaheuristic optimisation algorithm called simulated annealing is conducted, which well supports the theoretical predictions on the local search algorithms. In the successful region with no phase transition, the computational cost of the simulated annealing to reach the ground state is estimated as the third order polynomial of the model dimensionality. 
\end{abstract}

\section{Introduction} \Lsec{Introduction}
Compressed sensing is a technique used to recover a high-dimensional signal from a limited number of measurements by utilising the fact that the signal of interest has redundancy and thus can be ``sparse''; many of the coefficients are set to zero when described with an appropriate basis. This technique has a long history~\cite{Claerbout:73,Santosa:86,Donoho:89}, but it has recently attracted increased attention as its high performance has been demonstrated in recent influential papers~\cite{Donoho:06,Candes:05,Candes:06a,Candes:06b,Candes:08}. There has been a surge of research of compressed sensing, which is based on a general idea that the signal has a sparse representation on an appropriate basis, because the idea can be shared in many other contexts such as data compression, multivariate regression, and variable selection. This trend has triggered a major evolution in techniques of signal and information processing, which is gradually forming a new framework called ``sparse modelling''~\cite{Okada:13,Rish:14,Mairal:14,Hastie:15}. 

For clarity, we provide a concise mathematical form to the problem treated here. Suppose a data vector $\V{y}\in \mR^{M}$ is generated by the following linear process with a design matrix $A\in \mR^{M\times N}$ and a signal vector $\V{x}_0 \in \mR^{N}$ such that:
\be
\V{y}=A\V{x}_0+\V{\xi},
\Leq{generative}
\ee
where $\V{\xi}$ is a noise vector, the component of which is assumed to be an independent and identically distributed (i.i.d.) variable from the normal distribution with zero mean and variance $\sigma_{\xi}^2$, $\mathcal{N}(0,\sigma_{\xi}^2)$. In the context of compressed sensing, the design matrix $A$ represents the measurement process, and given $A$ and $\V{y}$, we try to infer $\V{x}_0$ for the situation $M<N$. This is an underdetermined problem and the sparsity assumption that the number of nonzero components of $\V{x}_0$ is smaller than $M$ is needed for solving it. With this assumption, the perfect reconstruction of $\V{x}_0$ is possible if the noise is absent. The most naive algorithm to achieve this is the exhaustive search, which selects the sparsest set of variables among error-free ones. This is clearly infeasible if the model dimensionality $N$ is large, and more efficient algorithms or approximations should be tailored. A common approximation is to relax the sparsity constraint. Many studies have been conducted along this direction, and some theoretical studies demonstrated that the perfect reconstruction of $\V{x}_0$ is possible under reasonable conditions even under such a relaxation~\cite{Donoho:09-1,Kabashima:09,Ganguli:10}. Associated efficient algorithms achieving perfect reconstruction in the noiseless case have been developed~\cite{Donoho:09-2,Donoho:10-1,Donoho:10-2,Rangan:10}.

Some degree of compromise such as the relaxation above appears to be unavoidable, because variable selection in the present problem is NP-hard in the worst case~\cite{Natarajan:95}. However, more recent works suggest that, even without such relaxation, variable selection can be achieved at reasonable expense for ``typical'' cases~\cite{Krzakala:12-1,Krzakala:12-2} where the design matrix is assumed to be i.i.d. from the normal distribution. Their formulation is based on the Bayesian framework assuming the signal's generative process is known. The algorithmic limit was computed by using non-rigorous statistical mechanical techniques, and an associated message-passing algorithm was developed and shown to achieve the limit, and those results have been supported from a firmer mathematical basis~\cite{Barbier:17}.  

However, the Bayesian framework is not always preferred in the context of signal processing. This is because the signal's generative process is not necessarily evident and is difficult to model in many practical situations. In such cases, it may be better to focus less on the signal sources and rely more on the less informative prior. According to this idea, in the context of data compression, the present authors recently proposed a variable selection criterion based on the following widely-used optimisation formulation~\cite{Nakanishi:16}:
\be
\hat{\V{x}}=\argmin_{\V{x}}||\V{y}-A\V{x}||_2^2~\mathrm{subj.~to}~||\V{x}||_0\leq K,
\Leq{L0optimization}
\ee
where $||\V{x}||_k=(\sum_{i}|x_i|^k)^{1/k}$ denotes the $\ell_k$ norm and the $\ell_0$ norm $||\V{x} ||_0$ is assumed to give the number of nonzero components of $\V{x}$. Our basic idea in~\cite{Nakanishi:16} is to compare all variable sets of size $K$ and to organise them as an ensemble in the statistical mechanical sense by regarding the fit error as energy. Our statistical mechanical analysis, again non-rigorous and performed under the same assumption on the design matrix as~\cite{Krzakala:12-1,Krzakala:12-2}, showed that the configuration space of the variable set is rather ``smooth'', implying that certain local search algorithms can efficiently find the minimum-error variable set. Based on this finding, we developed an algorithm based on the so-called simulated annealing (SA) algorithm~\cite{Kirkpatrick:83}, which is a Monte Carlo (MC)-based optimisation solver, and demonstrated that it can efficiently find the minimum-error set for a wide range of parameters~\cite{Obuchi:16-1,Obuchi:16-2}. These results again suggest that variable selection in the present setting can be efficiently achieved, even without relaxation or resorting to the Bayesian framework. 

The success of the MC based method further motivates us to analyse the property of the ensemble of the variable set in detail. The previous analysis~\cite{Nakanishi:16} was limited to the data compression context, and another analysis directly relevant to compressed sensing or multivariate regression is desirable. The present paper addresses this point. The main difference from~\cite{Nakanishi:16} is the presence of the true signal $\V{x}_0$ in \Req{generative}. This introduces other criteria on the reconstructed signal such as the prediction ability and the error to the true signal. We provide a quantitative analysis for these issues.

The remainder of the paper is organised as follows. In \Rsec{Problem}, we state the problem setting and the formulation which we employ in this paper. The meaning of the formulation in relation to the Bayesian framework is also explained. In \Rsec{Analytical}, we provide the analytical solution of the fit error and related quantities derived by the statistical mechanical formulation. In \Rsec{Numerical}, we present the results of numerical experiments using a careful MC method to support our analytical computations. The performance of the SA algorithm for finding the minimum-error set is also revisited. The final section concludes the study. 

\section{Problem and Formulation} \Lsec{Problem}
\subsection{Problem setting and notation} \Lsec{Problem setting}
As noted in the \Rsec{Introduction}, the data is supposed to be generated by the linear process \NReq{generative}.
We assume the true signal $\V{x}_0$ is $K_0$-sparse and $K_0$ is less than $M$: $||\V{x}_0||_0=\nnzero_0<M$. For notational convenience, we introduce the $\ell_0$ operator of a vector $\V{x}$ as $|\V{x}|_{0}$, which results in a binary vector whose component is $(|\V{x}|_{0})_i=1$ if $x_i\neq 0$, or $(|\V{x}|_{0})_i=0$ otherwise. The support of the true signal is represented by a support vector $\V{c}_0\equiv |\V{x}_0|_0$. 

We are interested in the fit quality to the data $\V{y}$ for a given set of variables described by a support vector $\V{c}\in\{0,1\}^N$, on a linear model basis. The fit quality is thus quantified by a mean squared error (MSE) for $\V{y}$ and the coefficients of the chosen variables are assumed to be optimised to describe $\V{y}$. The optimised coefficients are written in the following form:
\be
\hat{\V{x}}(\V{c}|\V{y},A)=\argmin_{\V{x}}||\V{y}-A(\V{c}\circ \V{x}) ||_2^2,
\Leq{x(c)}
\ee
where $(\V{c}\circ \V{x})_i=c_i x_i$ represents the Hadamard product. To eliminate an ambiguity in \Req{x(c)}, the coefficients of variables out of the support are set to zero, $c_i=0 \Rightarrow \hat{x}_i(\V{c})=0$, to provide consistency with the support operator: $\V{c}=|\hat{\V{x}}(\V{c})|_0$. We denote the corresponding MSE with $\V{y}$ by
\be
\MSEy(\V{c}|\V{y},A)=\frac{1}{2M}||\V{y}-A\hat{\V{x}}(\V{c}|\V{y},A) ||_2^2. \Leq{output MSE}
\ee
This is called the output MSE throughout this paper. In addition to the output MSE, we are interested in the MSE with the true signal defined by
\be
\MSEx(\hat{\V{x}}|\V{x}_0)
=
\frac{1}{2N}||\hat{\V{x}}-\V{x}_0||_2^2.
\Leq{input MSE}
\ee
Hereafter this is termed the input MSE, and the hat symbol is assumed to represent an estimator of the corresponding quantity. 

The primary object of our investigation is the histogram of $\MSEy(\V{c}|\V{y},A)$ when changing the set of variables $\V{c}$. There are two reasons for evaluating this quantity. One is related to the reconstruction of $\V{x}_0$. In this context, lower values of the output MSE are not always preferable and we need more global information regarding the set of variables to obtain a good solution. The other involves possible algorithmic implications. To find a small output-MSE configuration, we usually conduct a local recursive search from certain initial conditions. The performance of such local search algorithms is strongly affected by the structure of the configuration space of $\V{c}$. The histogram of $\MSEy(\V{c}|\V{y},A)$ provides the necessary information about the structure. 

More specifically, we evaluate the exponential rate of the histogram in the large size limit $N\to \infty$ while keeping $\alpha=M/N$ and $\rho=\nnzero/N=\sum_{i}c_i/N$ finite. This quantity is simply the statistical mechanical entropy defined by 
\be
s\lb \MSEp|\rho,\V{y},A \rb
=\frac{1}{N}
\log 
 \lb 	
  \#\{\bm{c}\ |\  \sum_{i}c_i=N\rho \wedge \MSEy(\V{c}|\V{y},A) =\MSEp \}
  \rb.
\Leq{entropy}
\ee
Entropies or similar thermodynamic functions associated with certain optimisation problems have provided algorithmic implications and benefits in several contexts such as information theory, computer science, and neural networks~\cite{Kabashima:00-1,Kabashima:00-2,Saad:01,Kudekar:11,Tanaka:02,Kabashima:03,Takeda:06,Mezard:02,Krzakala:07,Krauth:89,Obuchi:09,Huang:13}. We see this strategy actually works well in the present problem below.

\subsection{Outline of analysis}
We outline the analysis of the entropy. Evaluation of the entropy is replaced with an assessment of a generating function that is a Legendre transform of the entropy. Assumptions to enable this assessment are stated as well.

\subsubsection{Generating function: Legendre transform of entropy}\Lsec{Generating function: Legendre}
Direct computation of the entropy is not easy, and a more systematic method of evaluation is available using a Legendre transform of the entropy, which we call free entropy throughout this study~\cite{mezard2009information}. 

We introduce a partition function $G(\mu|\rho,\V{y},A)$ as
\be
G(\mu|\rho,\V{y},A) \equiv \lb \prod_{i=1}^{N}\sum_{c_i=0,1} \rb \delta\lb \sum_{i}c_i-N\rho \rb e^{-M\mu \MSEy(\V{c}|\V{y},A)},
 \Leq{G}
\ee
where $\delta(\cdot)$ is the delta function. The free entropy is represented by the exponential rate of $G$. Considering the definition of the entropy and assuming that the saddle-point method is applicable in the large $N$ limit, we can easily see that the following relation holds
\be
g(\mu|\rho,\V{y},A)\equiv \frac{1}{N}\log G(\mu|\rho,\V{y},A)=\max_{\MSEp \in \supp{s(\MSEp|\rho,\V{y},A)}}
\lbb
-\alpha \mu \MSEp + s(\MSEp|\rho,\V{y},A)
\rbb,
\Leq{g}
\ee
where $\supp{f(x)}$ denotes the support of the function $f(x)$. Hence $g$ and $s$ are the Legendre transforms of each other. Assuming $s(\MSEp)$ is concave with respect to $\MSEp$, then $g$ and $s$ have a one-to-one correspondence and are connected by the control parameter $\mu(\geq 0)$. This control parameter plays the role of ``inverse temperature'' in physics. The maximiser in \Req{g} and the corresponding entropy, $\MSEp(\mu|\rho,\V{y},A)$ and $s(\mu|\rho,\V{y},A)=s(\MSEp(\mu|\rho,\V{y},A)|\rho,\V{y},A)$, are thus parameterised as
\subbe
\Leq{parametric}
\be
&&
\MSEp(\mu|\rho,\V{y},A)=-\frac{1}{\alpha}\Part{}{\mu}{}g(\mu|\rho,\V{y},A),
\\
&&
s(\mu|\rho,\V{y},A)=g(\mu|\rho,\V{y},A)-\mu\Part{}{\mu}{}g(\mu|\rho,\V{y},A).
\ee
\subee
Employing these relations, we can easily handle the dominant output MSE and determine the corresponding entropy from $g$. 

However, there are two difficulties in the evaluation of $g$. One is the dependence on $\V{y}$ and $A$, and the other is the presence of the least squares problem in the definition of $\MSEy(\V{c})$.

The first problem is overcome as follows. The free entropy $g$ is a self-averaging quantity, as is the entropy. Typical values of self-averaging quantities are in accordance with their averaged values in the large $N$ limit. This means that we can calculate the averaged value of $g$ over $A$ and $\V{y}=A\V{x}_0+\V{\xi}$ instead of directly considering the entropy's dependence on those quantities. We denote the average over $\V{x}_0,\V{\xi}$ and $A$ by square brackets with appropriate subscripts as $[\cdots ]_{\V{x}_0,\V{\xi},A}$. Unfortunately, this average is not easy to calculate. The so-called replica method is a great aid in such a situation, and is symbolised by the following identity    
\be
g(\mu|\rho)=\lsb g(\mu|\rho,\V{y},A) \rsb_{\V{x}_0,\V{\xi},A}
=
\lim_{n\to 0}\frac{1}{Nn}\log \lsb G^n(\mu|\rho,\V{y},A) \rsb_{\V{x}_0,\V{\xi},A}.
\Leq{replica trick}
\ee
In addition to this identity, we assume that $n$ is a positive integer. This assumption enables us to compute the average $[\cdots ]_{\V{x}_0,\V{\xi},A}$. After computing this average, we take the limit $n\to 0$ by employing the analytical continuation from $n \in \mathbb{N}$ to $n \in \mathbb{R}$ under the so-called replica symmetric (RS) or the replica symmetry breaking (RSB) ansatz, which will be explained later. 

The second problem is solved by introducing a variable $\beta$ and taking a limit as follows
\be
e^{-M\mu \MSEy(\V{c}|\V{y},A)}
=\lim_{\beta \to \infty} 
e^{-\mu \mc{H}(\V{c}|\beta,\V{y},A) },
\ee
where 
\be
\mc{H}(\V{c}|\beta,\V{y},A) 
=-\frac{1}{\beta} \log \int\mathrm{d}\bm{x} 
\prod_{i}\lbb (1-c_i)\delta(x_i) + c_i \rbb
\mathrm{e}^{-\frac{\beta}{2}||\bm{y}-\bm{A}(\bm{c}\circ\bm{x})||_2^2},
\Leq{Hamiltonian}
\ee
where the factor $\prod_{i}\lbb (1-c_i)\delta(x_i) + c_i \rbb$ is introduced to make the integral well-defined and can be regarded as a prior for $\V{x}$. In the limit $\beta \to \infty$, only the contribution corresponding to the solution of the least squares problem in \Req{x(c)} survives the integration, and $\mc{H}(\V{c}|\beta,\V{y},\V{A}) \to M \MSEy(\V{c}|\V{y},\V{A})$. We further assume that $\nu=\mu/\beta$ is a positive integer as well as $n$ in \Req{replica trick}. This enables us to treat in parallel the summation over $\V{c}$ and the integration over $\V{x}$, as well as the average $[\cdots ]_{\V{x}_0,\V{\xi},A}$. The limit $\beta\to \infty \Leftrightarrow \nu \to 0$ is taken after those operations through the analytic continuation. 

These operations can be summarised in a line 
\be
&&
g(\mu|\rho)=
\lim_{n \to 0} \lim_{\nu \to 0} 
 \frac{1}{Nn} \log
  \lsb
   \lbb
    \Tr{\V{c}} 
      \lb 
    \Tr{\V{x}|\V{c}}        
        \mathrm{e}^{-\frac{1}{2}\frac{\mu}{\nu}||\V{y}-\V{A}(\V{c}\circ \V{x}) ||_2^2   }  
      \rb^{\nu}
   \rbb^n
  \rsb_{\V{x}_0,\V{\xi},\V{A}},
  \Leq{phi-replica}
\ee
with abbreviations $ \Tr{\V{c}}=\lb \prod_{i=1}^{N}\sum_{c_i=0,1} \rb \delta\lb \sum_{i}c_i-N\rho \rb $ and $ \Tr{\V{x}|\V{c}}=\int\mathrm{d}\V{x} \prod_{i}\lbb (1-c_i)\delta(x_i) + c_i \rbb$ . Overall, to calculate the entropy, we assess the free entropy $g$. The averages over $\V{x}_0,\V{\xi},A$ are taken through the replica method with a replica number $n$, and the internal variables $\V{x}$ are integrated with an additional replica number $\nu$. 

\subsubsection{Assumptions for theoretical computation}
The description above is generic, but for technical reasons we need additional assumptions to complete the computation. The most crucial assumption is applied to the distribution of $A$: Each component of $A$ is assumed to be i.i.d. from $\mc{N}(0,1/N)$. Relaxing this assumption, {\it i.e.} introducing correlations between components, makes the analysis much more complicated. Admittedly, this assumption is not necessarily realistic. However, the purpose of this paper is to provide an analytical basis to understand the variable-selection performance, and we consider that the random-matrix assumption can provide sufficiently nontrivial implications for this purpose.

Thus assuming the absence of correlations among components of $A$, we note that only the average behaviour of the signal components is relevant. According to this observation, without loss of generality, we may assume a factorised prior of the signal vector $\V{x}_0$ as 
\be
P(\V{x}_0)=\prod_{i=1}^{N}\lbb (1-\rho_0)\delta(x_{0i}) + \rho_0P_{0}(x_{0i})\rbb,
\Leq{factorised}
\ee
where $\rho_0=\nnzero_0/N$ is the density of nonzero components and $P_{0}(x)$ is a prior distribution for the nonzero component. Our theoretical computation can be performed for any prior $P_{0}(x)$ having no probability mass at $x=0$, and we keep it unspecified for a while.  

Usually, the entropy function $s(\MSEp)$ enjoys some useful properties such as non-negativity, boundedness, bounded support, concavity, and analyticity. We assume these properties, but as shown below, the concavity and analyticity are partially broken in the present problem. This causes problems in evaluating the parametric form~\NReq{parametric}, but they can be bypassed by some additional considerations when conducting the saddle-point method. The analyticity breaking of the entropy is actually related to algorithmic performances and is one of the central issues discussed in this paper.

\subsubsection{Probabilistic meaning and intrinsic hierarchy of the problem}
Our formulation has a probabilistic meaning which involves two different intrinsic hierarchies in the present problem. We can define the distribution for $\V{c}$ as
\be
P^{(1)}(\V{c}|\mu,\rho,\V{y},A)=\frac{1}{G(\mu|\rho,\V{y},A)}\delta\lb \sum_{i}c_i-N\rho \rb e^{-M\mu \MSEy(\V{c}|\V{y},A)},
\Leq{P1}
\ee
Let us denote the average over $P^{(1)}$ by angular brackets as $\Ave{\cdots}_{\V{c}}$. This distribution is clearly conditioned by $\V{y}$ and $A$. We also define another distribution for $\V{x}$ given $\V{c}$ as
\be
P^{(2)}(\V{x}|\beta,\V{c},\V{y},A)=
\frac{1}{Z}
\prod_{i}\lbb (1-c_i)\delta(x_i)+c_i \rbb
e^{-\frac{\beta}{2} ||\V{y}-A\lb \V{c}\circ\V{x} \rb||_2^2},
\ee
where
\be
Z=\int d \V{x}
\prod_{i}\lbb (1-c_i)\delta(x_i)+c_i \rbb
e^{-\frac{\beta}{2} ||\V{y}-A\lb \V{c}\circ\V{x} \rb||_2^2}
=e^{-\beta \mathcal{H}(\V{c}|\beta,\V{y},A)}
.
\ee
This distribution is conditioned by $\V{c}$ in addition to $\V{y}$ and $A$. Hence, we denote the average over $P^{(2)}$ by $\Ave{\cdots}_{\V{x}|\V{c}}$. The simultaneous average over both $P^{(1)}$ and $P^{(2)}$ is denoted by double angular brackets $\dAve{\cdots}$.

Recalling that $\V{y}$ and $A$ are also random variables, we note that there are three different hierarchies of random variables: $\V{x}$ is conditioned by $\V{c}$ which is conditioned by $\V{y}$ and $A$. This discrimination is a natural consequence of the structure of the present problem\footnote{There is an analogy between the 1st step RSB formulation and the present problem: $(\V{y},A)$ correspond to quenched variables; $\V{c}$ and $\V{x}$ are dynamical variables, but $\V{c}$ determines a pure state and $\V{x}$ is an active dynamical variable inside the pure state; the replica number $\nu$ corresponds to Parisi's breaking parameter.}.

\subsection{Relationship to Bayesian inference using sparsity-inducing priors}
Our formulation is related to a Bayesian framework. To demonstrate the relationship, we introduce the following prior distribution of $\V{x}$ given $\V{c}$:
\be
P(\V{x}|\V{c})=\prod_{i}\lbb (1-c_i)\delta(x_i)+c_i \phi(x_i) \rbb,
\Leq{P(x|c)}
\ee
where $\phi$ is the prior distribution of the nonzero components. As our purpose is to achieve a variable selection, i.e. choosing the best support $\V{c}$, the variable $\V{x}$ can be treated as a hidden variable. Hence in the Bayesian framework, the most rational approach is to sample $\V{c}$ from the following posterior distribution:
\be
P(\V{c}|\V{y})\propto P(\V{c}) \int d\V{x}~P(\V{x}|\V{c})P(\V{y}|\V{x},\V{c}),
\Leq{P(c|y)}
\ee
where $P(\V{y}|\V{x},\V{c})$ is our model distribution of data, which is derived through the noise distribution with variance $\sigma^2$:
\be
P(\V{y}|\V{x},\V{c})\propto e^{-\frac{1}{2\sigma^2}|| \V{y}-A\lb \V{c} \circ \V{x} \rb||_2^2   },
\Leq{P(y|c,x)}
\ee
and the prior distribution of $\V{c}$, $P(\V{c})$, is set to be a uniform distribution at a fixed $\nnzero=N\rho$
\be
P(\V{c})=\binom{N}{N\rho}^{-1}\delta \lb \sum_{i}c_i-N\rho \rb.
\Leq{P(c)}
\ee
This method is optimal when our parameters and model match the true generative process. This matching condition is called the Nishimori condition in physics. Performance at the optimality can be an issue to be studied further; however, in the present paper, we do not pursue this direction. Instead, we perform a maximum a posteriori (MAP) estimation for $\V{x}$ by assuming a (un-normalised) flat prior $\phi(x)=1,~(\forall{x}\in\mR)$. This yields the $\hat{\V{x}}(\V{c}|\V{y})$ expression defined in \Req{x(c)} as the MAP estimator. Hence, the MAP estimation approximates \Req{P(c|y)} as
\be
P(\V{c}|\V{y}) \approx 
P(\V{c})P(\hat{\V{x}}(\V{c}) |\V{c})P(\V{y}|\hat{\V{x}}(\V{c}),\V{c})
\propto
P^{(1)}(\V{c}|1/\sigma^2,\rho,\V{y},A).
\ee
Overall, the posterior distribution of $\V{c}$ defined in \Req{P1} can be regarded as a MAP estimation of \Req{P(c|y)} in the Bayesian framework\footnote{Insightful readers may doubt the probabilistic interpretation of $P(\V{x}|\V{c})$ because it is not possible to normalise $P(\V{x}|\V{c})$ due to the presence of flat prior $\phi$. This inconvenience can be solved by replacing the flat prior with $(1/\sqrt{2\pi L})e^{-x_i^2/(2L)}$ and taking the limit $L\to \infty$, although this causes another problem in model selection according to the marginal likelihood in the usual Bayesian framework. However in the following discussions, any model selection using this criterion is not performed and the entire treatment in the main text does not inherit any of these problems.}.
 
There are two reasons for treating the MAP estimator. The first reason is the computational cost of \Req{P(c|y)}. The computation of \Req{P(c|y)} requires integration with respect to $\V{x}$, which is computationally expensive in general even if we employ versatile approximations such as the MC method. In contrast, the MAP estimator allows us to skip this integration and provides a reasonable estimator \NReq{x(c)}, which is relatively easy to compute. The second reason is the plausibility in matching our inference model with the true generative process, as also discussed in~\Rsec{Introduction}. In many practical tasks for which our regression model is employed, it is not realistic to assume that we have precise knowledge regarding the generative process. Hence, certain mismatch between the inference and generative models is inevitable. This is a common criticism against applying the Bayesian approach to signal processing tasks. On the contrary, the MAP estimator with the uninformative flat prior $\phi(x)=1$ allows us to bypass this problem and can yield better performance than the Bayes estimator \NReq{P(c|y)} in certain mismatching cases. These reasons naturally motivate us to investigate \Req{P1} instead of \Req{P(c|y)}. 

\subsection{Related work}
Here we give a brief summary of several preceding work treating related problems and make it clear how the present paper is similar to or different from those work. 

In~\cite{Guo:05}, Guo and Verd\'u studied a random linear estimation problem in the context of code-division multiple access by using the replica method, as in~\cite{Tanaka:02}. Its MAP estimator was considered in~\cite{Rangan:12} again by using the replica method. Reeves and Gastpar treated similar linear models in the variable selection context as in this paper: They computed the trade-off relation between the measurement rate ($\alpha=M/N$ in our notation) and a distortion quantifying an error rate in the reconstruction of the true support, and derived rigorous lower and upper bounds of the measurement rate to achieve given value of distortion and compared it with the replica result~\cite{Reeves:12-1,Reeves:12-2,Reeves:13}. Another investigation by Reeves and Pfister succeeded to prove that the replica prediction is exact, by tightening the bounds under the assumption that the matrix $A$ is i.i.d. from the normal distribution under the Bayes optimal setting~\cite{Reeves:16}. In~\cite{Tulino:13}, the same problem was investigated by using the replica method except that the matrix $A$ is drawn from rotationally invariant ensembles. These preceding results employing the replica method were conducted under the RS ansatz, and Bereyhi et al. examined the RSB ansatz and showed that the $k$-step RSB with small $k$ can much improve the RS solution's inconsistency appearing when the MAP estimator with $\ell_0$-norm regularization is considered~\cite{Bereyhi:16,Bereyhi:17}. In \cite{Gamarnik:17}, a similar problem with a restriction such that the signal components take only binary values was studied by using the improved second moment method. From a wider viewpoint, generic glassy natures of MAP estimators in inference problems were examined in~\cite{Antenucci:18} by considering a rank-one matrix estimation as an example; the so-called survey propagation, which is a variant of message-passing algorithms taking into account glassy natures, was reported to fail in improving the inference accuracy than the standard message-passing algorithm~\cite{Donoho:09-2,Donoho:10-1,Donoho:10-2,Rangan:10}. In~\cite{Fyodorov:18}, a similar model was considered in the context of reconstructing encrypted signals and investigated by using the replica method; the full-step RSB ansatz was applied and thus the result is expected to be exact and tight.

These study have a connection to this paper in the basic problem setting, but we stress that our present formulation is very different from all of them. We again note that in this paper {\em all possibilities} of the support are examined by computing the entropy curve, while the usual replica analysis treats certain specific supports or estimators only. All the preceding work referred in the previous paragraph fall into this case. For example, the estimator given in \Req{L0optimization}, which is a subject of study in~\cite{Kabashima:09,Rangan:12,Reeves:12-1,Reeves:12-2,Reeves:13,Tulino:13,Bereyhi:16,Bereyhi:17}, corresponds to just one point in the entropy curve: The minimum $\MSEp$ point given $K$.  Meanwhile, the present formulation enables us to simultaneously compute all the other supports, or estimators optimizing the coefficients $\V{x}$ on the chosen support. This leads to the distribution of the MSE which is nothing but the entropy curve. Hence, our formulation provides more global information about the problem, yielding deeper insights both in the information theoretic and algorithmic perspectives, as shown in~\cite{Nakanishi:16} and below. A drawback of this is the analytical procedure much more complicated than the usual replica analysis, as explained in \Rsec{Generating function: Legendre}.

\section{Analytical results}\Lsec{Analytical}

\subsection{Summary of order parameters}
As shown below, the free entropy is characterised by a number of macroscopic order parameters and we summarise them here. The order parameters are defined as
\subbe
\Leq{order parameters}
\be
&&
m=\frac{1}{N}\sum_i  \dAve{ x_{0i} c_i x_i } ,
\\ &&
R=\frac{1}{N}\sum_i  \dAve{ c_i x_i^2 },
\\ &&
Q=\frac{1}{N}\sum_i  \Ave{c_i\Ave{ x_i }_{\V{x}|\V{c}}^2}_{\V{c}},
\\ &&
q=\frac{1}{N}\sum_i  \dAve{ c_ix_i }^{2}.
\ee
\subee
$m$ is the overlap with the true signal $\V{x}_0$ and is relevant to the reconstruction performance of $\V{x}_0$. $R$ and $Q$ describe the powers (per element) of the reconstructed signal, but the latter takes into account the ``thermal'' fluctuation that results from the introduction of $\beta$. These two quantities fall within the limit $\beta \to \infty$, but their infinitesimal difference yields an important contribution 
\be
\chi=\beta(R-Q).
\Leq{limit-chi}
\ee
This is $O(1)$ even in the limit $\beta \to \infty$. The last order parameter $q$ directly reflects the fluctuation of the support vector $\V{c}$ and exhibits the RSB in some parameter regions. 

Using these order parameters, the average value of the input MSE is
\be
\dAve{\MSEx(\V{x}|\V{x}_0)}=\rho_0\POWx-2m+R \to \rho_0\POWx-2m+Q,~(\beta \to \infty),
\Leq{MSEx_ave}
\ee
where $\POWx$ represents the typical power (per non-zero element) of the signal defined by 
\be
\POWx=\int dx_0~x_0^2P_0(x_0).
\ee
The output MSE is computed from \BReq{parametric} once the free entropy is obtained in terms of the order parameters. Hence, we can naturally compute both the input and output MSEs in the present formalism.

\subsection{Expressions of free entropy }\Lsec{Expressions of}

\subsubsection{RS solution}\Lsec{RS solution}
Postponing the details of analysis to \Rsec{g-calculation}, we present the resultant formulas of the free entropy and related quantities. The expression for $g$ in the RS level is given by 
\be
&&
g_{\mathrm{RS}}(\mu|\rho)=
\Extr{\Omega_{\mathrm{RS}}}
\Biggl\{
\tilde{\rho} \rho+\frac{1}{2}\tilde{Q}Q-\frac{\tilde{\chi}\chi}{2\mu}+\frac{1}{2}\tilde{q}q-\tilde{m}m
\no \\ &&
+\rho_0\int dx_0P_0(x_0) \int Dz \log \lb 1+Y^{\mathrm{RS}}_{\tilde{m}}\rb
+(1-\rho_0)\int Dz \log  \lb 1+Y^{\mathrm{RS}}_{0}\rb
\no \\ &&
+\frac{\alpha}{2}\lbb \log\frac{1+\chi}{D_{\mathrm{RS}}}-\frac{\mu (V+q)}{D_{\mathrm{RS}}}\rbb
\Biggr\},
\Leq{g_RS}
\ee
where for simplicity of notation we let $\Omega_{\mathrm{RS}}=\lbb \chi,Q,q,m,\tilde{\rho},\tilde{\chi},\tilde{Q},\tilde{q},\tilde{m} \rbb$, $Dz=dze^{-\frac{1}{2}z^2}/\sqrt{2\pi}$, $V=\rho_0 \POWx+\sigma_{\xi}^2-2m$, and
\be
&&
\Delta_{\mathrm{RS}}=Q-q,
\\ &&
D_{\mathrm{RS}}=1+\chi+\mu\Delta_{\mathrm{RS}}
\\ &&
h^{\mathrm{RS}}_{\tilde{m}}=\tilde{m}x_0+\sqrt{\tilde{q}}z,
\\ && 
Y^{\mathrm{RS}}_{\tilde{m}}=\sqrt{\frac{\tilde{\chi}+\tilde{Q}}{\tilde{Q}+\tilde{q}}}
e^{-\tilde{\rho}+\frac{1}{2}\frac{1}{\tilde{Q}+\tilde{q}}\lb h^{\mathrm{RS}}_{\tilde{m}} \rb^{2}}.
\Leq{hY_RS}
\ee
$Y^{\mathrm{RS}}_{0}$ is obtained by substituting $\tilde{m}=0$ into $Y^{\mathrm{RS}}_{\tilde{m}}$, and $h^{\mathrm{RS}}_0$ is defined similarly. The symbol $\Extr{\Omega}$ denotes the extremisation condition with respect to $\Omega$ coming from the saddle-point method. This extremisation condition yields the following equations of state (EOS):
\subbe
\Leq{EOS_RS}
\be
&&
\hspace{-10mm}
\tilde{\chi}=\alpha \lbb \frac{\mu^2\Delta_{\mathrm{RS}}}{(1+\chi)D_{\mathrm{RS}}}
+\frac{ \mu^2(V+q) }{D_{\mathrm{RS}}^2}\rbb,
\Leq{chitilde_RS}
\\ &&
\hspace{-10mm}
\tilde{Q}=\alpha \lbb \frac{\mu }{D_{\mathrm{RS}}}
-\frac{\mu^2 (V+q) }{D_{\mathrm{RS}}^2}\rbb,
\Leq{Qtilde_RS}
\\ &&
\hspace{-10mm}
\tilde{q}=\alpha \frac{\mu^2 (V+q)}{D_{\mathrm{RS}}^2},
\Leq{qtilde_RS}
\\ &&
\hspace{-10mm}
\tilde{m}= \frac{\alpha \mu}{D_{\mathrm{RS}}},
\Leq{mtilde_RS}
\\ &&
\hspace{-10mm}
\rho=\rho_0\int dx_0 P_0(x_0) \int Dz \frac{Y^{\mathrm{RS}}_{\tilde{m}}}{1+Y^{\mathrm{RS}}_{\tilde{m}}}+(1-\rho_0)\int Dz \frac{Y^{\mathrm{RS}}_{0}}{1+Y^{\mathrm{RS}}_{0}},
\Leq{rho_RS}
\\ &&
\hspace{-10mm}
\chi=\frac{\mu}{\tilde{\chi}+\tilde{Q}}
\lbb 
\rho_0\int dx_0 P_0(x_0)\int Dz \frac{Y^{\mathrm{RS}}_{\tilde{m}}}{1+Y^{\mathrm{RS}}_{\tilde{m}}}+(1-\rho_0)\int Dz \frac{Y^{\mathrm{RS}}_{0}}{1+Y^{\mathrm{RS}}_{0}}
\rbb,
\Leq{chi_RS}
\\ &&
\hspace{-10mm}
Q=\frac{\tilde{\chi}-\tilde{q}}{(\tilde{\chi}+\tilde{Q})(\tilde{Q}+\tilde{q})}
\lbb
\rho_0\int dx_0 P_0(x_0)\int Dz \frac{Y^{\mathrm{RS}}_{\tilde{m}}}{1+Y^{\mathrm{RS}}_{\tilde{m}}}+(1-\rho_0)\int Dz \frac{Y^{\mathrm{RS}}_{0}}{1+Y^{\mathrm{RS}}_{0}}
\rbb
\no \\ &&
\hspace{-10mm}
+\frac{1}{(\tilde{Q}+\tilde{q})^2}\Blbb
\rho_0
\int dx_0 P_0(x_0)\int Dz~
\frac{ \lb h^{\mathrm{RS} }_{\tilde{m}} \rb^2 Y^{\mathrm{RS} }_{\tilde{m}} }{1+Y^{\mathrm{RS}}_{\tilde{m}}}
+
(1-\rho_0)\int Dz~
\frac{\lb h^{\mathrm{RS} }_{0} \rb^2  Y^{\mathrm{RS}}_{0} }{1+Y^{\mathrm{RS}}_{0}}
\Brbb,
\Leq{Q_RS}
\\ && \hspace{-10mm}
q=
\frac{1}{(\tilde{Q}+\tilde{q})^2}
\Blbb
\rho_0\int dx_0 P_0(x_0)
\int Dz~  \lb \frac{ h^{\mathrm{RS} }_{\tilde{m}}  Y^{\mathrm{RS}}_{\tilde{m}}}{1+Y^{\mathrm{RS}}_{\tilde{m}}} \rb^2
+
(1-\rho_0)\int Dz~
\lb \frac{h^{\mathrm{RS} }_{0}  Y^{\mathrm{RS}}_{0}}{1+Y^{\mathrm{RS}}_{0}} \rb^2
\Brbb
,
\Leq{q_RS}
\\ &&
\hspace{-10mm}
m=
\frac{1}{\tilde{Q}+\tilde{q}}\lbb
\rho_0\int dx_0 P_0(x_0) \int Dz \frac{x_0 h^{\mathrm{RS}}_{\tilde{m}} Y^{\mathrm{RS}}_{\tilde{m}}}{1+Y^{\mathrm{RS}}_{\tilde{m}}}
\rbb
.
\Leq{m_RS}
\ee
\subee
Note that all tilde variables are conjugates of their respective variables and are introduced to expand delta functions with the Fourier transform as shown in \Rsec{g-calculation}. From \Req{EOS_RS}, we obtain some simple relations
\subbe
\be
&&
\chi=\frac{\rho}{\alpha-\rho},
\\ &&
\tilde{\chi}+\tilde{Q}=\mu (\alpha-\rho), 
\\ &&
\tilde{Q}+\tilde{q}=\frac{\alpha \mu}{D_{\mathrm{RS}}}=\tilde{m}, 
\\ &&
\tilde{\chi}-\tilde{q}=\frac{\alpha \mu^2 \Delta_{\mathrm{RS}}}{(1+\chi)D_{\mathrm{RS}}},
\\ &&
\MSEp(\mu|\rho)=-\frac{1}{\alpha}\Part{g(\mu|\rho)}{\mu}{}=\frac{\tilde{\chi}}{2\alpha \mu^2}.
\ee 
\Leq{simples_RS}
\subee

\subsubsection{RSB solution and the instability of the RS solution}\Lsec{RSB solution}
The RS solution can be inaccurate when the configuration space of $\V{c}$ exhibits spontaneous breaking into many locally separated components. In such a situation, the RSB ansatz should be adopted. The RSB solution is categorised by the level of the emerging hierarchical structure of the separation. In the simplest case called the 1st step RSB (1RSB) solution, only one level of hierarchy is taken and we examine this in the present paper. The 1RSB solution is actually sufficient to expose the instability of the RS solution and hence is sufficient to achieve the present purpose of obtaining implications to local search algorithms.

We postpone the detailed derivation of the 1RSB solution to \Rsec{g-calculation}. The explicit 1RSB formula involving $g$ is given as
\be
&& \hspace{-14mm}
g_{\mathrm{1RSB}}(\mu,\tau;\rho)=
\Extr{\Omega_{\mathrm{1RSB}}}
\Biggl\{
\tilde{\rho} \rho+\frac{1}{2}\tilde{Q}Q-\frac{\tilde{\chi}\chi}{2\mu}-\frac{1}{2}(\tau-1)\tilde{q}_1q_1+\frac{1}{2}\tau\tilde{q}_0q_0-\tilde{m}m
\no \\ && \hspace{-14mm}
+\frac{\rho_0}{\tau}\int dx_0P_0(x_0) \int Dz_0 \log \int Dz_1 \lb 1+Y^{\mathrm{1RSB}}_{\tilde{m}}\rb^{\tau}
+\frac{1-\rho_0}{\tau}\int Dz_0 \log  \int Dz_1 \lb 1+Y^{\mathrm{1RSB}}_{0}\rb^{\tau}
\no \\ && \hspace{-14mm}
+\frac{\alpha}{2}
\lbb 
\log\frac{1+\chi}{D_1}
+\frac{1}{\tau}\log\frac{D_1}{D_0}
-\frac{\mu (V+q_0)}{D_0}
\rbb
\Biggr\},
\Leq{g_1RSB}
\ee
where $\tau$ is Parisi's breaking parameter, $\Omega_{\mathrm{1RSB}}=\lbb \chi,Q,q_1,q_0,m,\tilde{\rho},\tilde{\chi},\tilde{Q},\tilde{q}_1,\tilde{q}_0,\tilde{m},\tau \rbb$, and
\be
&&
\Delta_1=Q-q_1,~\Delta_0=q_1-q_0,
\\ &&
D_1=1+\chi+\mu\Delta_{1},
\\ &&
D_0=1+\chi+\mu\Delta_{1}+\tau\mu\Delta_0,
\\ &&
h^{\mathrm{1RSB}}_{\tilde{m}}=\tilde{m}x_0+\sqrt{\tilde{q}_1-\tilde{q}_0}z_1+\sqrt{\tilde{q}_0}z_0.
\\ && 
Y^{\mathrm{1RSB}}_{\tilde{m}}=\sqrt{\frac{\tilde{\chi}+\tilde{Q}}{\tilde{Q}+\tilde{q}_1}}
e^{-\tilde{\rho}+\frac{1}{2}\frac{1}{\tilde{Q}+\tilde{q}_1}\lb h^{\mathrm{1RSB}}_{\tilde{m}} \rb^{2}},
\ee
As in the RS case, $h^{\mathrm{1RSB}}_0$ and $Y^{\mathrm{1RSB}}_{0}$ are given by inserting $\tilde{m}=0$ into $h^{\mathrm{1RSB}}_{\tilde{m}}$ and $Y^{\mathrm{1RSB}}_{\tilde{m}}$, respectively. The corresponding EOS are involved and are given in \Rsec{g-calculation}.

By examining the 1RSB solution, we can determine the instability points of the RS solution. Empirically, two types of instabilities are known to appear in a wide range of systems:
\begin{description}
\item[Global instability]{The RS solution is locally stable but there emerges another solution involving exponentially many metastable states, which induces the so-called random first-order transition (RFOT). We thus call the associated instability RFOT instability in this paper.}
\item[Local instability]{The local instability of the RS solution, which can be signaled by expanding the 1RSB solution with respect to $\Delta_0$ and observing its coefficient. This is also known as the de Almeida-Thouless (AT) instability. }
\end{description}
According to these empirical facts, we derive two instability conditions below. 

The RFOT instability is known to emerge at $\tau=1$ and can be detected in an easy manner as follows. For $\tau=1$, we can identify $q_0$ and $\tilde{q}_0$ with $q$ and $\tilde{q}$ in the RS solution, respectively, because their EOS formally accord with each other. As well, the 1RSB EOS of all other order parameters except for $q_1$ and $\tilde{q}_1$ become identical to their corresponding RS EOS. Hence, we should compute $q_1$ and $\tilde{q}_1$ on top of the RS solution and examine whether the nontrivial solution $q_1\neq q_0=q$ exists or not. The equations to be solved are 
written in terms of $\Delta_0$ and $\tilde{\Delta}_0\equiv \tilde{q}_1-\tilde{q}_0=\tilde{q}_1-\tilde{q}$ as follows
\be
&&
\tilde{\Delta}_0=\frac{\alpha \mu^2}{1+\chi+\mu(Q-q)}\frac{\Delta_0}{1+\chi+\mu(Q-q-\Delta_0)},
\Leq{Delta0tilde_1RSB}
\\ &&
\Delta_0=\frac{1}{\lb \tilde{Q}+\tilde{q}+\tilde{\Delta}_0 \rb^2}
\Blbb
\rho_0\int dx_0 P_0(x_0)
\int Dz_0~  
\frac{\int Dz_1  
\lb h^{\mathrm{1RSB} }_{\tilde{m}} Y^{\mathrm{1RSB}}_{\tilde{m}} \rb^2
(1+Y^{\mathrm{1RSB}}_{\tilde{m}})^{-1}
}{
1+Y^{\mathrm{RS}}_{\tilde{m}}
}
\no \\ &&
+
(1-\rho_0)\int Dz_0~
\frac{\int Dz_1  
\lb h^{\mathrm{1RSB} }_{0} Y^{\mathrm{1RSB}}_{0} \rb^2
(1+Y^{\mathrm{1RSB}}_{0})^{-1}
}{
1+Y^{\mathrm{RS}}_{0}
}
\Brbb
-q.
\Leq{Delta0_1RSB}
\ee
In these equations, we should read $\tilde{q}_1=\tilde{q}+\tilde{\Delta}_0$ in $h^{\mathrm{1RSB} }$ and $Y^{\mathrm{1RSB} }$. The trivial RS solution $\Delta_0=0$ always exists and the question is whether a nontrivial solution $\Delta_0\neq 0$ exists or not. Such a nontrivial solution is absent for the low $\mu$ region but is present at sufficiently large values of $\mu$. The lowest value of $\mu$ for which the nontrivial solution exists defines the RFOT point $\mu_{\rm RFOT}$.

The AT instability is observed by examining the presence of a nontrivial solution around $\Delta_0=0$. This can be accomplished by expanding the right hand side of \Req{Delta0_1RSB} with respect to $\Delta_0$ up to the first order after inserting \Req{Delta0tilde_1RSB} into $\tilde{\Delta}_0$. If the coefficient of the first-order term is greater than unity, a nontrivial solution emerges. This condition is written using the RS solution only, because the small $\Delta_0$ limit implies that the order parameters $q_1$ and $q_0$ of the 1RSB solution can be identified as $q$ in the RS solution, and the corresponding tilde variables can also be identified. The explicit stability condition of the RS solution against the AT instability is given by
\be
&&
\alpha > \rho_0\int dx_0 P_0(x_0) \int Dz 
\lb 
\frac{Y^{\mathrm{RS}}_{\tilde{m}}}{1+Y^{\mathrm{RS}}_{\tilde{m}}}
+
\frac{\lb h^{\mathrm{RS}}_{\tilde{m}}\rb^2}{\tilde{Q}+\tilde{q}}
\frac{ Y^{\mathrm{RS}}_{\tilde{m}}}{\lb 1+Y^{\mathrm{RS}}_{\tilde{m}} \rb^2}
\rb^2
\no \\ &&
+(1-\rho_0)\int Dz 
\lb 
\frac{Y^{\mathrm{RS}}_{0}}{1+Y^{\mathrm{RS}}_{0}}
+
\frac{\lb h^{\mathrm{RS}}_{0} \rb^2}{\tilde{Q}+\tilde{q}}
\frac{ Y^{\mathrm{RS}}_{0} }{\lb 1+Y^{\mathrm{RS}}_{0} \rb^2}
\rb^2.
\Leq{AT}
\ee
This condition always holds for sufficiently low values of $\mu$. The lowest $\mu$ value violating \Req{AT} indicates the AT transition point $\mu_{\rm AT}$. 

These two instabilities are known to affect the performance of local search algorithms. The origin of this affliction by the RFOT transition is clear---there emerge exponentially many local minima and thus the search/dynamics is easily trapped in one of those states; the typical trapping state will be the most numerous one and will be far from the true global minimum. Each trapping state in this case is separated by high energy barriers and hence escaping will take an exponentially long time~\cite{mezard2009information}. Meanwhile, the influence of the AT instability is less trivial than the RFOT. According to the standard physical picture, when the AT instability occurs, the structure of the configuration space of $\V{c}$ has many saddle-point-like structures, which leads to a complicated critical slowing down of the dynamics and thus the performance of local search algorithms will be strongly degraded. However, this degradation will be less serious compared to that caused by the RFOT transition, because in the AT case it is considered that there exist certain directions along which the system can escape from a local saddle point. This may take a long time, as the energy landscape will be very flat along the escape directions owing to the saddle-point nature, but will be shorter than the RFOT case, where an exponentially long time is required. These descriptions have actually been supported by numerical simulations of some metaheuristic algorithms in several optimisation problems~\cite{mezard1987spin,Cugliandolo:94,Cugliandolo:95,Cugliandolo:08,Montanari:04,Krzakala:13}. We, however, stress that there are still many unclear points about the dynamics around the AT instability and further investigations using concrete algorithms, such as Monte-Carlo methods or message-passing algorithms, are desired.

As will be seen in the next section, we have several characteristic regions depending on the parameters $\rho$ and $\sigma_{\xi}^2$. In some regions, the RSB transitions induced by AT and RFOT instabilities occur, which makes it difficult for the system's dynamics to converge to the equilibrium distribution. In other regions, the RS solution is always stable when $\mu$ is changed. However, the RS-stable regions are separated into two small regions, one has no phase transitions and thus the metaheuristic algorithm can work well, and the other has another 1st order transition which prevents the algorithm from approaching the global minimum. These descriptions will be actually confirmed by numerical experiments of a metaheuristic algorithm in \Rsec{Monte Carlo-based}. 

\subsubsection{Some simple limits}\Lsec{Some simple limits}
To check our replica results, we summarise some simple solutions obtained at particular limits below.
\paragraph{High temperature solution}
A trivial solution in the high temperature limit, $T=\mu^{-1} \to \infty$, is derived from \Req{EOS_RS}. From simple algebra based on \Reqs{EOS_RS}{simples_RS}, we obtain
\be
&&
\frac{Y^{\mathrm{RS}}_{\tilde{m}}}{1+Y^{\mathrm{RS}}_{\tilde{m}}}=\frac{Y^{\mathrm{RS}}_{0}}{1+Y^{\mathrm{RS}}_0}=\frac{e^{-\tilde{\rho}}}{1+e^{-\tilde{\rho}}}=\rho,
\Leq{rho-mu=0}
\ee
Accordingly,
\subbe
\Leq{OP-mu=0}
\be
&&
m= \rho \rho_0 \POWx,
\\ &&
q=\frac{\rho^2}{\alpha-\rho^2}\lbb (1+\alpha-2\rho)\rho_0\POWx+\sigma_{\xi}^2 \rbb,
\\ &&
Q=\frac{\rho}{\alpha-\rho}\lbb (1+\alpha-2\rho)\rho_0\POWx+\sigma_{\xi}^2 \rbb.
\ee
\subee
Using the relation $\tilde{\rho}=-\log (\rho/(1-\rho))$, we have
\subbe
\Leq{mu=0}
\be
&&
g(\mu=0,\rho)=-\rho \log \rho -(1-\rho)\log (1-\rho)\equiv H_2(\rho),
\\ &&
\MSEp(\mu=0,\rho)=\frac{1}{2}\frac{\alpha-\rho}{\alpha} \lb (1-\rho)  \rho_0\POWx +\sigma_{\xi}^2 \rb,
\\ &&
s(\mu=0,\rho)=H_2(\rho),
\ee
\subee
where $H_2(\rho)$ is the binary entropy, giving a reasonable result. 

The local stability of this solution can be checked by substituting \Reqs{rho-mu=0}{OP-mu=0} into \Req{AT}, which yields 
\be
\alpha > \rho^2.
\ee
This always holds in the meaningful setup of $\rho\leq 1$ and $\rho\leq \alpha$, and hence the high temperature solution is stable.

\paragraph{Perfect reconstruction in the noiseless limit}
Of particular interest for the noiseless limit $\sigma_{\xi}=0$ is whether we can achieve the perfect reconstruction of $\V{x}_0$. We call the corresponding solution the perfect reconstruction (PR) solution, which is defined by
\be
c_i=1,~(\forall{i}~{\rm s.t.}~|x_{0i}|_0=1).
\Leq{PRsupport}
\ee
Note that the support components outside the true support may take unity: $c_i$ can be $1$ even if  $|x_{0i}|_0=0$. This is because the coefficients $\hat{\V{x}}(\V{c})$ outside the true support become automatically zero as a result of the optimization if all the components of the true support is covered as \Req{PRsupport}, leading to the vanishing MSEs. In other words, the PR solution always exists if the estimated nonzero density is greater than or equal to the true one, $\rho \geq \rho_0$. This seemingly indicates that it is safer to overestimate the nonzero density $\rho$. This is, however, not necessarily true because larger estimates of $\rho$ tend to involve unfavourable phase transitions, as shown below. 

The output MSE $\MSEp$ of the PR solution is zero and the entropy becomes
\be
&&
s(\MSEp=0)=\frac{1}{N}\log \binom{N(1-\rho_0)}{N(\rho-\rho_0)}
\no \\ &&
=
(1-\rho_0)\log(1-\rho_0)-(\rho-\rho_0)\log (\rho-\rho_0)
-(1-\rho)\log(1-\rho)
\equiv s_{\rm PR}.
\Leq{s_PR}
\ee
We can actually find this PR solution in the limit $\mu \to \infty$ of our RS formula \NReq{EOS_RS}. To derive this, we have to carefully treat the scaling of $V+q$ and $\Delta_{\mathrm{RS}}$ within that limit. We first assume the following scaling:  
\be~
\mu (V+q) \to 0,~
\mu \Delta_{\mathrm{RS}} \to 0,~
{\rm and }~\tilde{\rho}=O(1),~(\mu \to \infty).
\Leq{PRA}
\ee
The consistency of this assumption is confirmed after the computation. Using this and \Req{simples_RS}, we can easily determine the following limits
\be
&& 
\frac{   \tilde{\chi}+\tilde{q} }{ \tilde{Q}+\tilde{q} } \to 1,
\frac{ \lb h_{\tilde{m}}^{\mathrm{RS}} \rb^2  }{ \tilde{Q}+\tilde{q} } 
\to (\alpha-\rho)\mu x_0^2  +2\sqrt{ (\alpha-\rho)(V+q) }\mu x_0 z ,~
\frac{ \lb h_{\tilde{0}}^{\mathrm{RS}} \rb^2  }{ \tilde{Q}+\tilde{q} } \to 0.
\Leq{factors-perfect}
\ee
These relations mean that $Y^{\mathrm{RS}}_{\tilde{m}}$ diverges or vanishes depending on the values of $x_0$ and $z$ while $Y^{\mathrm{RS}}_{0}$ converges to $e^{-\tilde{\rho}}$. The vanishing region of $Y^{\mathrm{RS}}_{\tilde{m}}$ is expressed in terms of $z$ as
\be
z > \frac{1}{2}\sqrt{\frac{\alpha}{(1+\chi)(V+q)}}x_0  & (x_0>0),    \\
z < \frac{1}{2}\sqrt{\frac{\alpha}{(1+\chi)(V+q)}}x_0   &   (x_0<0).   
\ee
As we have assumed $(V+q)\to 0~(\mu \to \infty)$, this vanishing region rapidly shrinks and does not provide any meaningful contribution. Hence, we may treat $Y^{\mathrm{RS}}_{\tilde{m}}$ as a diverging factor in all contributing regions. This consideration yields, from \Req{rho_RS}:
\be
e^{-\tilde{\rho}} \to \frac{\rho-\rho_0}{1-\rho}.
\Leq{rho-perfect}
\ee
Additionally, some algebraic operations from \Reqs{EOS_RS}{simples_RS} lead to
\be
&&
(V+q)=
\frac{\rho_0}{\alpha-\rho_2}
\Blbb
\alpha
\int dx_0 P_0(x_0)x_0^2 \int Dz (1+Y^{\rm RS}_{\tilde{m}})^{-2}
\no \\ &&
+2\sqrt{\alpha(V+q)}
\int dx_0 P_0(x_0) \int Dz~x_0 zY_{\tilde{m}}(1+Y^{\rm RS}_{\tilde{m}})^{-2}
\Brbb,
\ee
where 
\be
\rho_2\equiv 
\rho_0 \int dx_0P_0(x_0) \int Dz z^2 \lb \frac{Y^{\rm RS}_{\tilde{m}}}{1+Y^{\rm RS}_{\tilde{m}}}\rb^2
+
(1-\rho_0)\int Dz z^2 \lb \frac{Y^{\rm RS}_{0}}{1+Y^{\rm RS}_{0}}\rb^2,
\Leq{rho_2}
\ee
which yields a finite contribution in the limit $\mu \to \infty$. The divergence of $Y^{\mathrm{RS}}_{\tilde{m}}$ implies that 
\be
(V+q) \propto \int dx_0 P_0(x_0) x_0^2 \int Dz \lb Y^{\rm RS}_{\tilde{m}} \rb^{-2}.
\ee
The precise scaling of the right hand side strongly depends on the choice of the prior distribution $P_0(x_0)$. For examples, if it is Gaussian $P_0(x_0)=\mathcal{N}(0,\POWx)$, then $(V+q)=O(\mu^{-\frac{3}{2}})$; if the signal strength is a constant at $P_0(x_0)=\delta(x_0-C)$ with $C\neq 0$, then $(V+q)$ exponentially decays as $\mu$ increases. Similar calculations apply to $\Delta_{\mathrm{RS}}$, which involves the same scaling of $\Delta_{\mathrm{RS}}$ as $(V+q)$. These scalings are consistent with the assumed ones \NReq{PRA}. 

Summarising these calculations in conjunction with \Reqs{EOS_RS}{simples_RS}, we determine that
\be
&&
\MSEp(\mu)=O\lb V+q \rb \to 0.
\Leq{eps_scaling}
\ee
This implies that the free entropy $g$ coincides with the entropy in the limit $\mu\to \infty$. Substituting the scalings obtained so far into \Req{g_RS}, we find that
\be
\lim_{\mu\to \infty}s(\mu)=\lim_{\mu\to \infty}g(\mu)=s_{\rm PR}.
\ee
The limiting behaviours of $\Delta_{\mathrm{RS}}$ and $V+q$ cause the input MSE to vanish:
\be
V+q\to V+Q=\dAve{\MSEx} \to 0.
\ee
Hence, the PR solution is successfully derived.

The local stability~\NReq{AT} of this PR solution should be checked. Inserting \Reqs{factors-perfect}{rho-perfect} into \Req{AT} results in the stability condition
\be
(\alpha-\rho_0)(1-\rho_0)> (\rho-\rho_0)^2.
\Leq{PRstability}
\ee
Hence, the PR solution is stable as long as $\rho_0 \leq \rho \leq \alpha$ where the PR solution exists. Furthermore, this stability is considered to be rather ``robust''. It is physically reasonable that the RS solution is stable even for $\rho < \rho_0$ if $\rho$ is sufficiently close to $\rho_0$ because the inequality \NReq{PRstability} is safely satisfied even at $\rho=\rho_0$. This will be demonstrated in the phase diagram shown below. Note that \Req{PRstability} is just a necessary condition and not a  sufficient one. 

The stability of the RS solution at and around the PR solution has an algorithmic implication to the $\ell_0$-minimisation approach~\cite{Wang:11,Chartrand:13,Zheng:15}. Namely, local search algorithms such as the message-passing algorithm will not be degraded by the rugged energy landscape if the initial condition is sufficiently close to the PR solution and hence the predictions based on the RS computation will be precise.

\subsection{Entropy curve and phase diagram}\Lsec{Entropy curve}
To obtain a concrete result, in the following we set the prior distribution of the nonzero component as a Gaussian:
\be
P_0(x_0)=\mathcal{N}(0,\POWx).
\Leq{Gaussian prior}
\ee
Owing to this assumption, the double integrations with respect to $x_0$ and $z$ in \Req{EOS_RS} can be merged into a Gaussian integration by a variable transform $(\tilde{m}x_0+\sqrt{\tilde{q}}z)\to \sqrt{\tilde{q}+\tilde{m}^2\POWx}z$. The EOS \NReq{EOS_RS} should be appropriately transformed. Except for \Req{m_RS}, this can be accomplished by neglecting the integration $\int dx_0 P_0(x_0)$ and reading 
\be
&&
h^{\mathrm{RS}}_{\tilde{m}}=\sqrt{\tilde{q}+\tilde{m}^2\POWx}z,
\ee
in the EOS and $Y^{\mathrm{RS}}_{\tilde{m}}$. \BReq{m_RS} involves the cross term of $x_0$ and $h^{\mathrm{RS}}_{\tilde{m}}$ so some special care is needed. A slight calculation yields
\be
m=
\frac{\tilde{m}\POWx}{ \tilde{Q}+\tilde{q} }\rho_0\int Dz \frac{z^2 Y^{\mathrm{RS}}_{\tilde{m}}}{1+Y^{\mathrm{RS}}_{\tilde{m}}}
\Leq{m_RS2}.
\ee
Furthermore, the signal variance $\POWx$ is set to be $\POWx=1/\rho_0$ to fix the per-element signal power to unity, $\rho_0\POWx=1$.

\subsubsection{Noiseless case}
We start from the noiseless limit $\sigma_{\xi}=0$. \Rfig{entropy_noiseless} shows the plot of entropy $s(\MSEp)$ versus $\MSEp$ for several different values of $\rho$. Other parameters are common and are set as $\alpha=0.5$ and $\rho_0=0.2$. The drawn curves are based on the RS solution, and the RSB solution is only used to indicate the respective instability point. 
\begin{figure}[htbp]
\begin{center}
 \includegraphics[width=0.48\columnwidth]{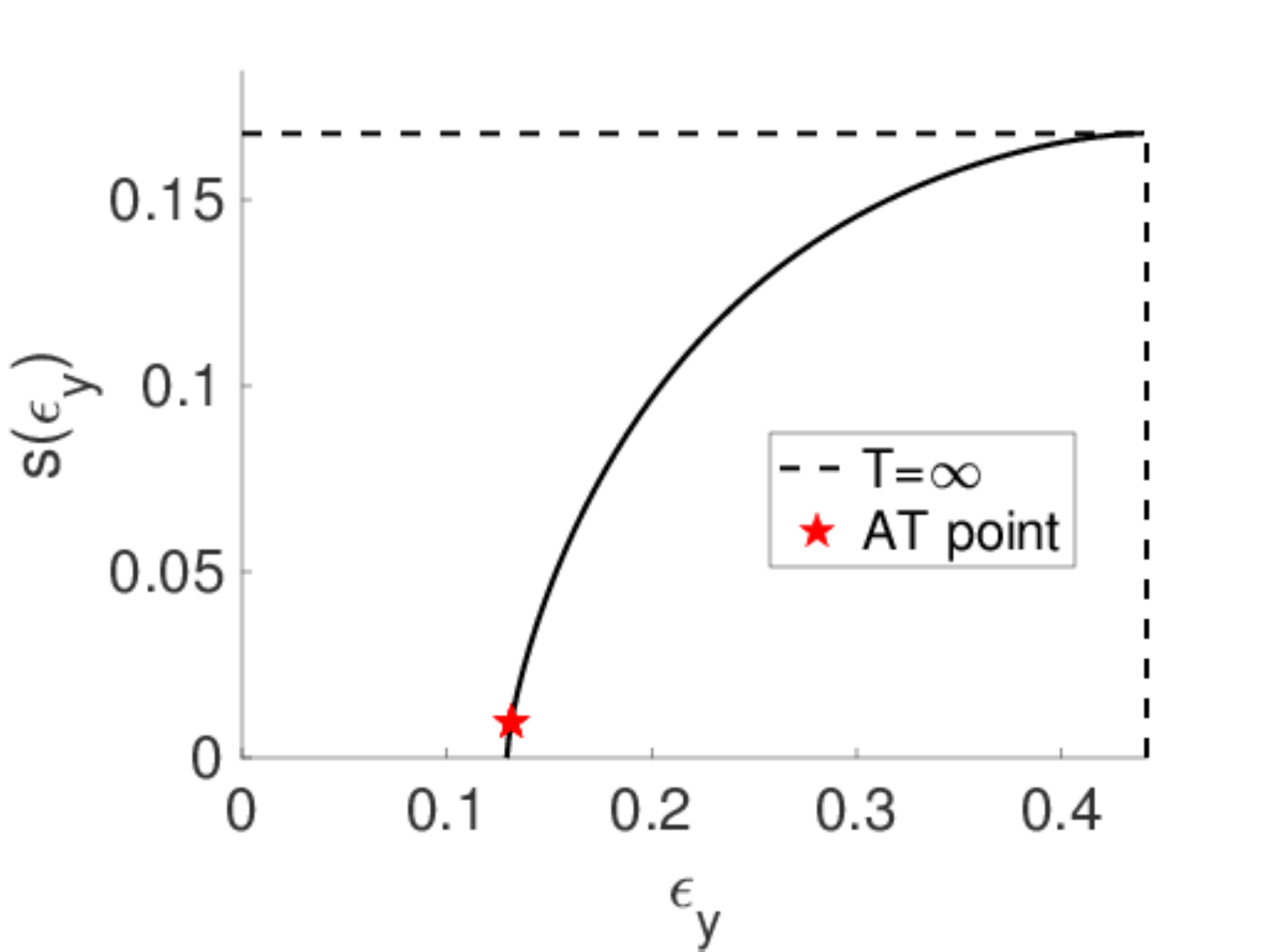}
 \includegraphics[width=0.48\columnwidth]{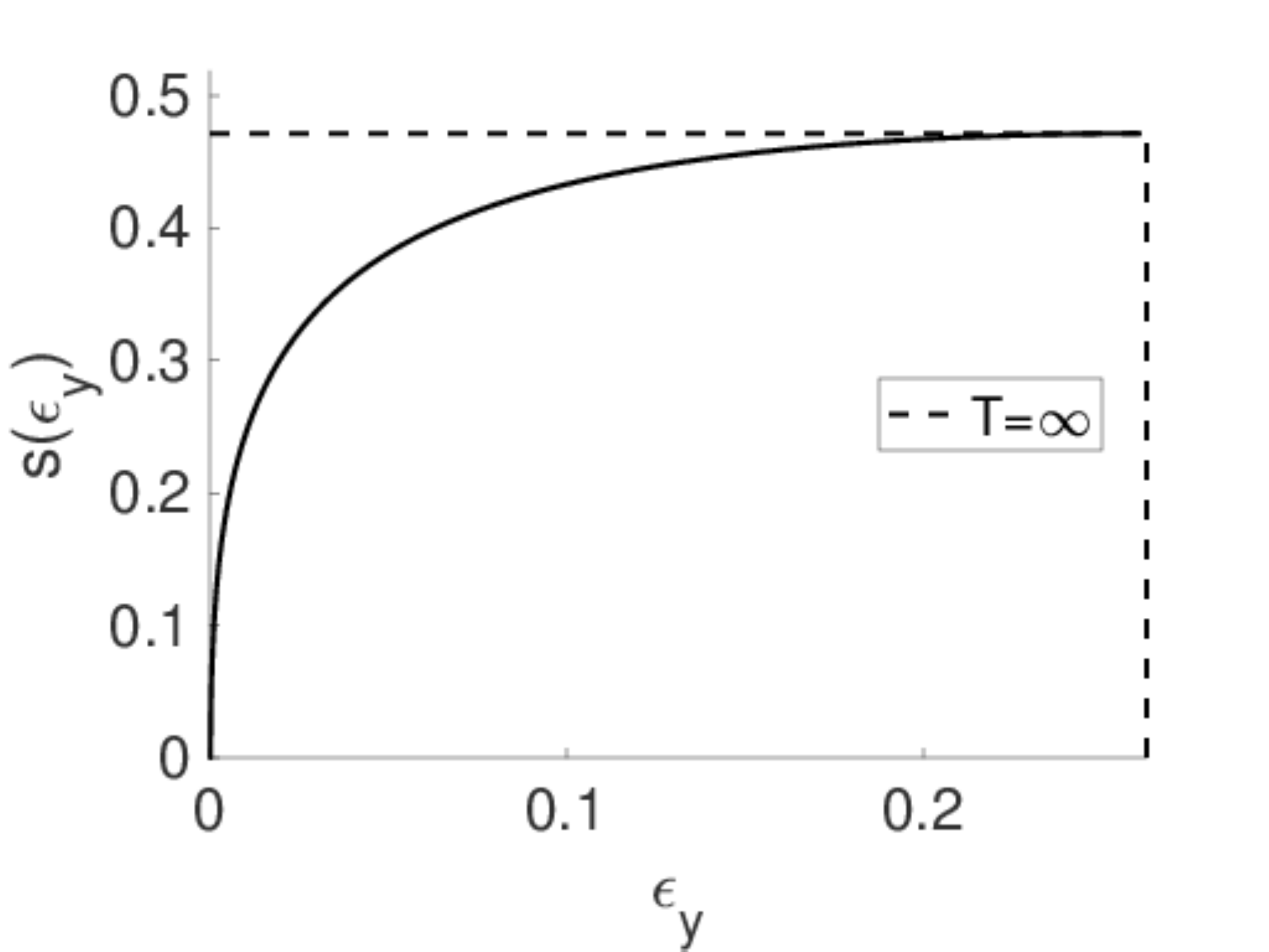}
 \includegraphics[width=0.48\columnwidth]{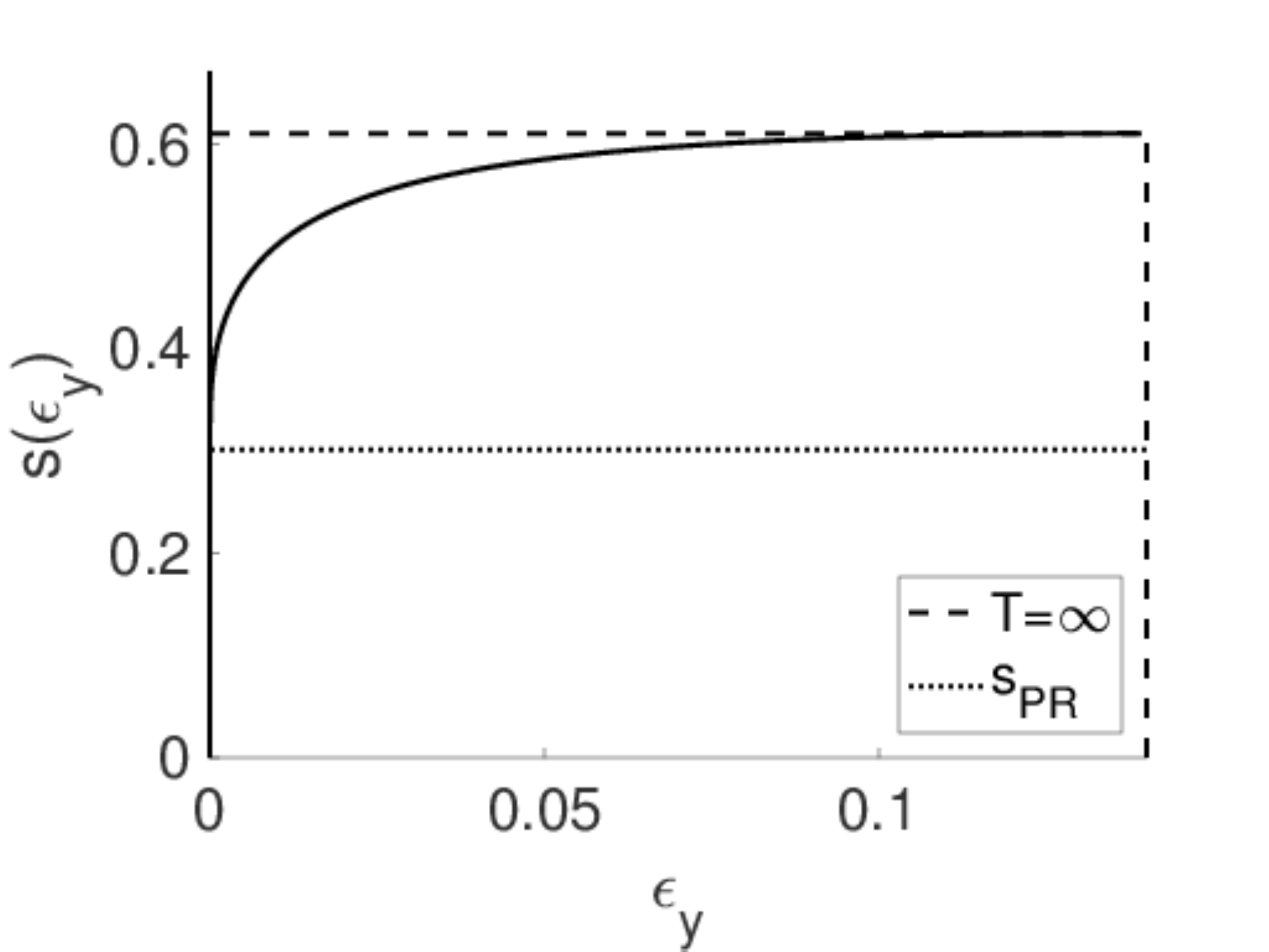}
 \includegraphics[width=0.48\columnwidth]{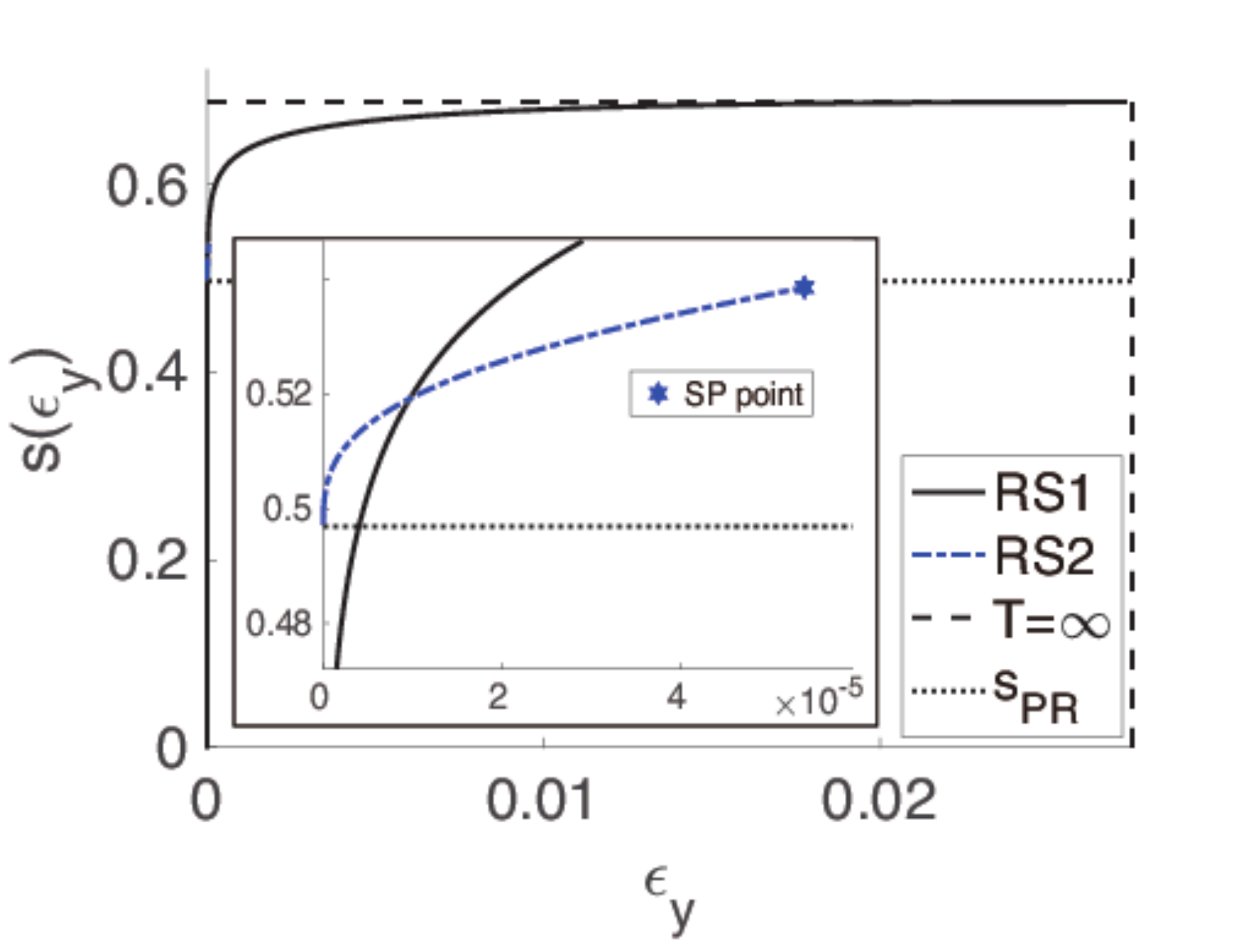}
\caption{Entropy plotted against the output MSE $\MSEp$, which is represented by the solid black line, for $\alpha=0.5$ and $\rho_0=0.2$ in the noiseless limit $\sigma_{\xi}=0$. The nonzero density is different among the four panels: $\rho=0.04<\rho_{\mathrm{AT}}$ (upper left), $\rho_{\mathrm{AT}}< \rho=0.18 <\rho_0$ (upper right), $\rho_{0}< \rho=0.3 < \rho_{\rm SP}$ (lower left), and $\rho_{\rm SP}< \rho=0.45$ (lower right). In the last case, there are two entropy curves denoted by the solid and dashed dotted lines, which correspond to two different phases, and the inset is a close-up around $(\MSEp,s(\MSEp))=(0,s_{\mathrm{PR}})$. Broken and dotted lines denote the high temperature solution \NReq{mu=0} and the PR solution \NReq{s_PR}, respectively. }
\Lfig{entropy_noiseless}
\end{center}
\end{figure}
As seen from \Rfig{entropy_noiseless}, we observe four different characteristic behaviours of $s(\MSEp)$. For small $\rho$, the AT instability occurs at small $\MSEp$, where the full-step RSB will be needed to correctly describe that region. This RS unstable region vanishes for larger $\rho$, defining a critical value $\rho_{\mathrm{AT}}(<\rho_0)$. In the region $\rho_{\mathrm{AT}} < \rho < \rho_0$, the RS solution is stable for the entire $\MSEp$ region having the nonnegative entropy $s(\MSEp)\geq 0$. A somewhat surprising fact in this region is that the entropy crisis (EC), $s(\MSEp)=0$, occurs at a finite critical value of $\mu_{\rm EC}(<\infty)$. As it approaches $\rho_0$, this critical value $\mu_{\rm EC}(\rho)$ diverges and the entropy curve is continuously connected to the PR solution in the region $\rho_0 \leq \rho$. For a wide range of $\rho(\geq \rho_0)$, the RS solution is again stable for the entire region of $\MSEp$. However, at larger values of $\rho$, there emerges a new phase transition, defining another critical value, $\rho_{\rm SP}$. For $\rho_{\rm SP} \leq \rho$, the PR solution is detached from the high temperature limit, and there are two different branches for the small $\MSEp$ or large $\mu$ region. Two critical temperature points are accordingly defined -- the spinodal point $T_{\rm SP}$ at which the low-temperature branch connected to the PR solution vanishes, and the first-order transition point $T_{\rm F}$ at which $g$ values of the two branches coincide. To locate the corresponding critical temperatures at a given $\rho(>\rho_{\rm SP})$, we plot $g$ against $T=1/\mu$ in \Rfig{g_noiseless}. As a reference, the output MSE $\MSEp$ is also plotted against $T$ around the critical temperatures in the right panel. 
\begin{figure}[htbp]
\begin{center}
 \includegraphics[width=0.49\columnwidth]{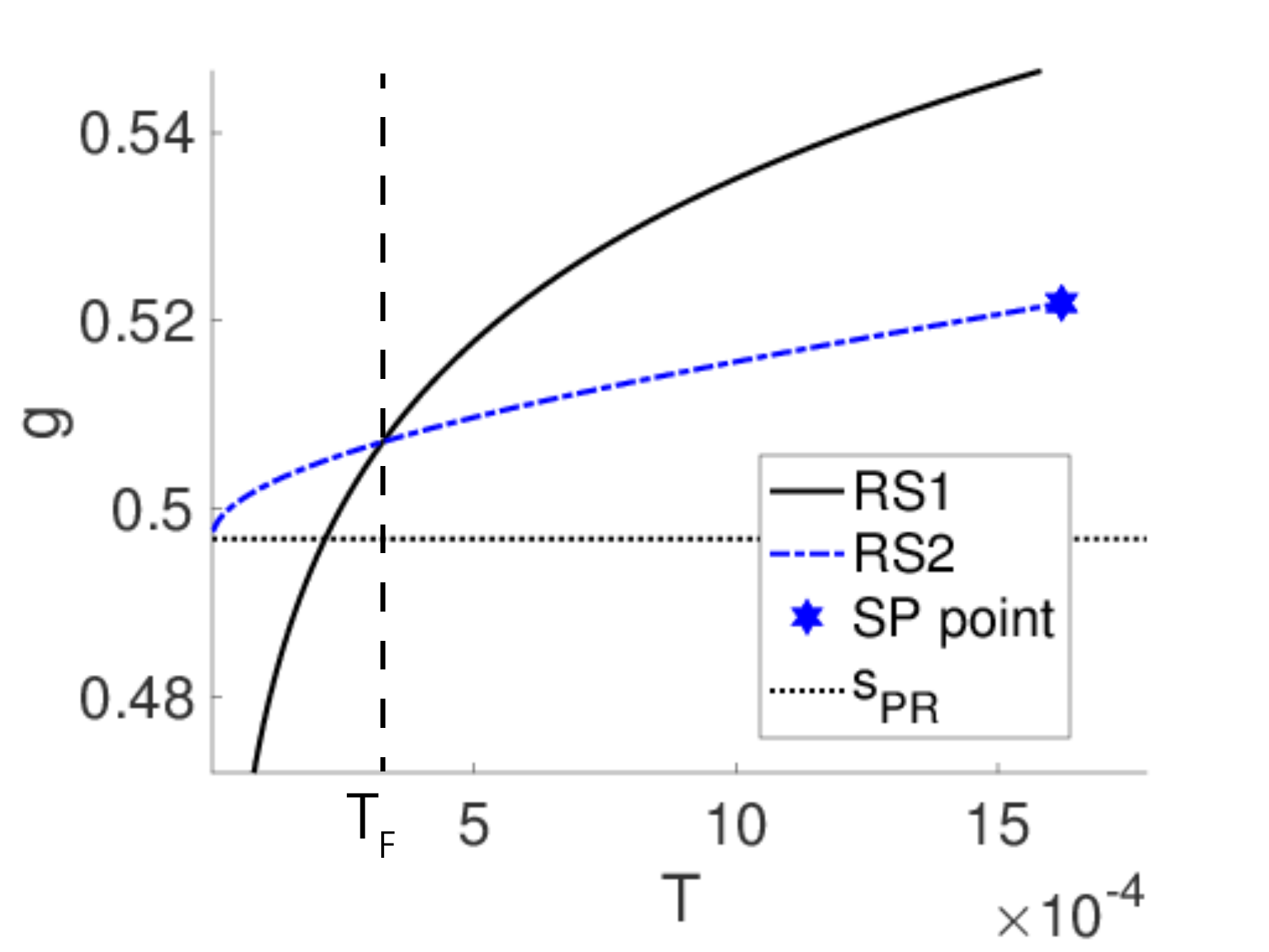}
 \includegraphics[width=0.49\columnwidth]{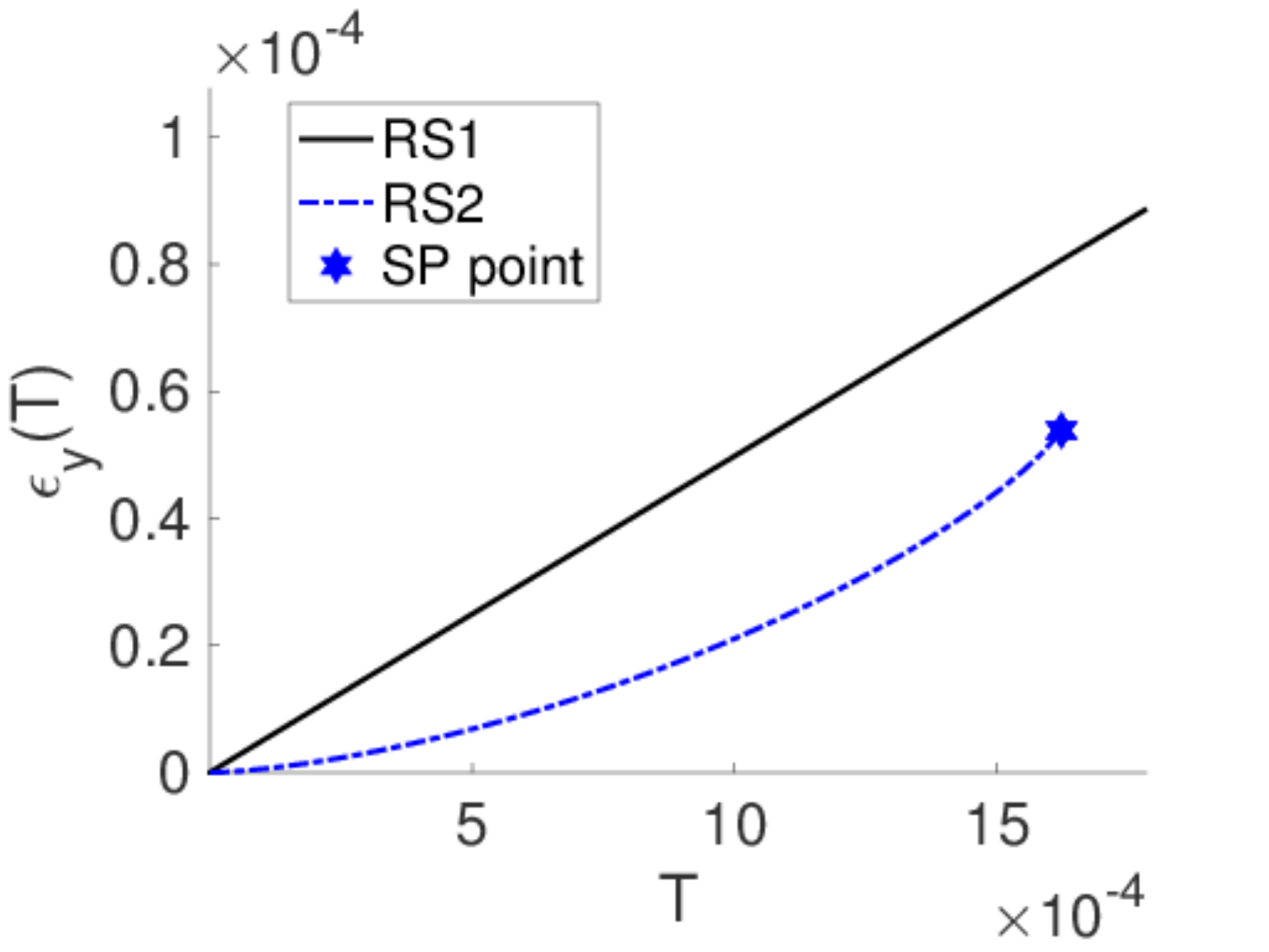}
\caption{Plots of the free entropy $g$ (left) and the output MSE $\MSEp$ (right) against the temperature $T=1/\mu$ around $T_{\rm F}\approx 3.3\times 10^{-4}$ (vertical broken line) defined by the intersection of two branches of $g$ represented by black solid and blue dashed dotted lines. The parameters are $\rho=0.45$, $\alpha=0.5$, and $\rho_0=0.2$. The output MSE of the right panel shows that both branches seem to produce vanishing $\MSEp$ in the limit $T\to 0$, though Branch 2 connected to the PR solution yields lower values.  }
\Lfig{g_noiseless}
\end{center}
\end{figure}
The presence of the first-order phase transition implies that the simple SA algorithm with a rapid annealing schedule will fail to find the PR solution in $\rho_{\rm SP}\leq \rho$. The system's dynamics goes along the branch connected to the high temperature limit and cannot move to the PR solution. Actually, this strongly affects the SA performance, as the values of input MSEs $\MSEx$ are quite different between the two branches, despite the output MSEs $\MSEy$ both being small, as shown in the right panel of \Rfig{g_noiseless}. This will be demonstrated in \Rsec{Monte Carlo-based}. 

Summarising the above findings, we draw a phase diagram in the $\rho$-$T$ plane for $\alpha=0.5$ and $\rho_0=0.2$ in \Rfig{PD_noiseless1}. 
\begin{figure}[htbp]
\begin{center}
 \includegraphics[width=0.49\columnwidth]{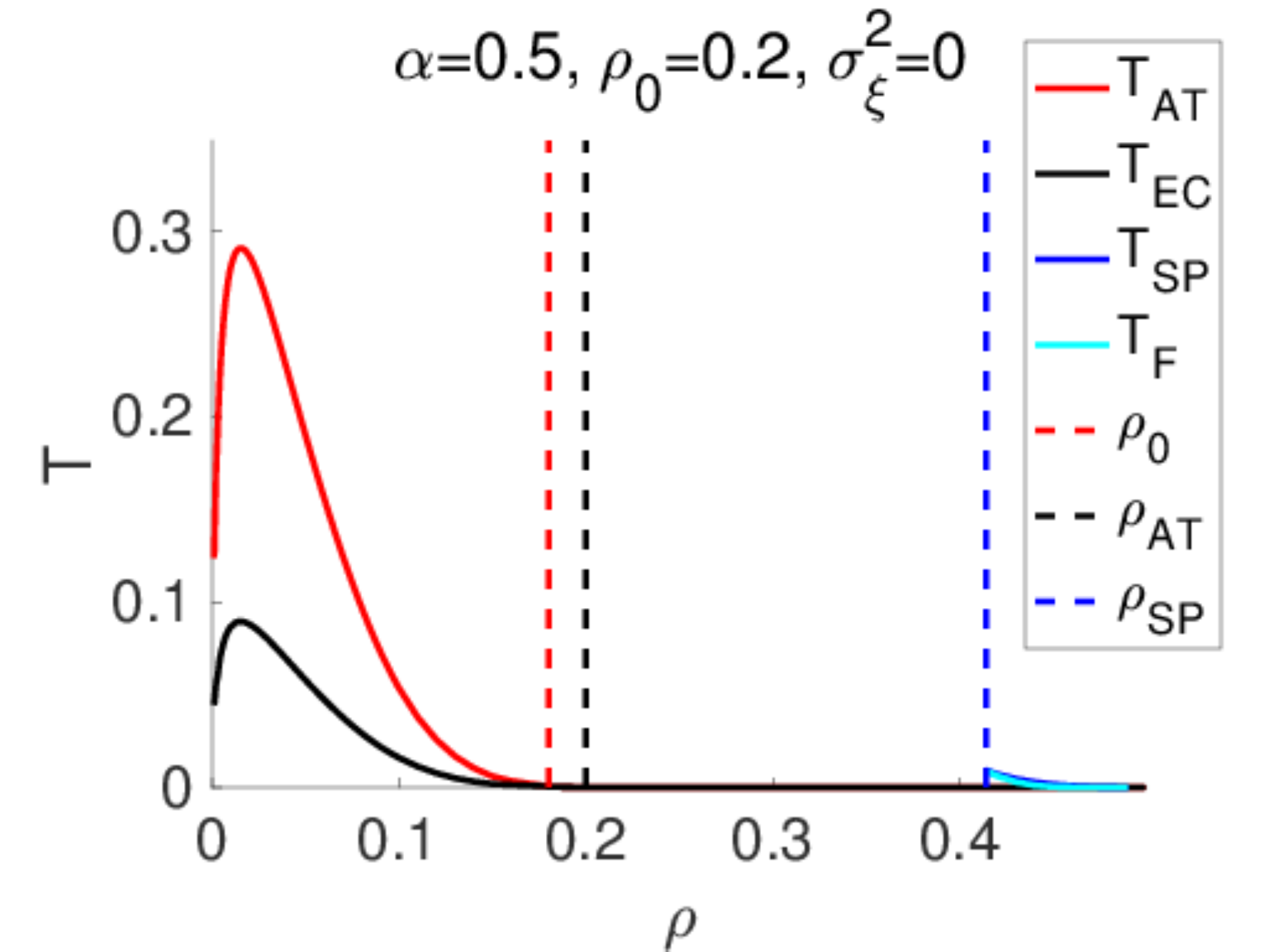}
 \includegraphics[width=0.49\columnwidth]{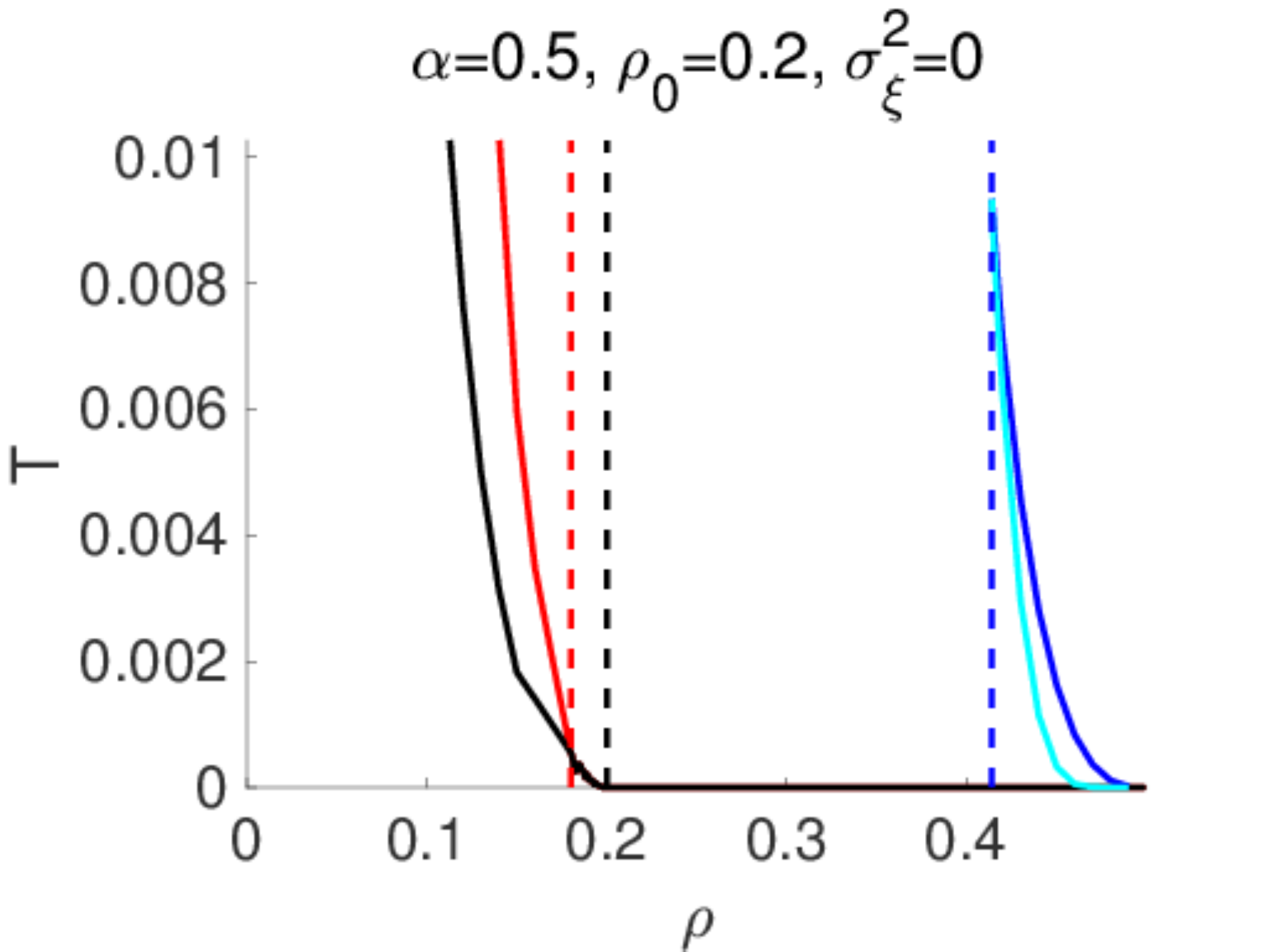}
\caption{Phase diagram in the $\rho$-$T$ plane for the noiseless case $\sigma_{\xi}=0$. The parameters are $\alpha=0.5$ and $\rho_0=0.2$. The right panel is a close-up of the left one in the low $T$ region, used to focus on the spinodal and the first-order transition lines, $T_{\rm SP}$ and $T_{\rm F}$, which are rather small compared to the other critical temperatures $T_{\rm AT}$ and $T_{\rm EC}$ in the region $\rho<\rho_0$. The vertical dashed lines denote the critical values of $\rho$: From left to right: $\rho_{\rm AT}\approx 0.17$, $\rho_{0}=0.2$, and $\rho_{\rm SP}\approx 0.43$. }
\Lfig{PD_noiseless1}
\end{center}
\end{figure}
Note that the entropy-crisis line $T_{\rm EC}(\rho)$ below the AT line $T_{\rm AT}(\rho)$ has no direct physical consequence because the RS solution is unreliable in that region. The exact entropy-crisis line would be derived by the full step RSB solution, but this is beyond the scope of the present paper.

\Rfig{PD_noiseless1} implies that finding the ground state is easy in the region $\rho_{\rm AT}< \rho < \rho_{\rm SP}$, while it is difficult in other regions: $\rho\le \rho_{\rm AT}$ and $\rho_{\rm SP}\le \rho$. We focus on how the easy region behaves when changing the external parameters $\alpha$ and $\rho_0$. We examine the parameters and find that the easy region shrinks as $\rho_0$ increases against a fixed $\alpha$ and finally vanishes at a certain critical value of $\rho_0$. As an example, the $\rho$-$T$ phase diagram for $\alpha=0.5$ and $\rho_0=0.3$ is shown in \Rfig{PD_noiseless2}. 
\begin{figure}[htbp]
\begin{center}
 \includegraphics[width=0.49\columnwidth]{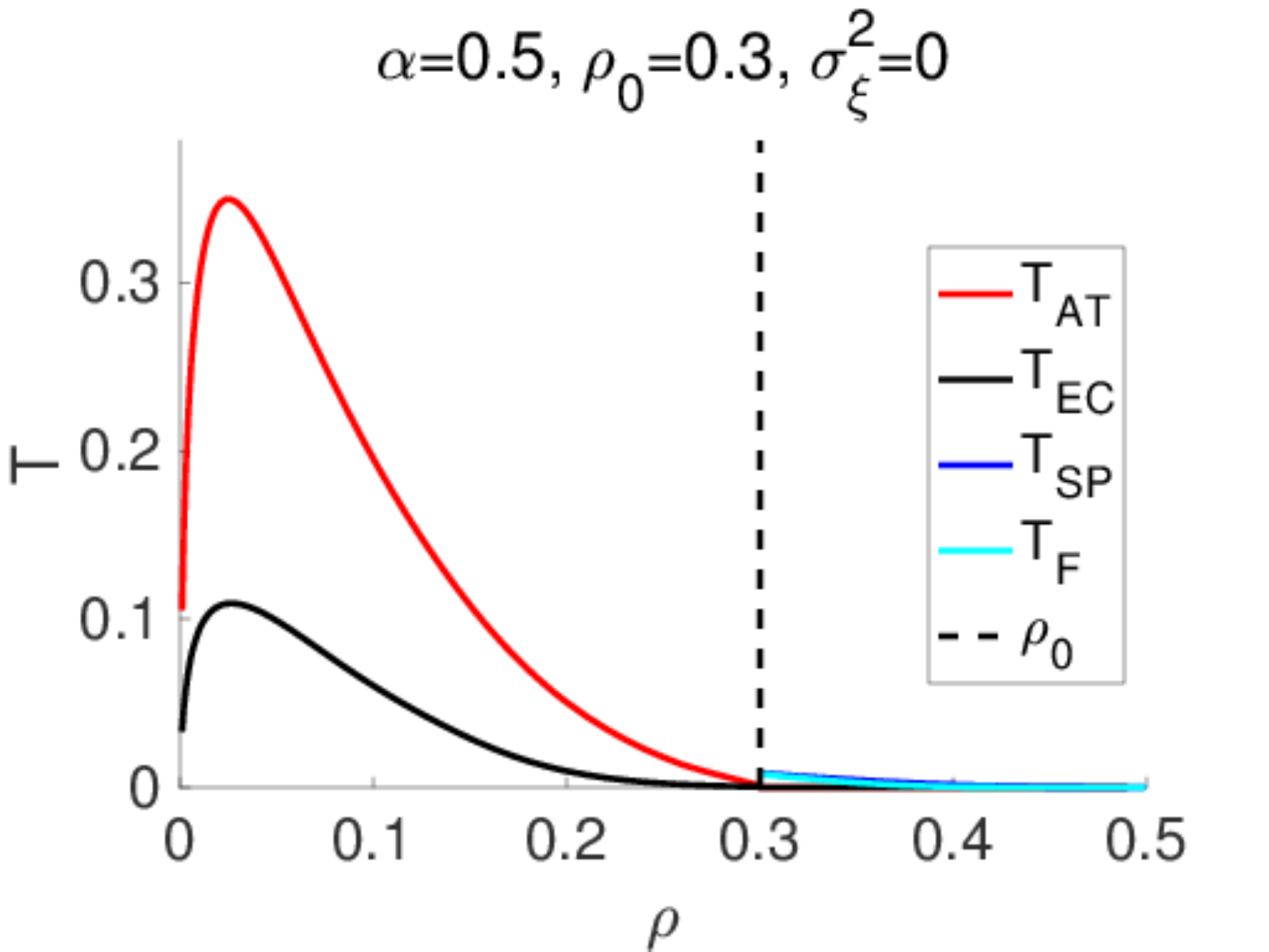}
 \includegraphics[width=0.49\columnwidth]{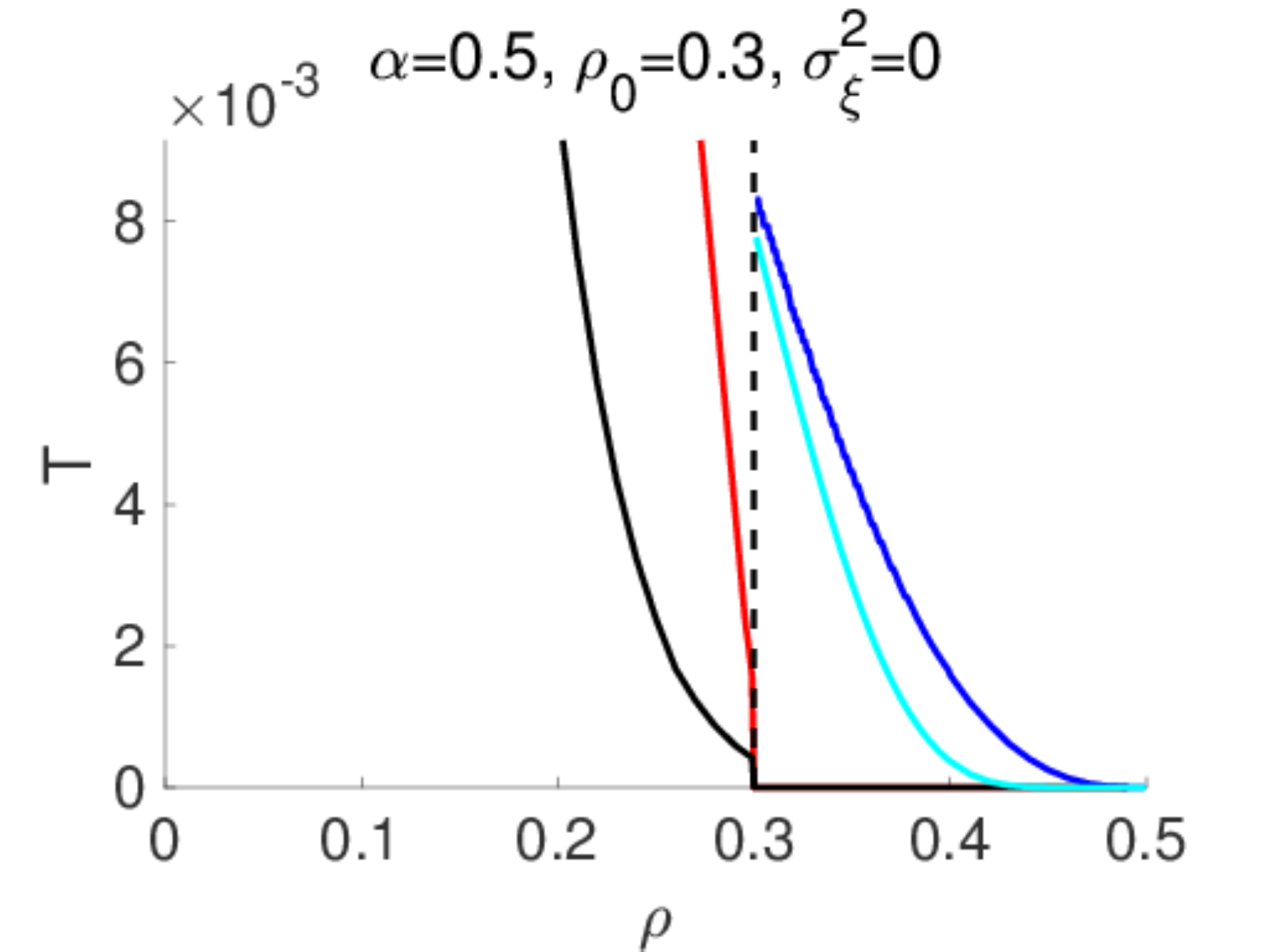}
\caption{Phase diagram in the $\rho$-$T$ plane for the noiseless case $\sigma_{\xi}=0$. The parameters are $\alpha=0.5$ and $\rho_0=0.3$. The right panel is a close-up of the left one in the low $T$ region. The spinodal and first-order transition lines cover the entire $\rho_0\leq \rho$ region, implying that local search algorithms will fail to find the PR solution for any $\rho$.}
\Lfig{PD_noiseless2}
\end{center}
\end{figure}
The spinodal and first-order transition lines cover the entire $\rho_0< \rho$ region and hence the easy region disappears, implying that local search algorithms cannot find the PR solution for any $\rho$ in this case. This vanishing of the easy region thus defines the algorithmic limit. By searching all parameter regions of $\rho_0$ and $\alpha$, we can draw a phase diagram of the algorithmic limit, which is shown in \Rfig{Algorithmic limit}.
\begin{figure}[htbp]
\begin{center}
 \includegraphics[width=0.49\columnwidth]{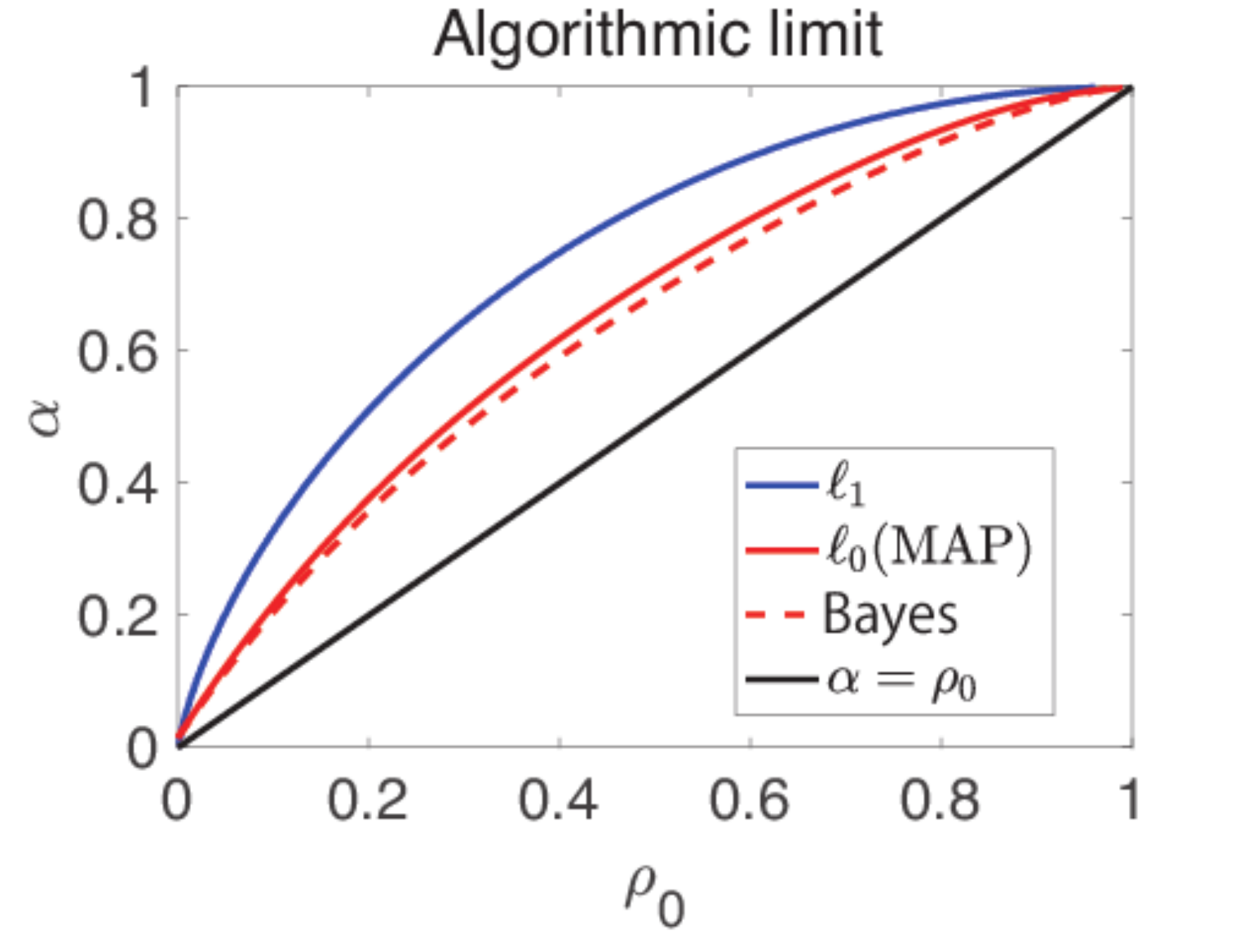}
\caption{The algorithmic limits for the perfect reconstruction of the planted solution $\V{x}_0$. The red solid line is the algorithmic limit derived here. Two other boundaries, the blue solid and red broken ones, indicate the $\ell_1$ relaxation derived in~\cite{Kabashima:09} and the Bayesian inference shown in~\cite{Krzakala:12-1}, respectively. Our result, which is regarded as an MAP approximation of the Bayesian inference, is competitive with the Bayesian result. }
\Lfig{Algorithmic limit}
\end{center}
\end{figure}
The performance of our formulation is clearly better than the $\ell_1$ relaxation~\cite{Kabashima:09} and is competitive with the Bayesian result~\cite{Krzakala:12-1}. This implies that the present formulation, which can be regarded as a MAP estimation in the Bayesian framework, does not significantly lose its reconstruction performance despite discarding the signal source information. This is encouraging the use of the present formulation in the context of signal recovery and is one of the main results of this paper. 

\subsubsection{Noisy case}
A new behaviour specific to the noisy case is the presence of the RFOT transition for strong noises at middle values of $\rho$. As an example, the entropy curves for $\sigma_{\xi}^2=10$ are given in \Rfig{entropy_noise}. 
\begin{figure}[htbp]
\begin{center}
 \includegraphics[width=0.32\columnwidth]{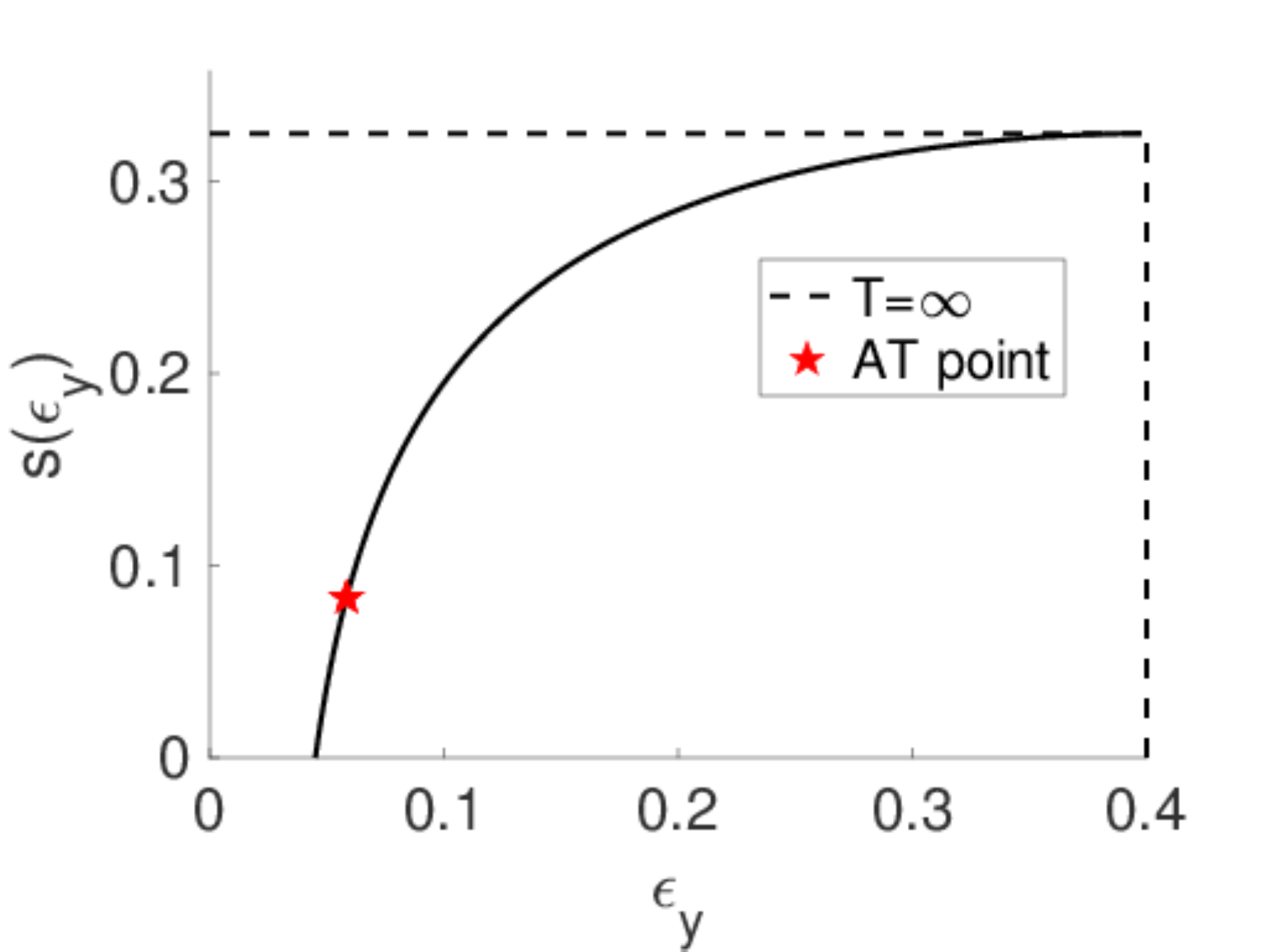}
 \includegraphics[width=0.32\columnwidth]{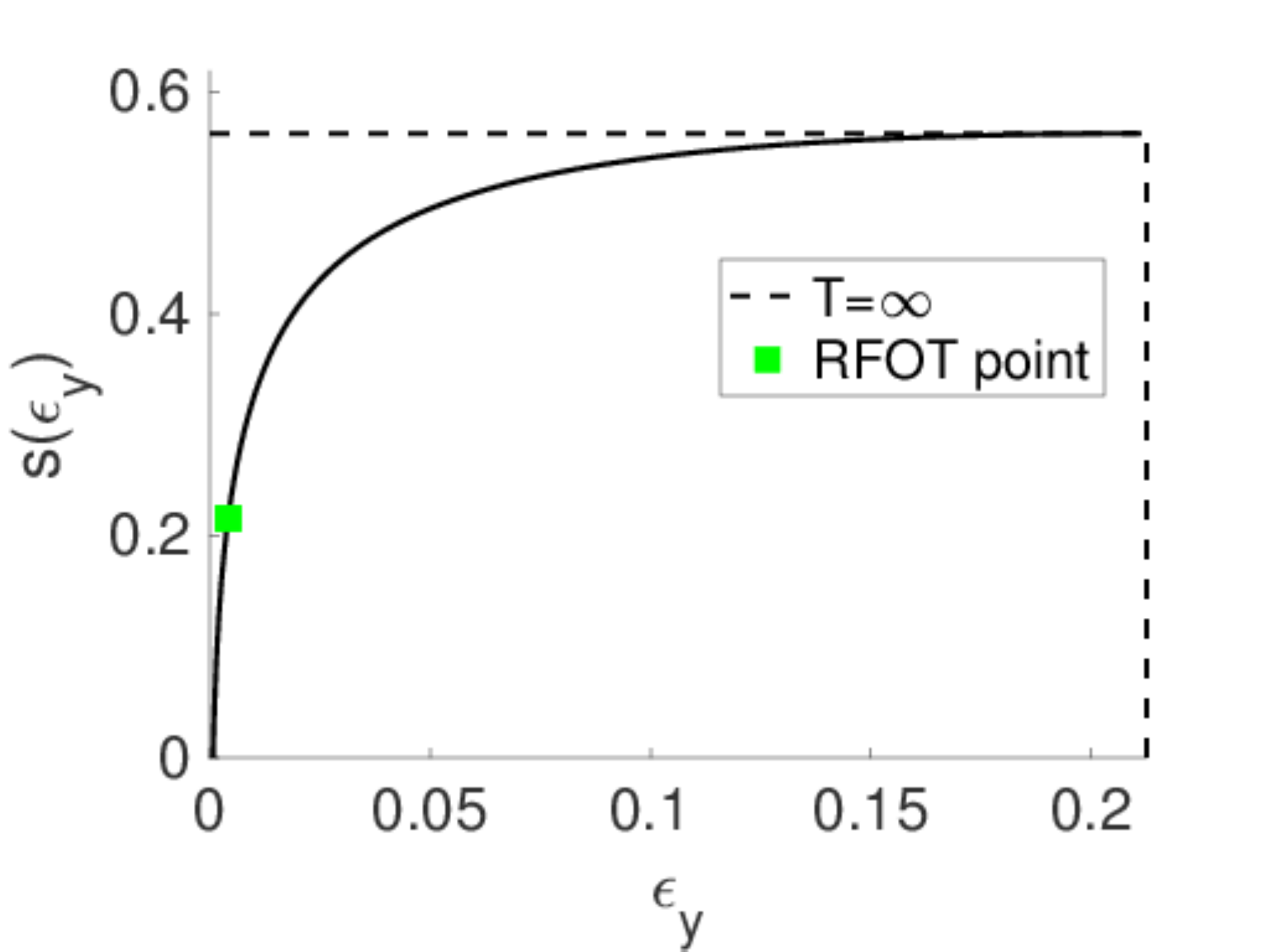}
 \includegraphics[width=0.32\columnwidth]{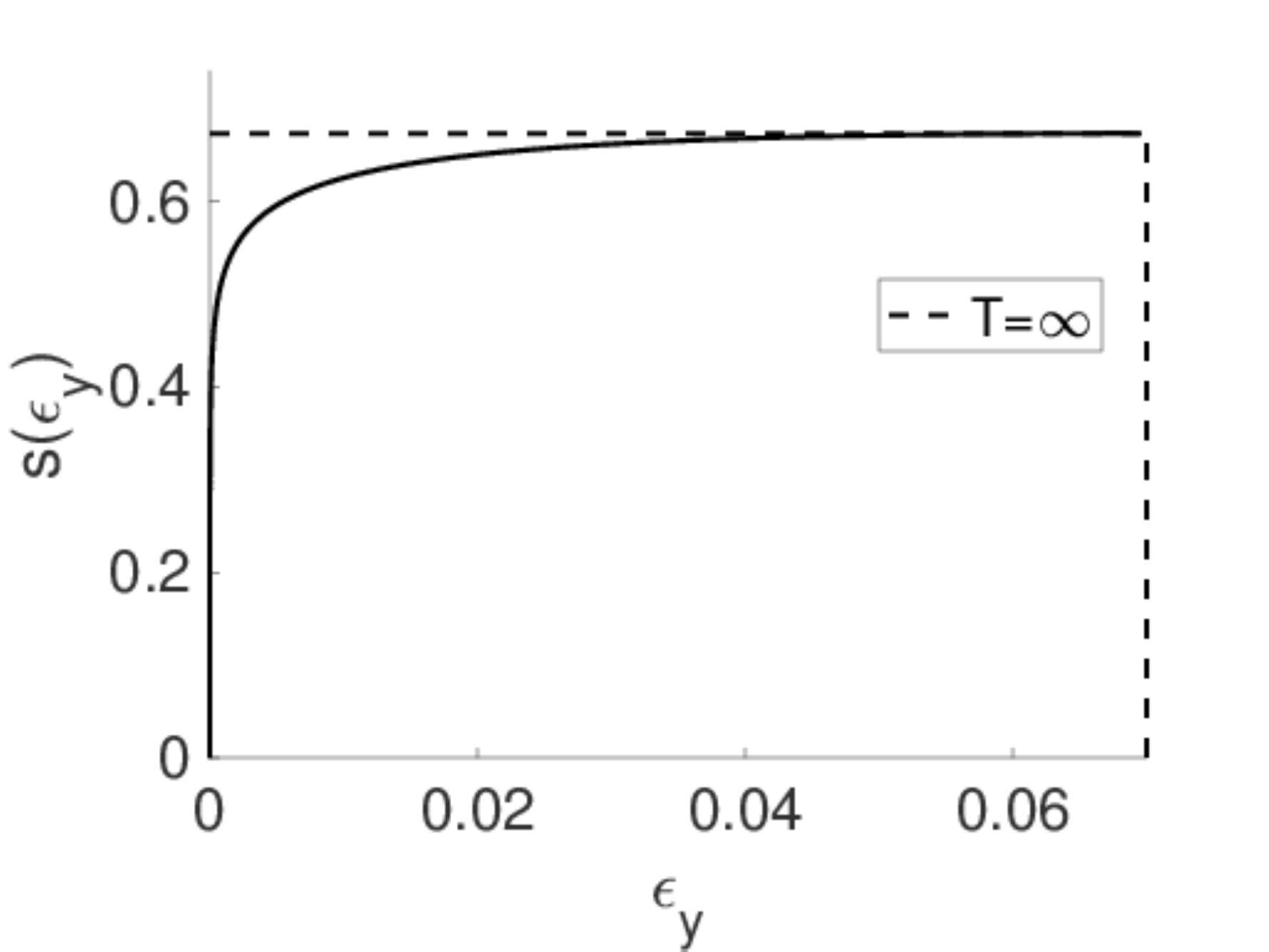}
\caption{The solid black line denotes the entropy curve plotted against the output MSE $\MSEp$ for $\alpha=0.5, \rho_0=0.2$ and $\sigma_{\xi}^2=10$. The non-zero component density $\rho$ is $0.1,~0.25$ and $0.4$ from left to right. The RFOT transition, which is absent in the noiseless case, appears at middle values of $\rho$. }
\Lfig{entropy_noise}
\end{center}
\end{figure}
Two critical $\rho$ values, $\rho_{\rm AT}$ and $\rho_{\rm RFOT}$, accordingly emerge. For $\rho\leq \rho_{\rm AT}$, the AT instability first occurs as the temperature decreases. For $\rho_{\rm AT} < \rho \leq \rho_{\rm RFOT}$, the RFOT begins to emerge above the AT instability temperature. For larger $\rho$, the RS solution is accurate for all temperature regions. The RS EC occurs at finite $T$ in that region, which is somewhat similar to the Ising perceptron problem~\cite{Krauth:89,Obuchi:09}. Note that $\rho_{\rm RFOT}$ does not exist if the noise is sufficiently weak. 

Summarising the above findings, we show phase diagrams for three noise strengths, $\sigma_{\xi}^2=0.001,~0.1$ and $10$ in \Rfig{PD_noisy}. 
\begin{figure}[htbp]
\begin{center}
 \includegraphics[width=0.32\columnwidth]{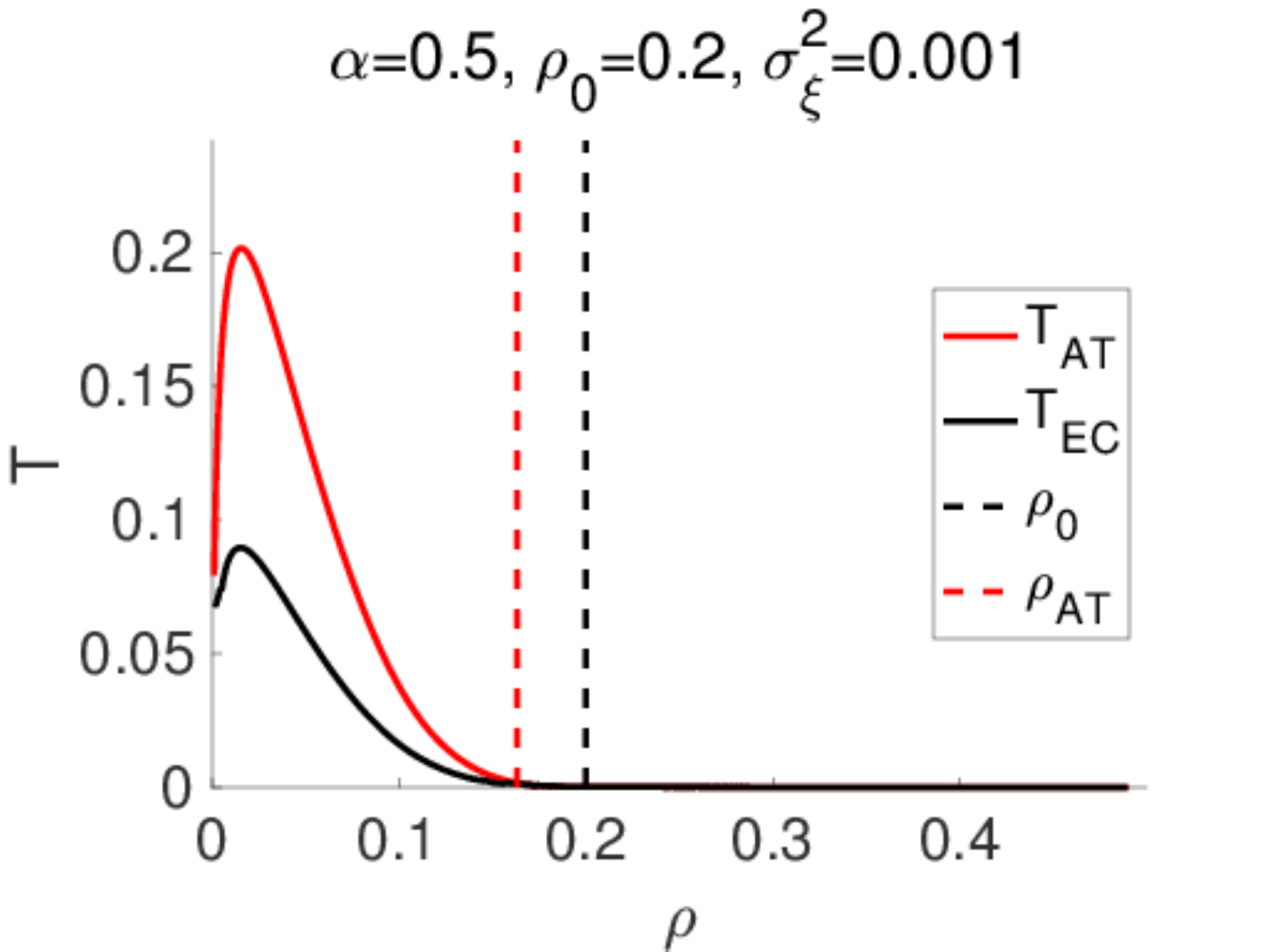}
 \includegraphics[width=0.32\columnwidth]{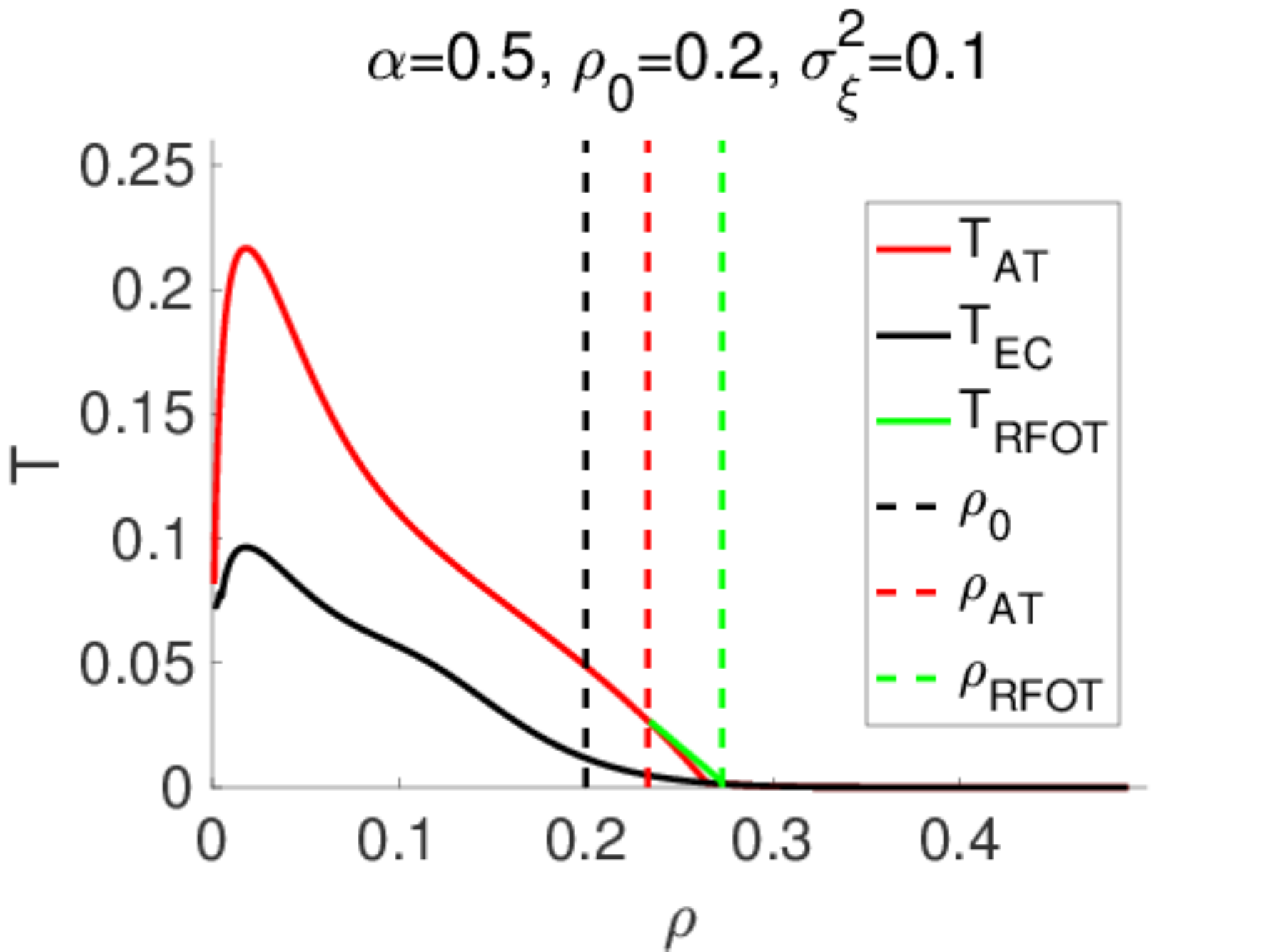}
 \includegraphics[width=0.32\columnwidth]{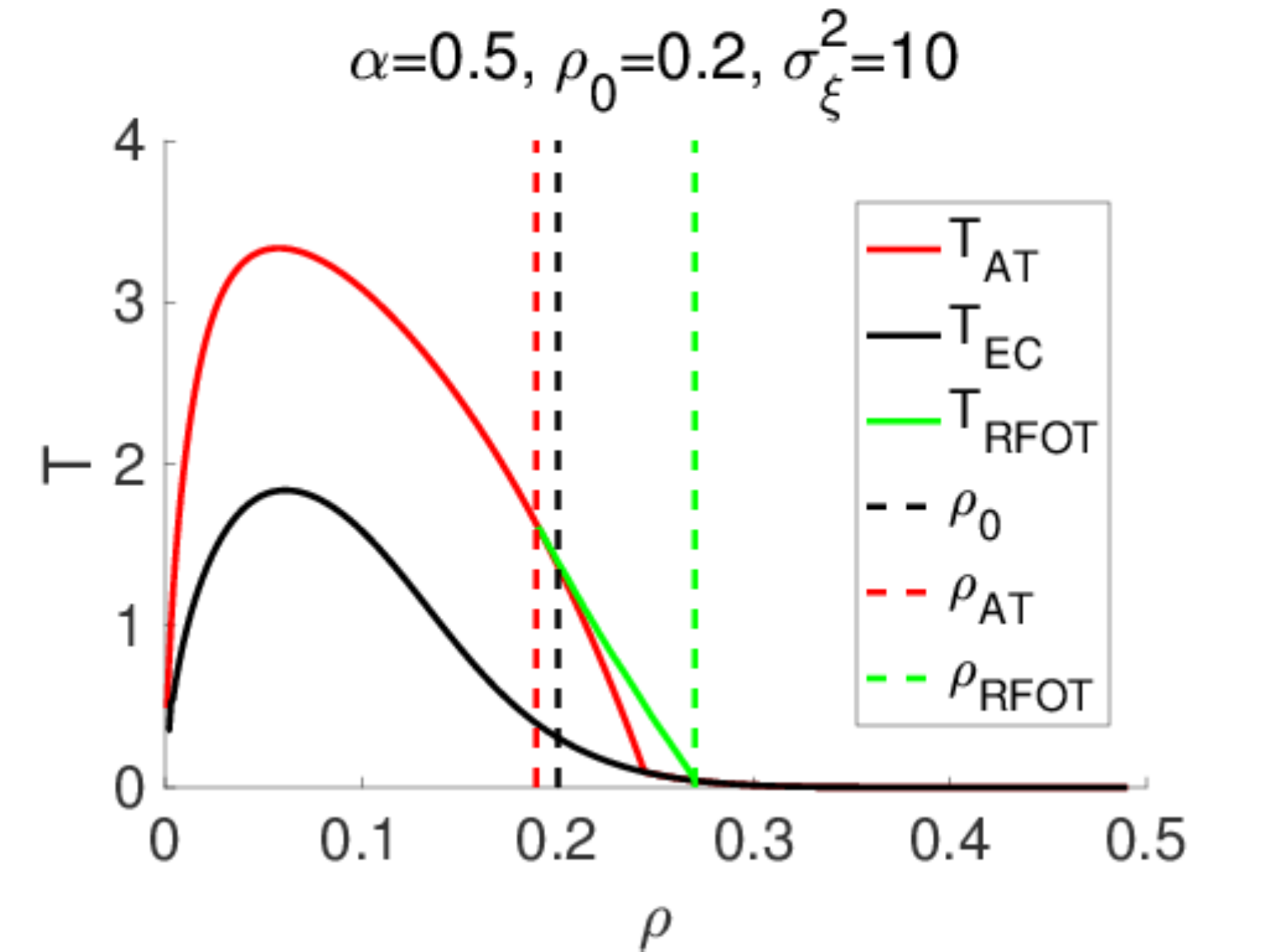}
\caption{Phase diagrams for three different noise strengths: $\sigma_{\xi}^{2}=0.001,~0.1$ and $10$ from left to right. The other parameters are $\alpha=0.5$ and $\rho_0=0.2$. The phase boundaries are denoted by solid lines with different colours (green: $T_{\rm RFOT}$, red: $T_{\rm AT}$, black: $T_{\rm EC}$). For the weak noise case (left), we could not locate the RFOT region. The vertical dashed lines represent the critical values of $\rho$.}
\Lfig{PD_noisy}
\end{center}
\end{figure}
The first-order transition in the noiseless case is quite fragile and disappears for very weak noise as seen in the left panel of \Rfig{PD_noisy}. We attempted to capture the critical values of $\sigma_{\xi}$ for the disappearance, but it proved numerically difficult and we did not pursue this point. As the noise increases, the phase boundary's shape expand more and more from the noiseless limit and the RSB region, $\rho<\rho_{\rm RFOT}$, seems to grow. We consider whether this RSB region will cover the low-temperature region completely as the noise is very large. To answer this, we tested the no signal case $\rho_0=0$ and $\sigma_{\xi}^{2}=1$ and observed that $\rho_{\rm RFOT}$ takes a value similar to the one in the right panel of \Rfig{PD_noisy}. Hence, an RS region exists even in the strong noise case, which is consistent with our previous analysis in the data compression context~\cite{Nakanishi:16}. 

\section{Numerical simulations}\Lsec{Numerical}

\subsection{Monte Carlo evaluation of entropy curves}\Lsec{Monte Carlo evaluation}
Here we examine the analytical results by comparing with the numerical simulations. Our simulations calculate the free entropy by the exchange Monte Carlo (MC) sampling \cite{Hukushima:96}. Estimation of the free entropy $g$ is accomplished by using the multi-histogram method \cite{Ferrenberg:88}.

Our MC sampling is based on the Metropolis criterion, where an MC move $\V{c}\to \V{c}'$ is accepted according to the probability
\be
p_{\rm accept}(\V{c}\to \V{c}')=\min(1,e^{-M\mu\lb \MSEy(\V{c}') -\MSEy(\V{c})\rb } ).
\ee
During the update, we would like to keep the non-zero components density $\rho=\sum_{i}c_i/N$ constant. For this, we generate trial moves $\V{c}\to \V{c}'$ by ``pair flipping'' of two support indicators, one equal to $0$ and the other equal to $1$. Namely, by choosing an index $i$ of the support indicator from ${\rm ONES}\equiv \{k|c_{k}=1 \}$ and another index $j$ from ${\rm ZEROS}\equiv \{k|c_{k}=0  \}$, we set $\V{c}'=\V{c}$, except for the counterpart of $(c_i,c_j)=(1,0)$, which is given as $(c'_i,c'_j)=(0,1)$. The pseudocode of our MC algorithm is given in \Rcode{MC}.
\alglanguage{pseudocode}
\begin{algorithm}[htbp]
\caption{MC update with pair flipping}\Lcode{MC}
\begin{algorithmic}[1]
\Procedure{MCpf}{$\V{c},\mu,\V{y},A$}\Comment{MC routine  with pair flipping}
	\State ${\rm ONES} \lA \{k|c_k=1\},~{\rm ZEROS} \lA \{k|c_k=0\}$
	\State randomly choose $i$ from ONES and $j$ from ZEROS
	\State $\V{c}' \lA \V{c}$
	\State $(c'_i, c'_j) \lA (0,1)$
	\State $(\MSEp,\MSEp')\lA (\MSEp(\V{c}|\V{y},A),\MSEp(\V{c}'|\V{y},A))$ 
	\State $p_{\rm accept} \lA \min(1,e^{-M\mu\lb \MSEp' -\MSEp\rb } )$
	\State generate a random number $r\in [ 0,1]$
	\If {$ r < p_{\rm accept} $}
		\State $\V{c} \lA \V{c}'$
	\EndIf       
	\State \Return $\V{c}$
\EndProcedure
\end{algorithmic}
\end{algorithm}
We define one MC step (MCS) as $N$ trials of pair flipping for each system at every temperature point. The exchange of every pair of neighboring temperature points is conducted after every $1/N$ MCS, which is a rather frequent exchange than conventions. 


In all simulations, we set $\alpha=0.5$, $\rho_0=0.2$ and $\rho_0\POWx=1$. The configurational average is calculated by taking the median over 1000 different samples of $(\V{x}_0,\bm{\xi},\bm{A})$. The error bars are estimated by the Bootstrap method. The examined system sizes are $N=30,40,\cdots,100$. The equilibration is checked by monitoring the convergence of all measured quantities $(g,s,\epsilon_y)$ to stable values by changing the total MCSs; for reference, we note that $256\times 10^2$ MCSs are needed for equilibration when $N=100$, $\rho=0.3$, $\sigma_\xi^2=10$. For burn-in, the first half of the total MCSs is discarded. 

\subsubsection{Simulation in noiseless case}\Lsec{Simulation in noiseless}
The free-entropy values evaluated by numerical simulations and the extrapolation to the infinite size limit of the noiseless case $\sigma_{\xi}^2=0$ are presented in \Rfig{extrapolation-noiseless}. 
\begin{figure}[htbp]
\begin{center}
 \includegraphics[width=0.45\columnwidth]{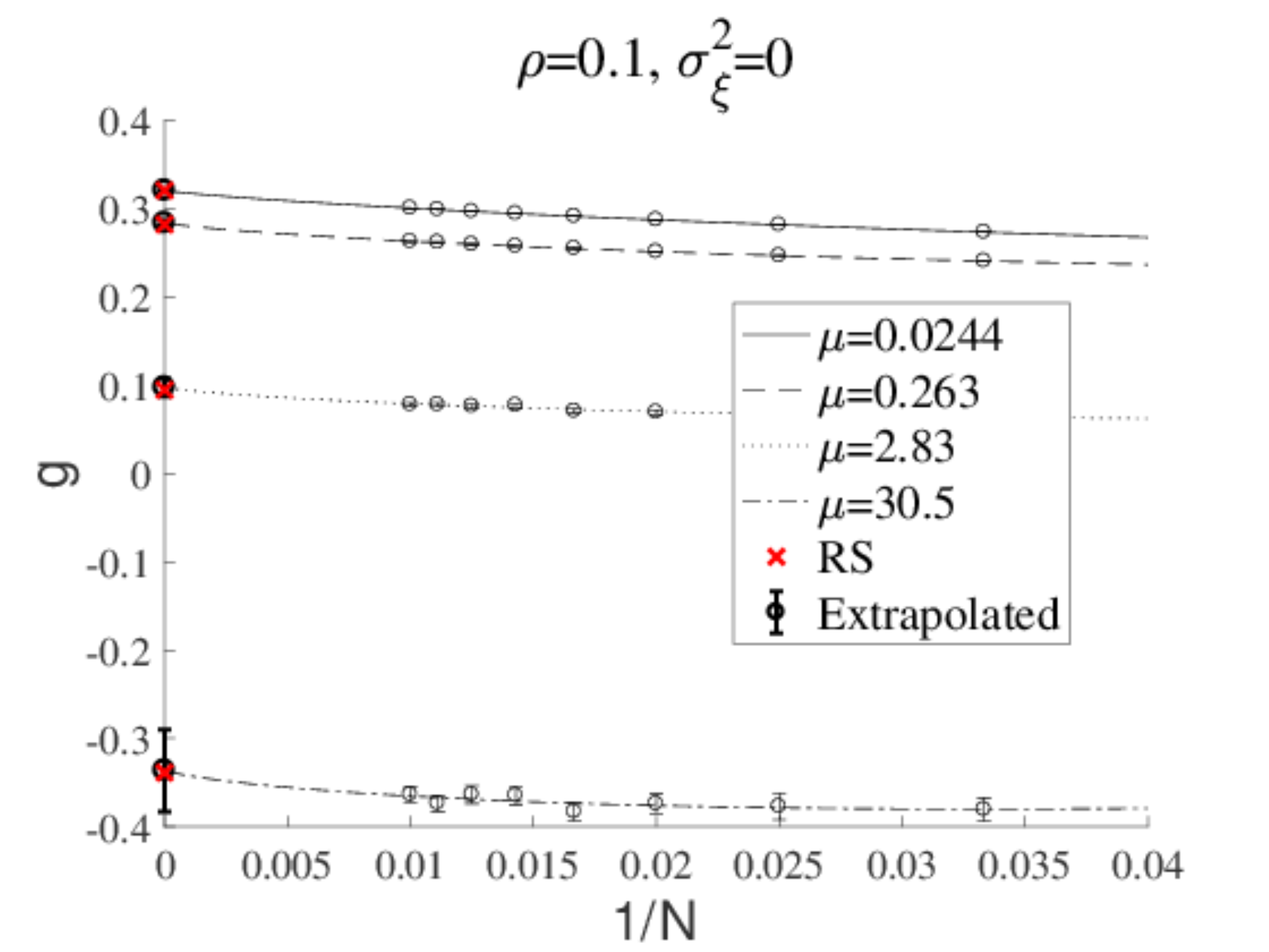}
 \includegraphics[width=0.45\columnwidth]{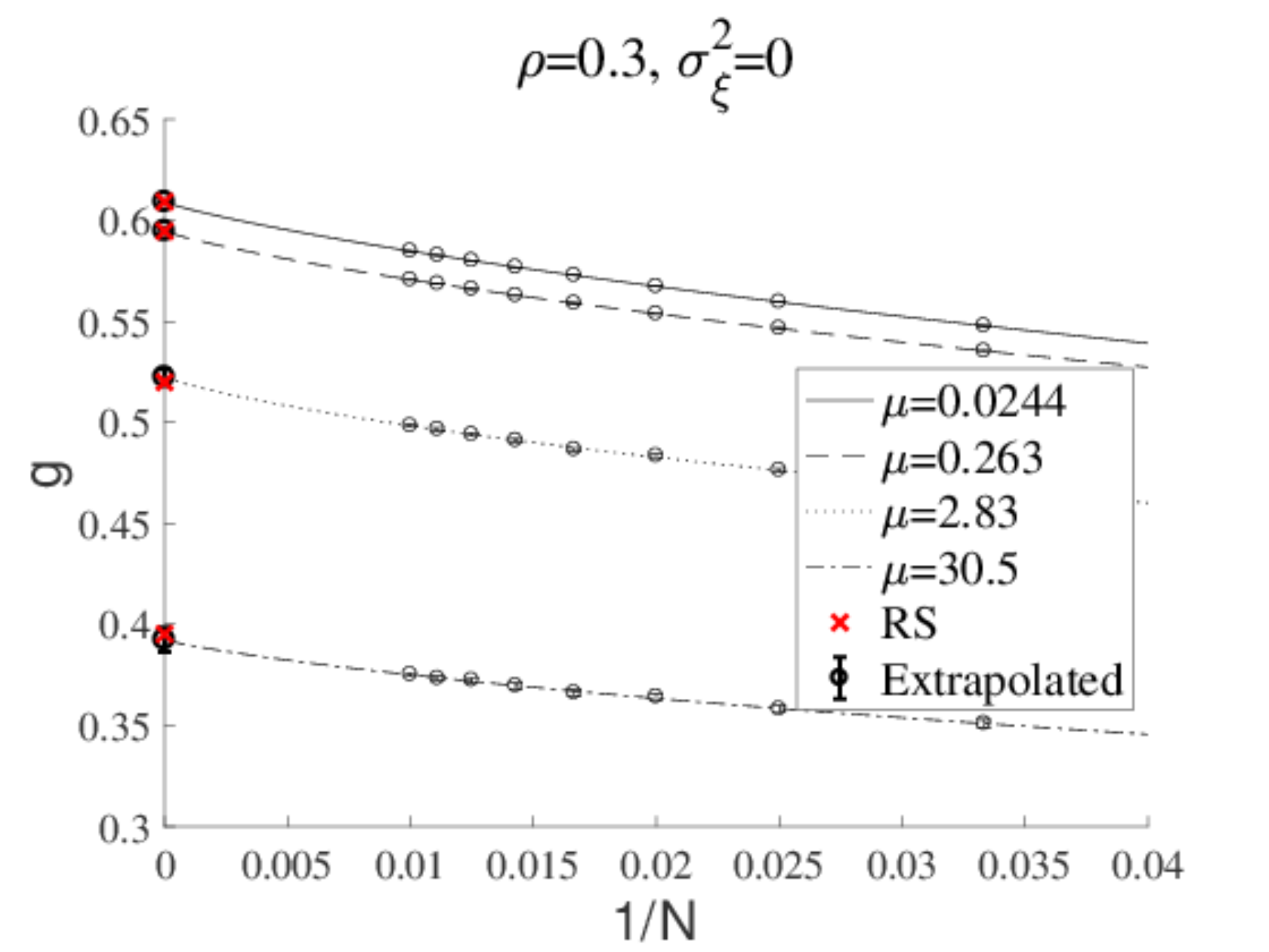}
\caption{Plots of $g$ against $1/N$ at several values of $\mu$ for $\rho=0.1$ (left) and $0.3$ (right) in the noiseless case $\sigma_{\xi}^2=0$. The lines are produced by linear regression based on \Req{regression}. On the vertical axis, the black circles and red crosses represent the extrapolated and analytical values in the $N \to\infty$ limit, respectively. }
\Lfig{extrapolation-noiseless}
\end{center}
\end{figure}
The extrapolation lines result from linear regression using an asymptotic form $g\approx a+bN^{-1}+cN^{-1}\log N^{-1}$. The regression is conducted by applying the least squares method as follows:
\be
\min_{a,b,c}\sum_N \left(a+b\frac{1}{N}+c\frac{1}{N}\log\frac{1}{N}-g(N)\right)^2.
\Leq{regression}
\ee
This asymptotic form is based on Stirling's formula and is exact at $\mu=0$, which motivates us to use the form even when $\mu\not=0$. The same asymptotic form is used for obtaining the extrapolated values of the output MSE $\MSEp$ and the entropy $s$. Using these values for the limit $N \to \infty$, we present the curves $g(T)$ and $s(\MSEp)$ in \Rfig{MCplot-noiseless}. The lines represent the RS analytical results. The circles represent the extrapolated values obtained from the numerical results. The extrapolated values show fairly good agreement with the RS analytical ones, justifying our analytical results.
\begin{figure}[htbp]
\begin{center}
 \includegraphics[width=0.45\columnwidth]{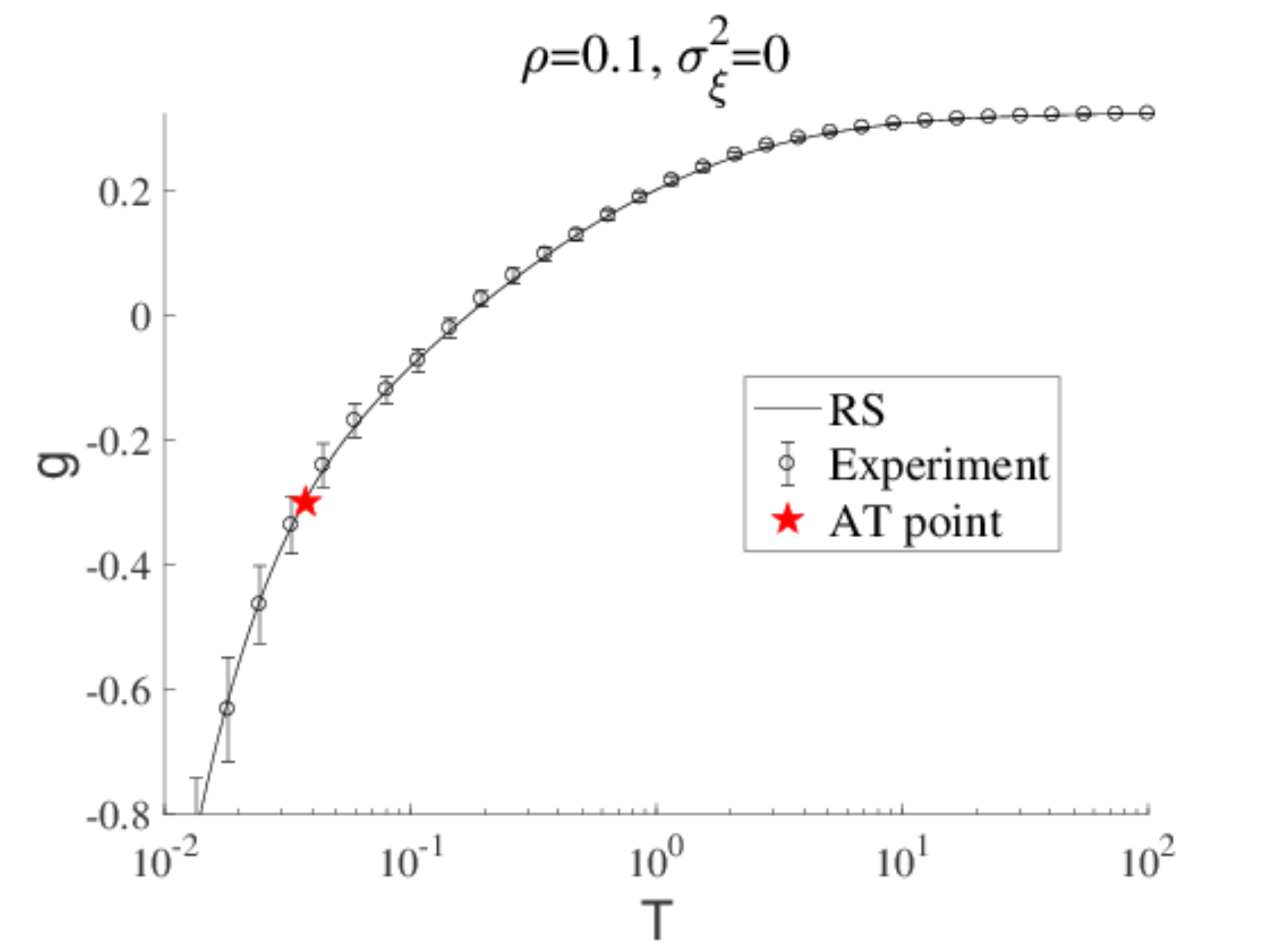}
 \includegraphics[width=0.45\columnwidth]{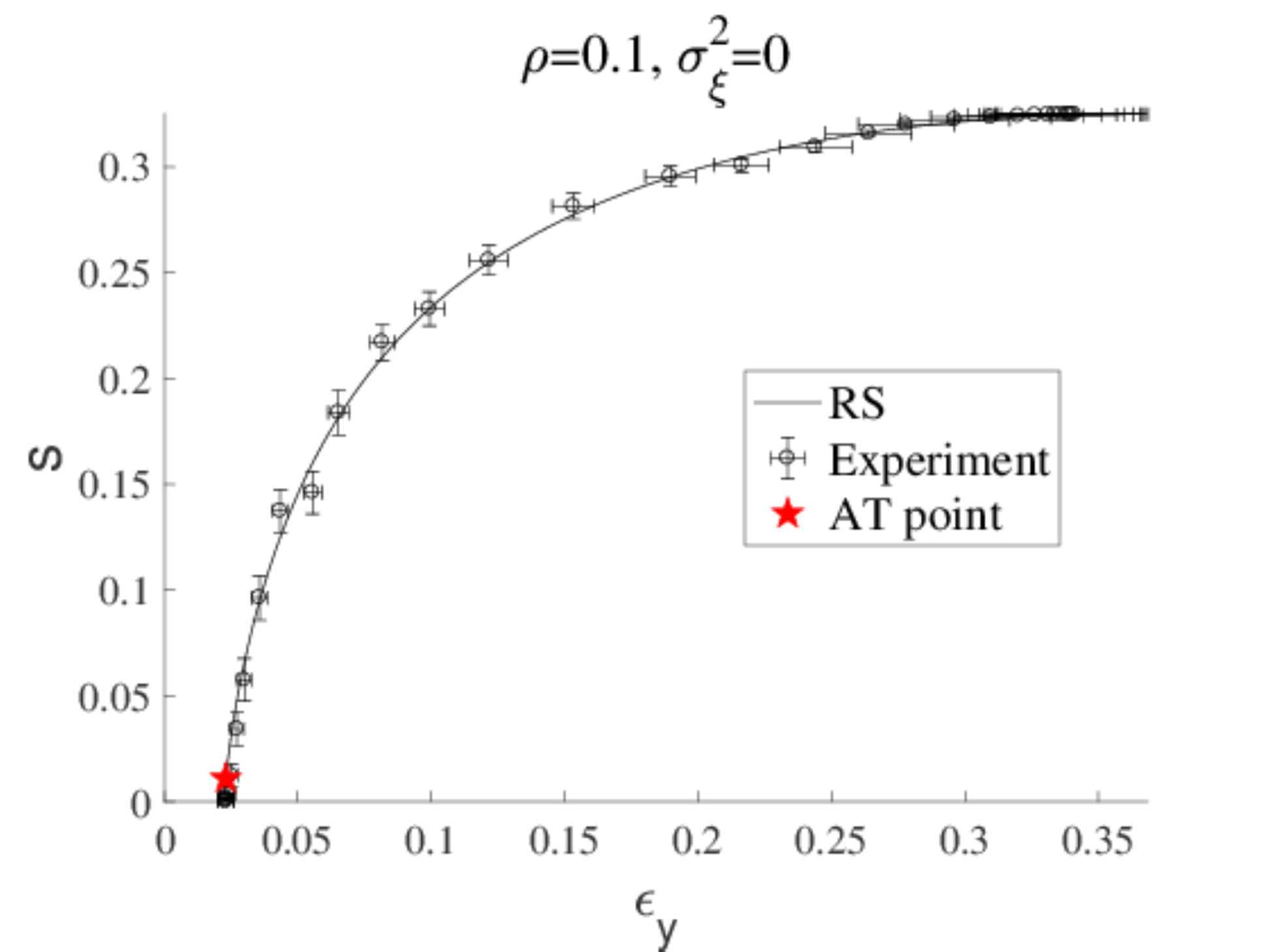}
 \includegraphics[width=0.45\columnwidth]{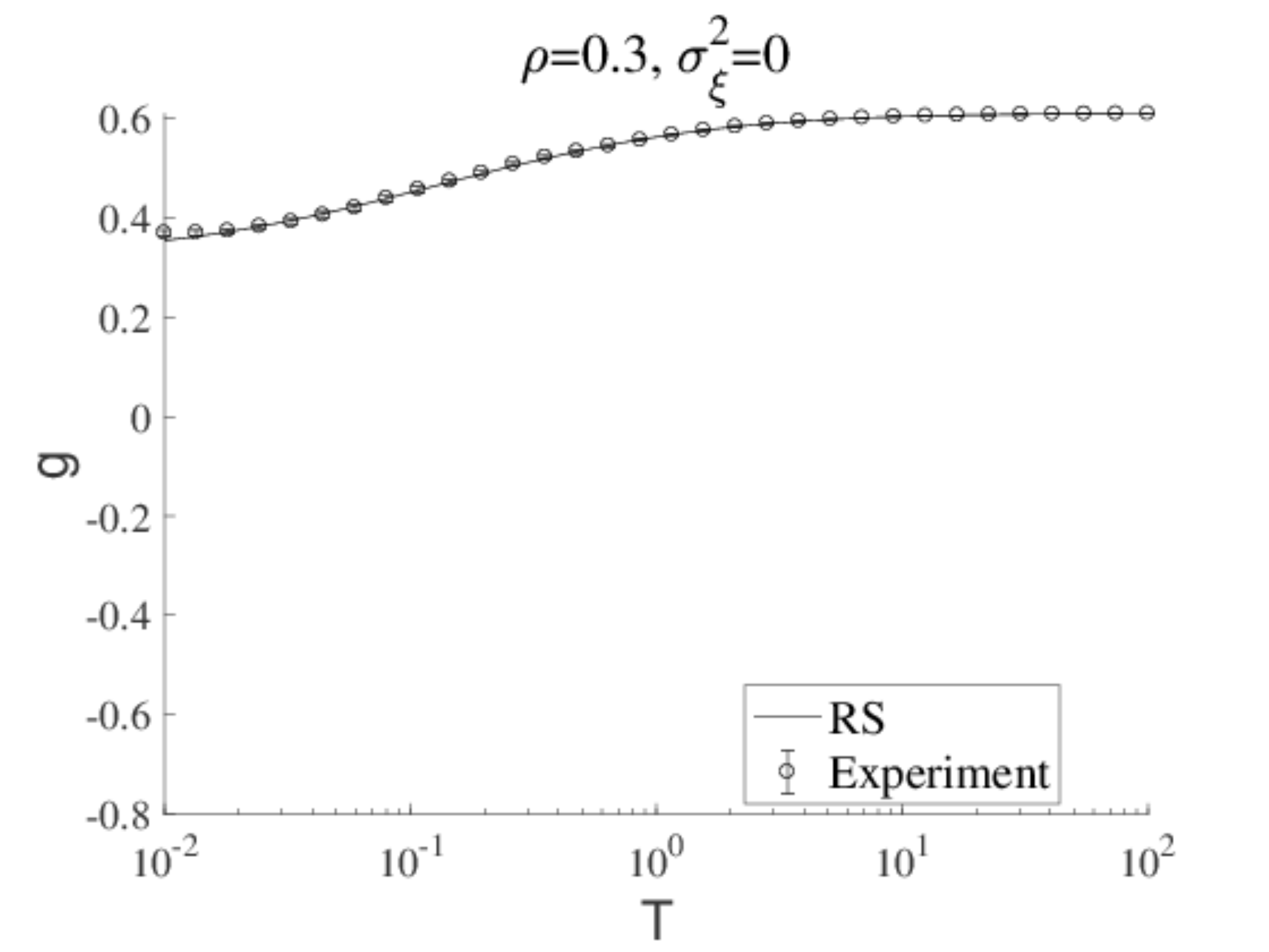}
 \includegraphics[width=0.45\columnwidth]{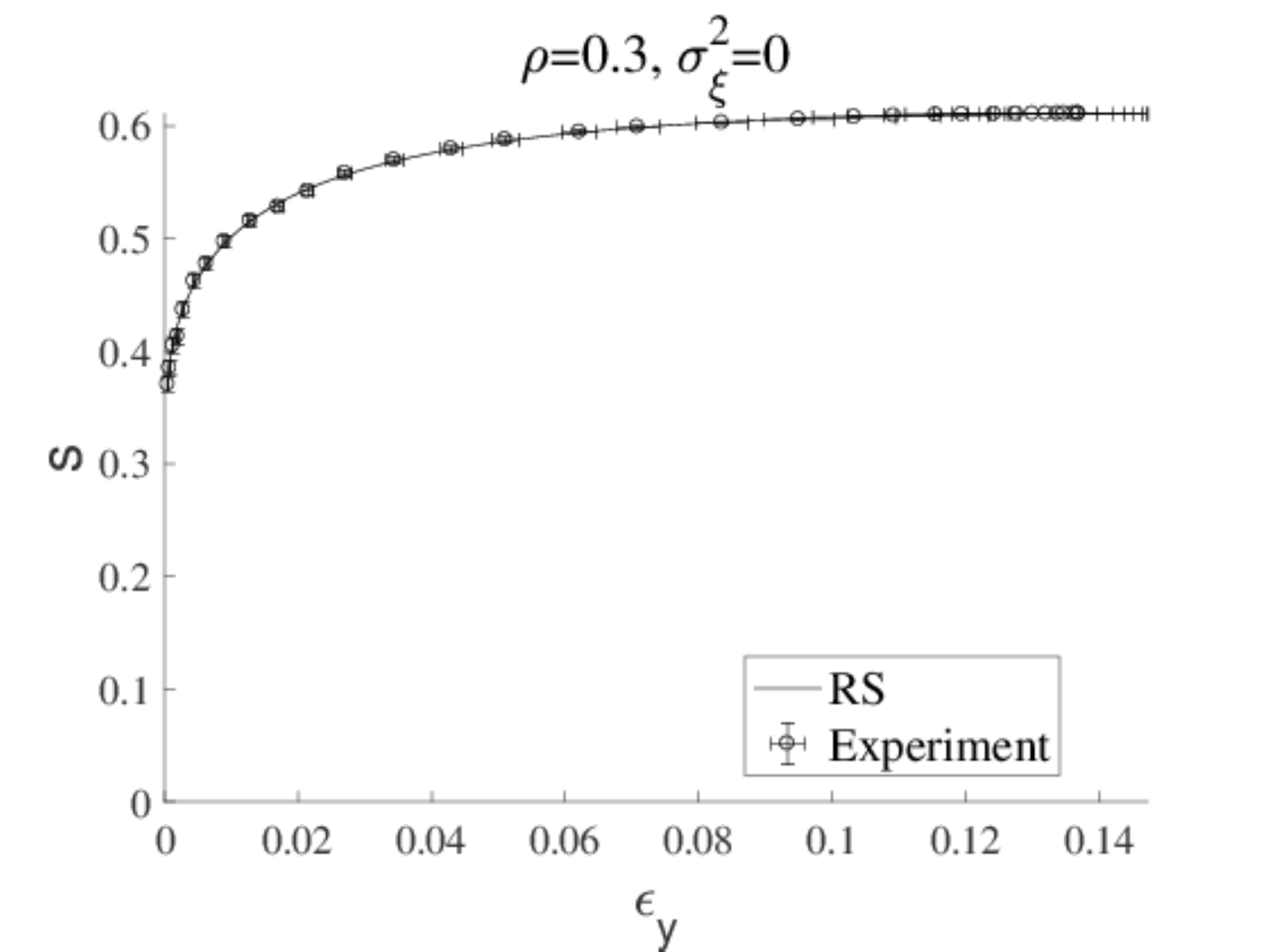}
\caption{The free-entropy curve $g(T)$ (left) and the entropy curve $s(\MSEp)$ (right) for $\rho=0.1$ (upper) and $0.3$ (lower) in the noiseless case $\sigma_{\xi}=0$.}
\Lfig{MCplot-noiseless}
\end{center}
\end{figure}
For the case of $\rho=0.1$, the AT instability occurs at $T\approx 0.04$, but even below this temperature, the agreement between the RS analytical result and the numerical one is fairly good, suggesting a weak RSB effect on $s$ and $\MSEp$. This is, however, not the case for the input MSE $\MSEx$, as demonstrated in \Rsec{Monte Carlo-based} below. 

\subsubsection{Simulation in noisy case}\Lsec{Simulation in noisy}
A similar analysis for the case of strong noise ($\sigma_{\xi}^2=10$) is performed and the results are shown in \Rfig{extrapolation-noisy} and \Rfig{MCplot-noisy}. 
\begin{figure}[htbp]
\begin{center}
 \includegraphics[width=0.45\columnwidth]{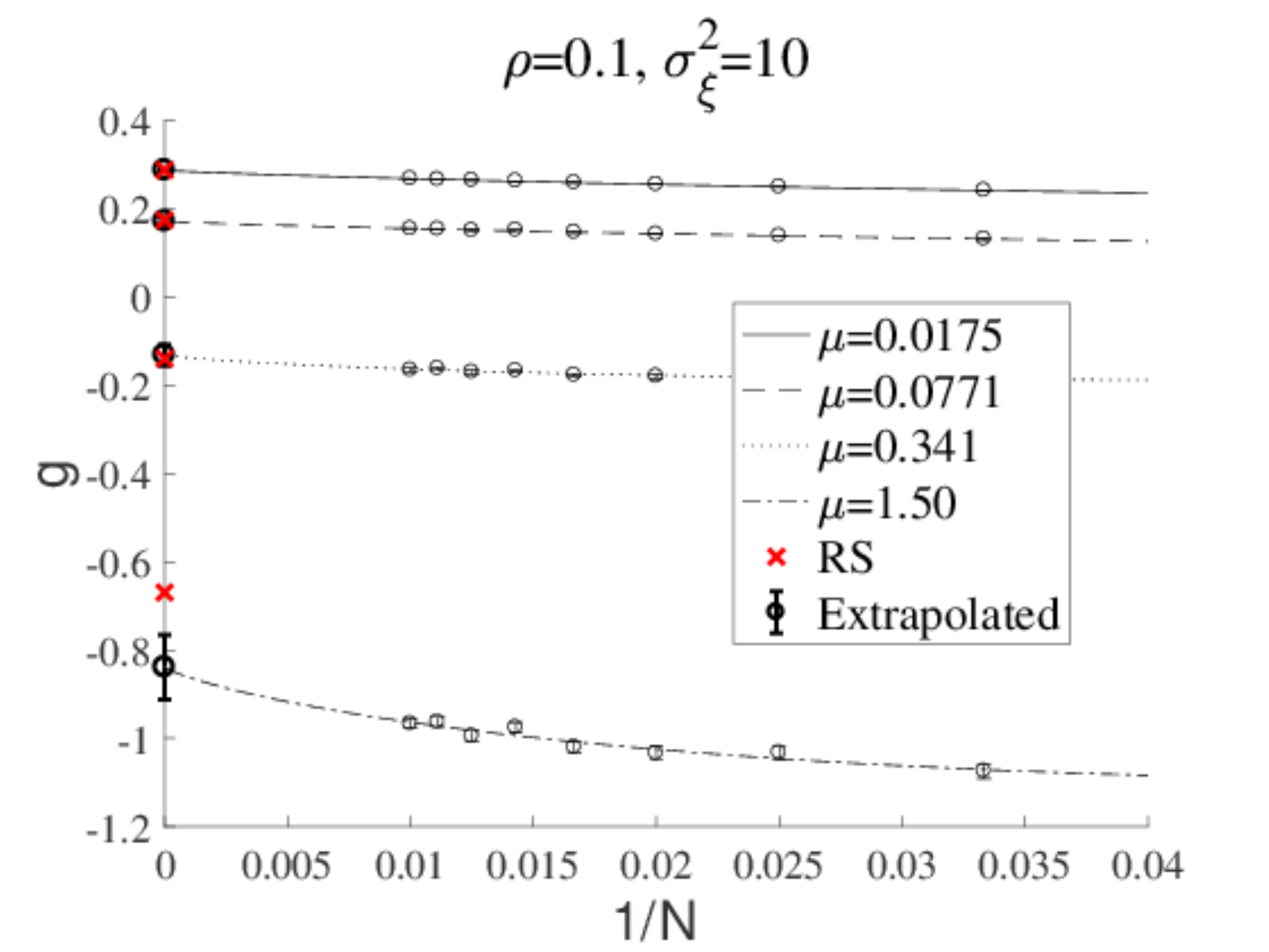}
 \includegraphics[width=0.45\columnwidth]{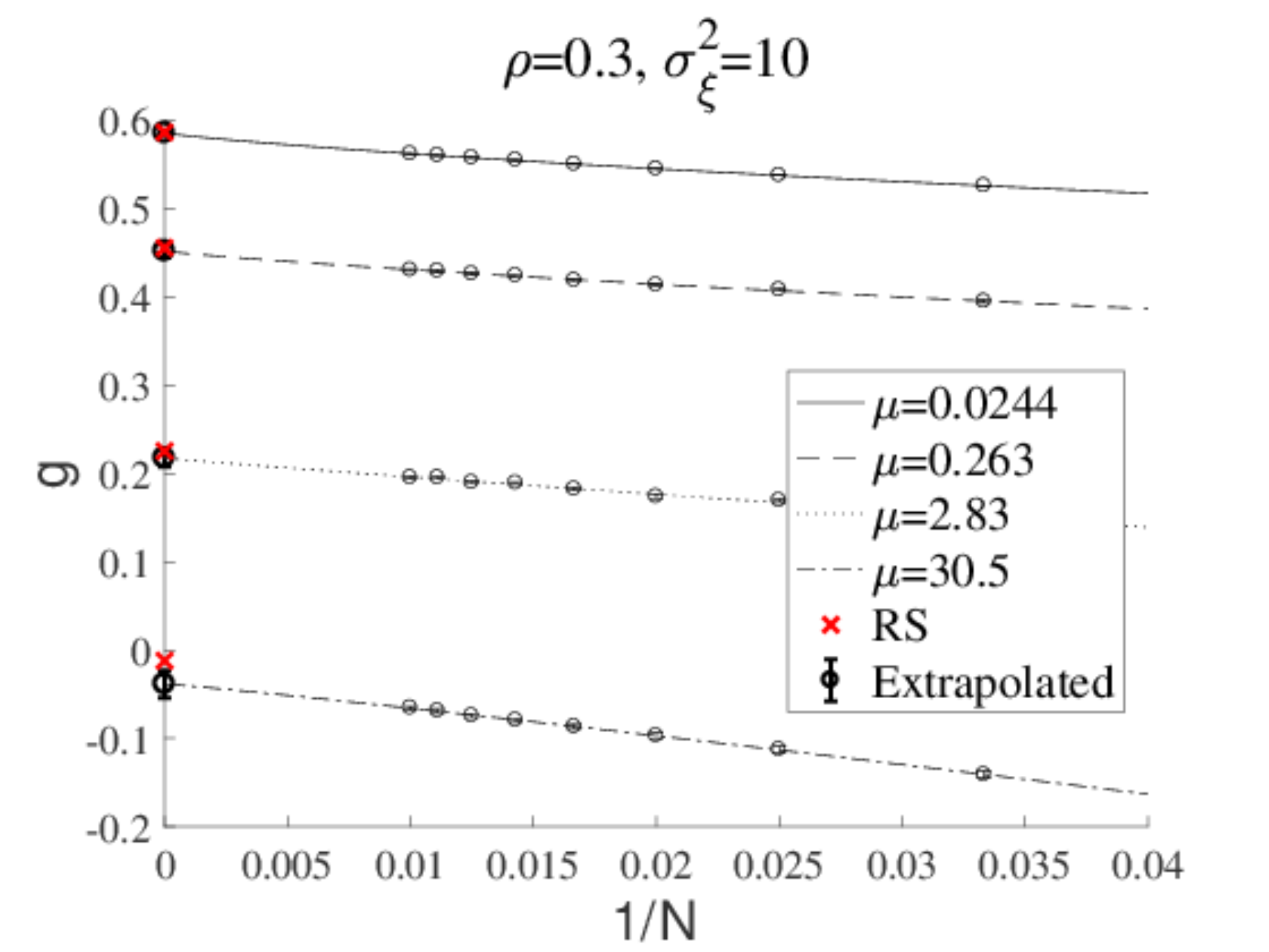}
\caption{Plots of $g$ versus $1/N$ at several values of $\mu$ for $\rho=0.1$ (left) and $0.3$ (right) in the strong noise case $\sigma_{\xi}^2=10$. The gap between the RS and extrapolated results at $\mu=1.5$ in the left panel is probably caused by the RSB effect. }
\Lfig{extrapolation-noisy}
\end{center}
\end{figure}
\begin{figure}[htbp]
\begin{center}
 \includegraphics[width=0.45\columnwidth]{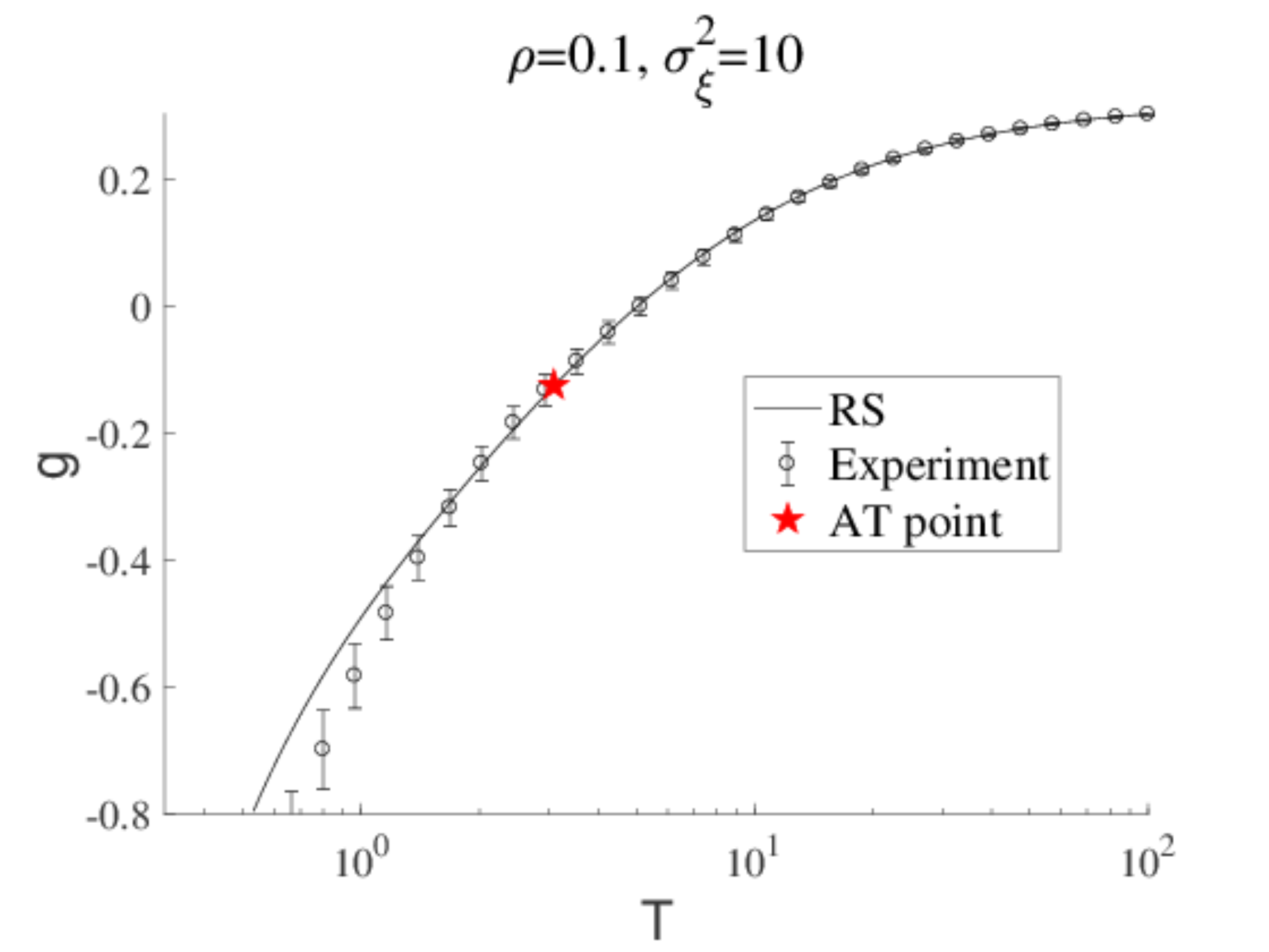}
 \includegraphics[width=0.45\columnwidth]{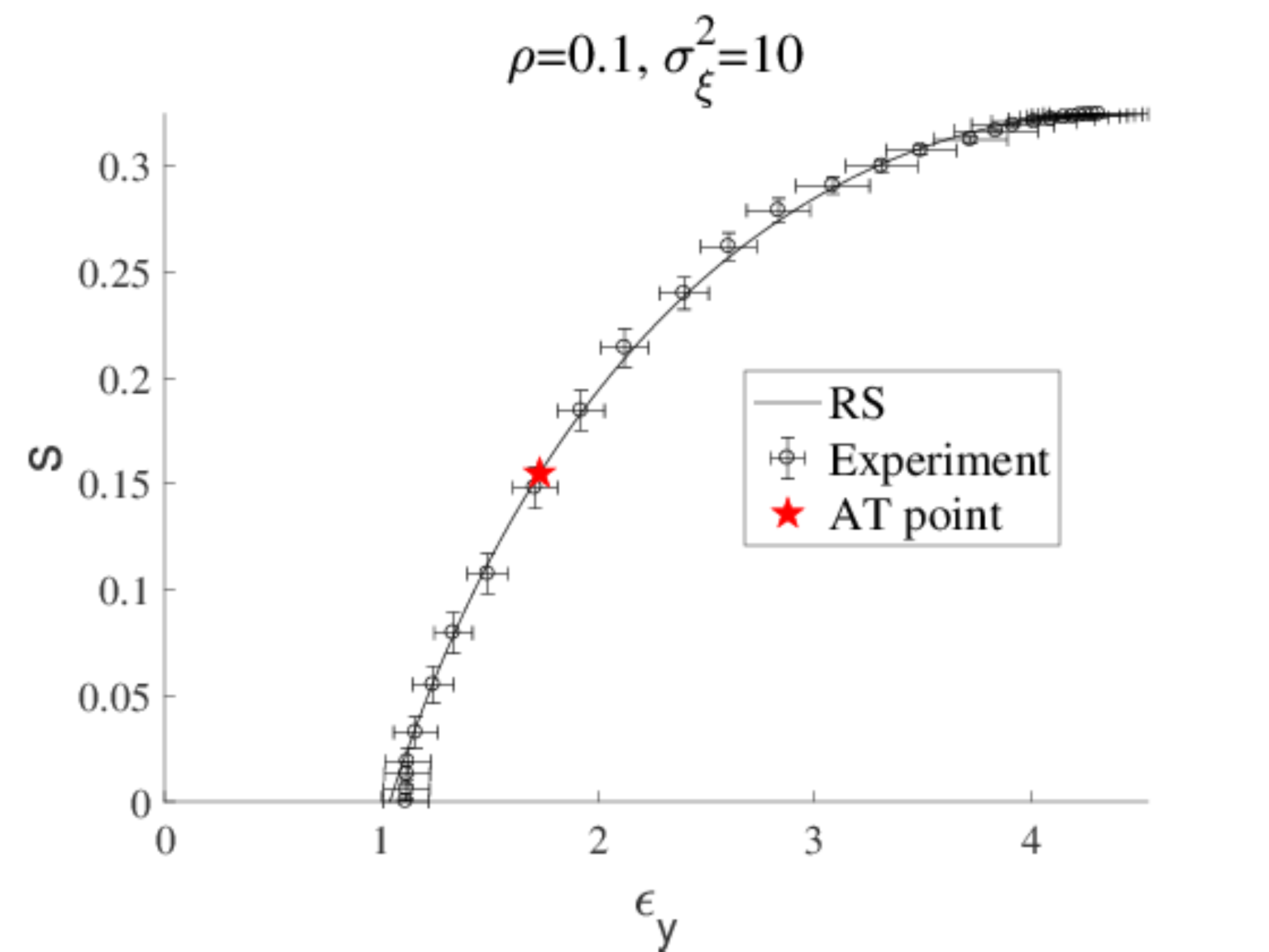}
 \includegraphics[width=0.45\columnwidth]{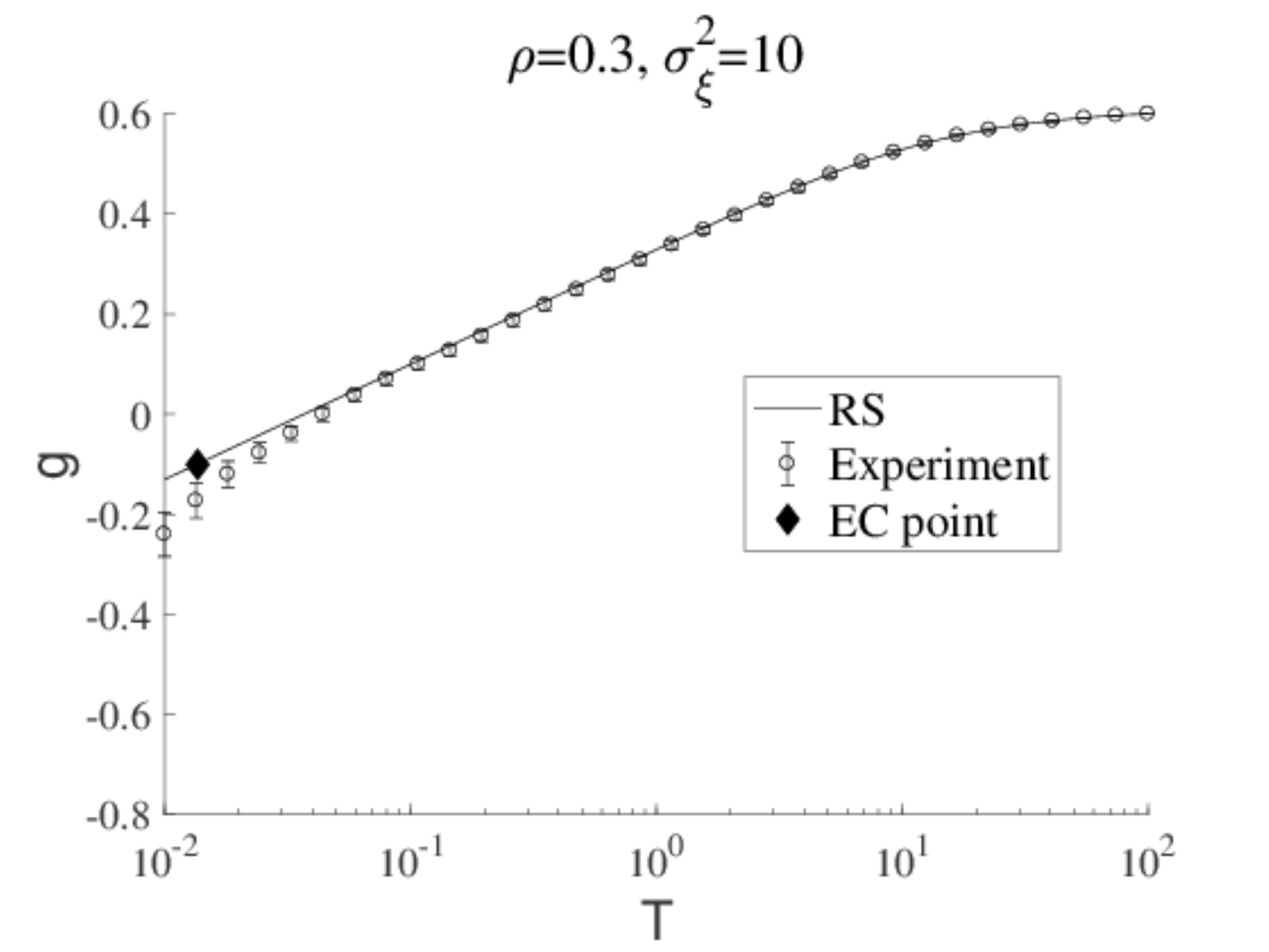}
 \includegraphics[width=0.45\columnwidth]{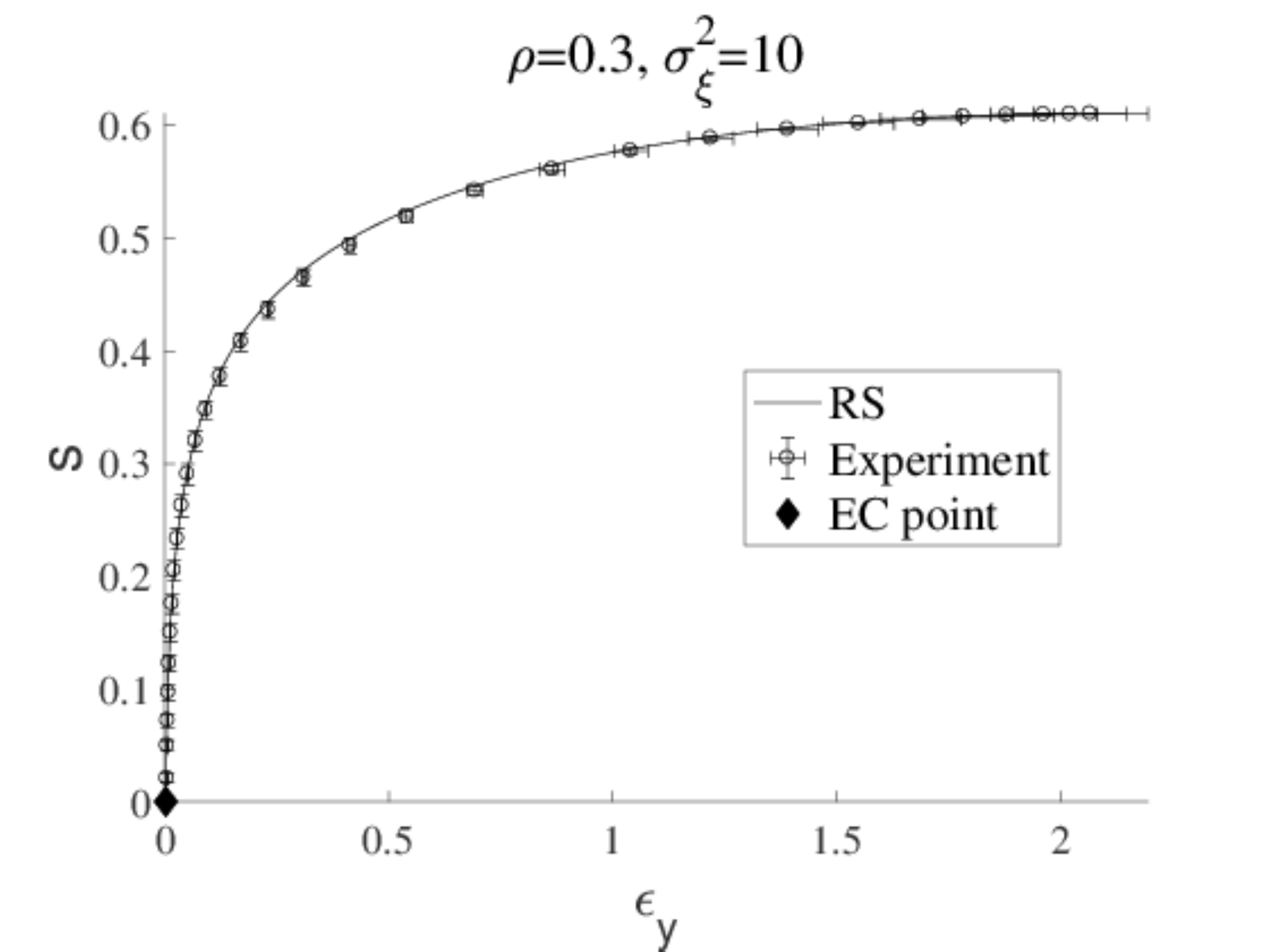}
\caption{The free-entropy curve $g(T)$ (left) and the entropy curve $s(\MSEp)$ (right) for $\rho=0.1$ (upper) and $0.3$ (lower) in the strong noise case $\sigma_{\xi}=10$. A deviation between the RS and the extrapolated values below the AT point in the left upper panel is considered to be caused by the RSB effect. }
\Lfig{MCplot-noisy}
\end{center}
\end{figure}
These figures again demonstrate good agreement between the RS and numerical results as long as the RSB does not occur. Below the RSB transition point, we observe a deviation between them, as shown in the left upper panel of \Rfig{MCplot-noisy}. In such a situation, generally speaking, the RS entropy curve can be regarded as an upper bound of the entropy values~\cite{Obuchi:10}, although a meaningful difference is not observed in the right upper panel of \Rfig{MCplot-noisy}. Again, the RSB effect on $s$ and $\MSEp$ appears to be weak.

We have a noteworthy remark to make on the EC phenomenon for $\rho=0.3$. We stress that this EC phenomenon can be described in the RS level and occurs at a {\em finite} temperature. This may be somewhat surprising for readers familiar with other models exhibiting similar EC phenomena, because in most of such systems the 1RSB treatment is needed to describe the EC phenomenon. We note that the energy levels of the present system around the ground state can be very {\em dense} and the energy gap between the ground and excited states can be extremely small, which can be argued by the fact that the PR solution in the noiseless case has numerous degeneracies for $\rho >\rho_0$. This gap is supposed to vanish in the $N\to \infty$ limit, presumably enabling the RS EC phenomenon to appear at a finite $\mu$. The agreement between the RS and numerical results strongly argues in favour of this description. Note that the EC phenomenon at small values of $\rho$ is in a different situation and its RS description is not accurate. This is because the AT instability occurs at higher temperatures in that region and hence the full step RSB treatment is needed. This is in contrast to the large-$\rho$ region in which no instability occurs at higher temperatures than $T_{\rm EC}$. 

\subsection{Monte Carlo-based optimisation and its performance}\Lsec{Monte Carlo-based}
The SA is a metaheuristic solver of generic optimisation problems based on the MC method. A variant of the SA for the present problem was proposed in~\cite{Obuchi:16-1} and its performance was examined in a limited parameter region of the present synthetic model and in a real astronomical dataset~\cite{Obuchi:16-1,Obuchi:16-2}. We re-examine this over a wider range of parameters to provide more quantitative information.

Our SA algorithm is summarised in \Rcode{SA}.
\begin{algorithm}[htbp]
\caption{SA for variable selection in sparse linear regression}\Lcode{SA}
\begin{algorithmic}[1]
\Procedure{SA}{$\{\mu_a,\tau_a \}_{a=1}^{L_{\mu}},\rho,\V{y},A$} 
	\State Generate a random initial configuration $\V{c}$ with $\sum_{i}c_i=N\rho$
	\For {$a=1:L_{\mu}$} \Comment{Changing temperature}
   		\For {$t=1:\tau_{a} $} 	\Comment{Sampling at $\mu=\mu_a$}
	   		\For {$i=1:N$} 		\Comment{Extensive number of updates}
   				\State $\V{c}\lA {\rm MC_{PF}}(\V{c},\mu_a,\V{y},A)$   \Comment{MC update with pair flipping}
	   		\EndFor
			\State \# Calculate the MSEs $\MSEx(\V{c}_t), \MSEp(\V{c}_t)$ of the current support vector $\V{c}_t=\V{c}$
  		\EndFor 
		\State \# Calculate the average as $\dAve{\MSEp} \approx (1/\tau_a)\sum_{t=1}^{\tau_a}\MSEp(\V{c}_t)$.  
	\EndFor
	\State \Return $\V{c}$ 
\EndProcedure
\end{algorithmic}
\end{algorithm}
The lines marked with \# are not necessarily needed for SA, but have been inserted for later convenience. In \Rcode{SA}, we have a set of inverse temperature points $\{ \mu_a \}_{a=1}^{ L_{\mu} }$ arranged in ascending order $(0=)\mu_{1}<\mu_{2}<\cdots<\mu_{L_{\mu}}(\gg1)$ and the waiting times $\{\tau_a \}_{a}$ at those points. Hence, as the algorithm proceeds, the temperature of the system $T=1/\mu$ decreases step by step. It is theoretically guaranteed that if the schedule of the decreasing temperature is slow enough, then the SA can find the optimal solution~\cite{Geman:84}. However, the guaranteed schedule is usually overcautious and in many practical situations we may choose a faster one. The actual schedule examined below consists of $ L_{\mu}=200$ temperature points chosen as
\be
&&
\mu_{a}=
\left\{
\begin{array}{cc}
 0.02 \cdot a &   (a=1,\cdots,50)  \\ 
 10^{0.04 \cdot (a-50)} & (a=51,\cdots,200=L_{\mu})     
\end{array}
\right.,
\ee
The first linear region of the schedule is simply inserted for visibility in the plots shown below and the important point is that the schedule is exponentially increasing as $a$ grows. The final temperature is very low, $T_{ L_{\mu}}=\mu_{ L_{\mu}}^{-1}=10^{-6}$. The waiting time at each temperature point is kept constant, at $\tau_a=\tau~(\forall{a})$, for simplicity. We show below that this rapid schedule works very efficiently for a wide range of parameters and discuss that the performance is closely related to the system's property at equilibrium, which was already calculated in \Rsec{Analytical}.

A noteworthy remark applies to the computational cost of this SA algorithm. This cost can be formally written as $O(L_{\mu}\tau N C_{\rm MC})$, where the last factor is the computational cost of each MC update. The most expensive operation is the matrix inversion required to calculate the energy of the output MSEs. If we use simple multiplication and Gauss elimination in the inversion process for each step, then $C_{\rm MC}=O(M(N\rho)^2+(N\rho)^3)$. However, we employ pair flipping in each update and the change in the relevant matrices in each update is small and successive. Using this fact and the matrix inversion formula, the total cost of each MC update can be reduced to $C_{\rm MC}=O((N\rho)^2+MN\rho)=O(N^2\alpha \rho)$, as explained in~\cite{Obuchi:16-1}. Hence, if $L_{\mu}$ and $\tau$ do not scale with $N$ and can be kept constant, the total computational cost is $O(L_{\mu}\tau N C_{\rm MC})=O(L_{\mu}\tau \alpha \rho N^3)$ and is scaled as the third order polynomial of the system size $N$. This is comparable with the versatile algorithms solving the $\ell_1$ relaxation and thus the present algorithm solves the $\ell_0$ problem with fairly reasonable computational cost. The assumption of constant $L_{\mu}$ and $\tau$ is not trivial, but it appears to be correct, i.e. sufficient to find the PR solution in the successful region, in the region we have numerically searched. Hence, we adopt this constant assumption below. 

\subsubsection{Reconstruction performance of simulated annealing}\Lsec{Reconstruction performance of}
\paragraph{Noiseless case}
Let us begin by showing the results for the noiseless case. The MSEs at $\alpha=0.5$ and $\rho_0=0.2$ are plotted against $T$ in \Rfig{SA-noiseless}. 
\begin{figure}[htbp]
\begin{center}
 \includegraphics[width=0.32\columnwidth]{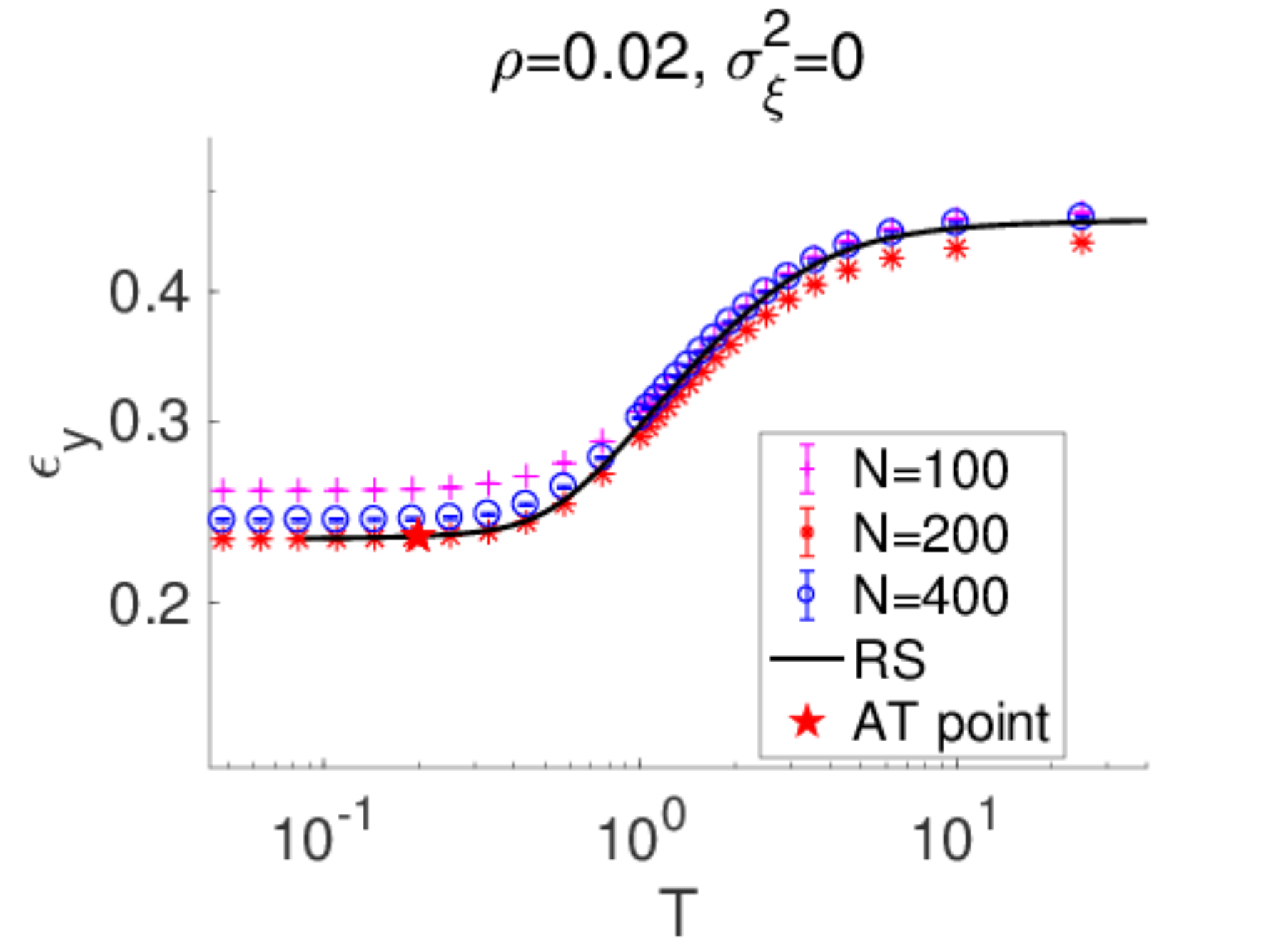}
 \includegraphics[width=0.32\columnwidth]{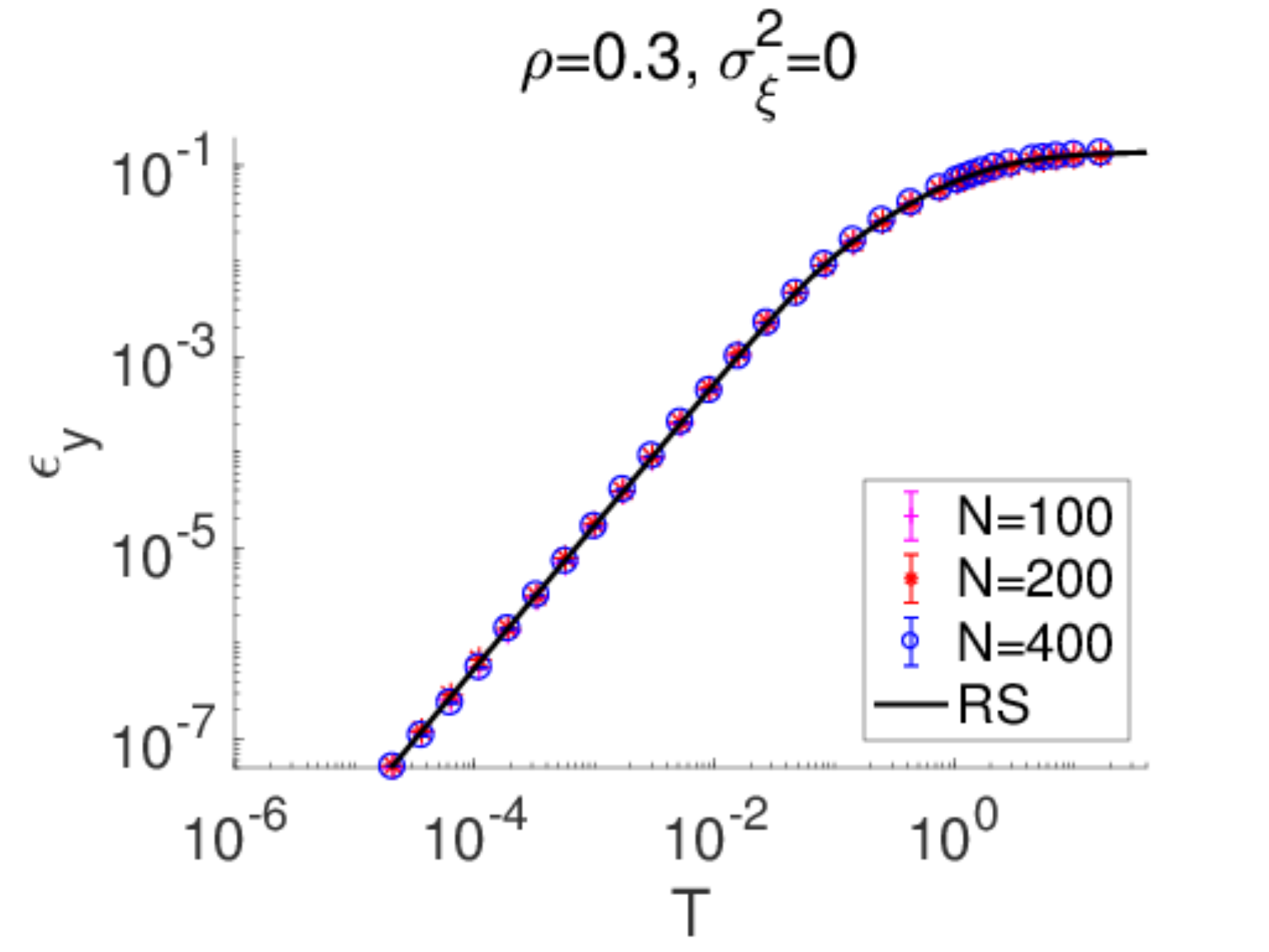}
 \includegraphics[width=0.32\columnwidth]{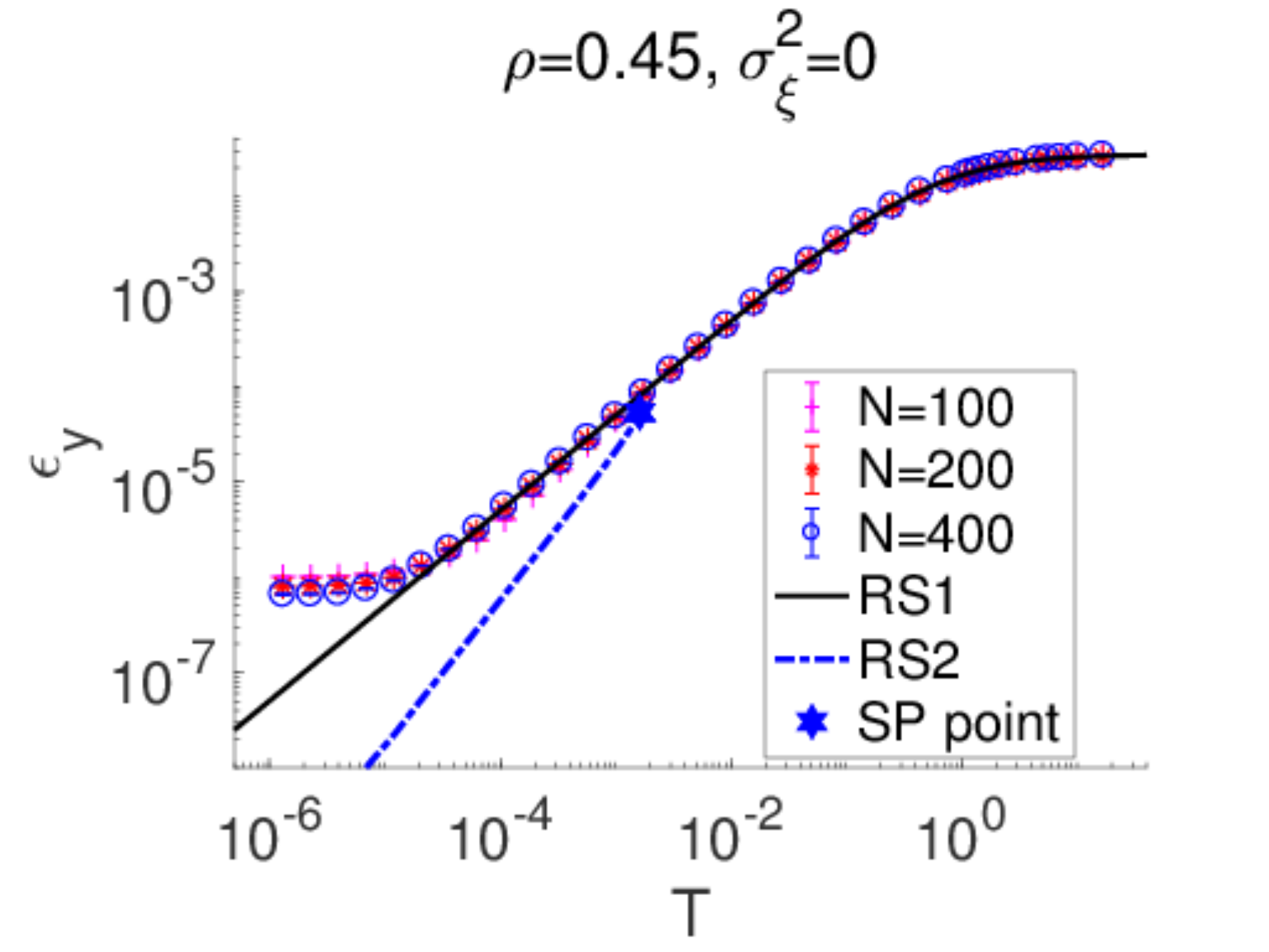}
 \includegraphics[width=0.32\columnwidth]{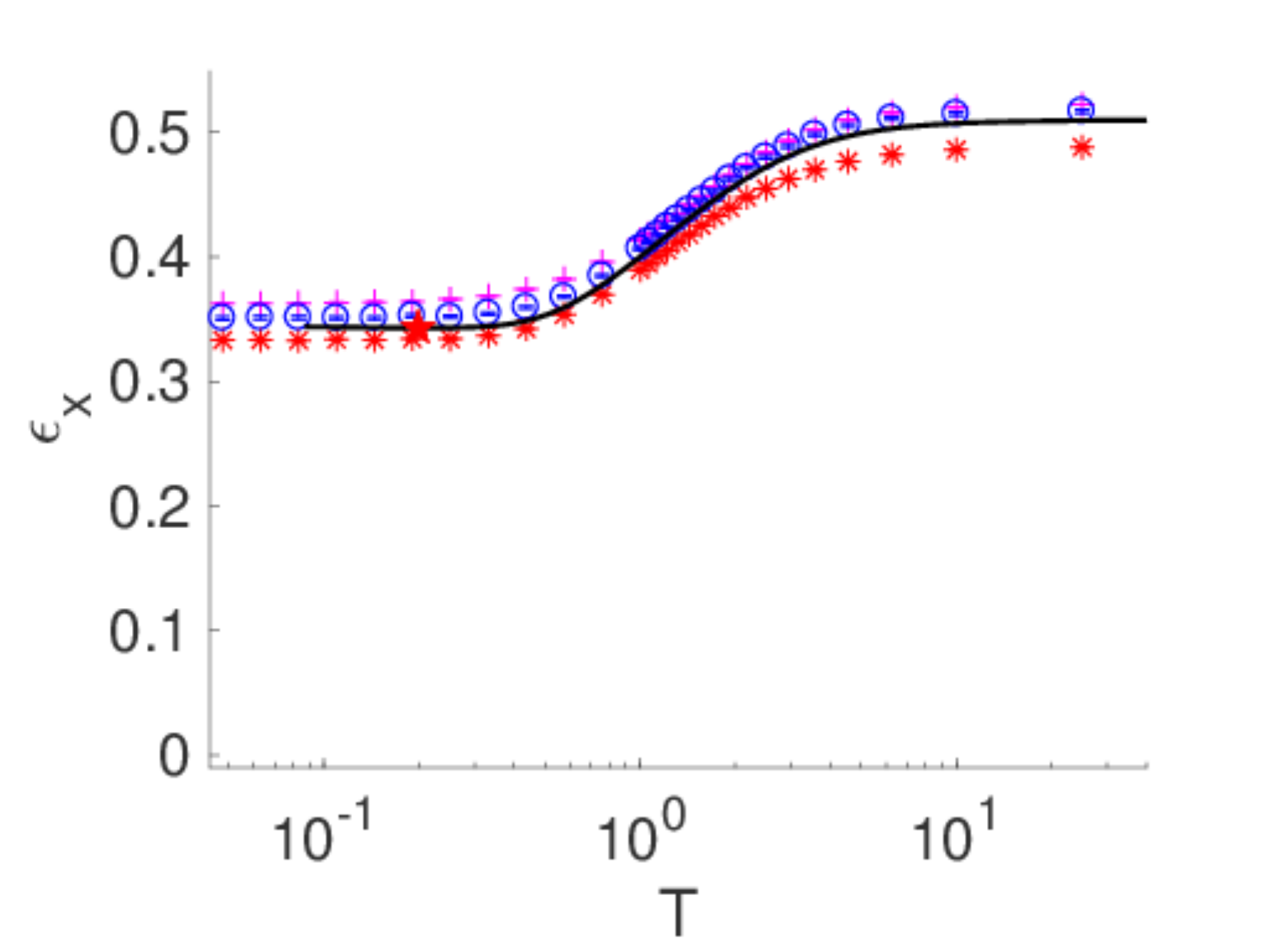}
 \includegraphics[width=0.32\columnwidth]{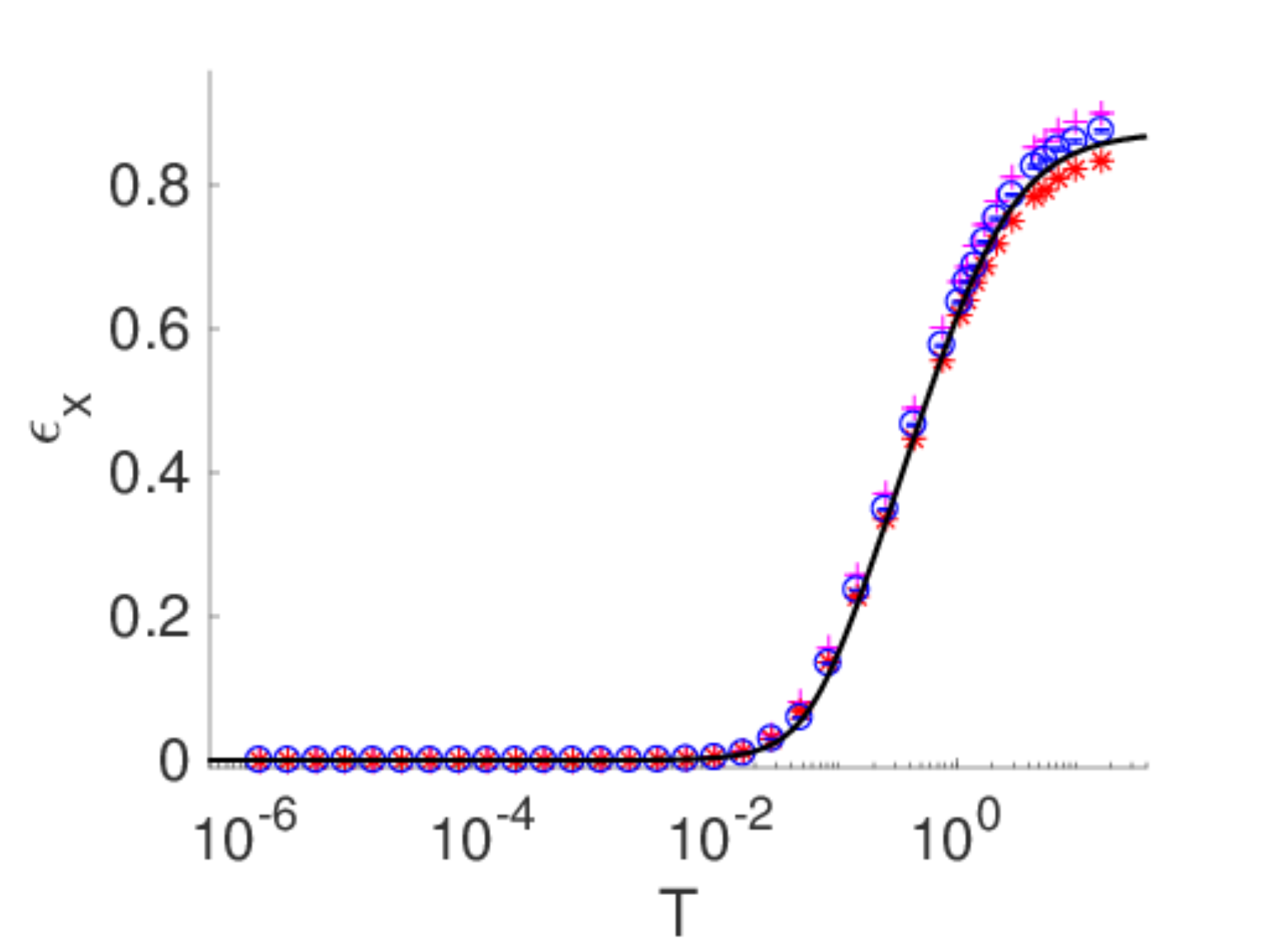}
 \includegraphics[width=0.32\columnwidth]{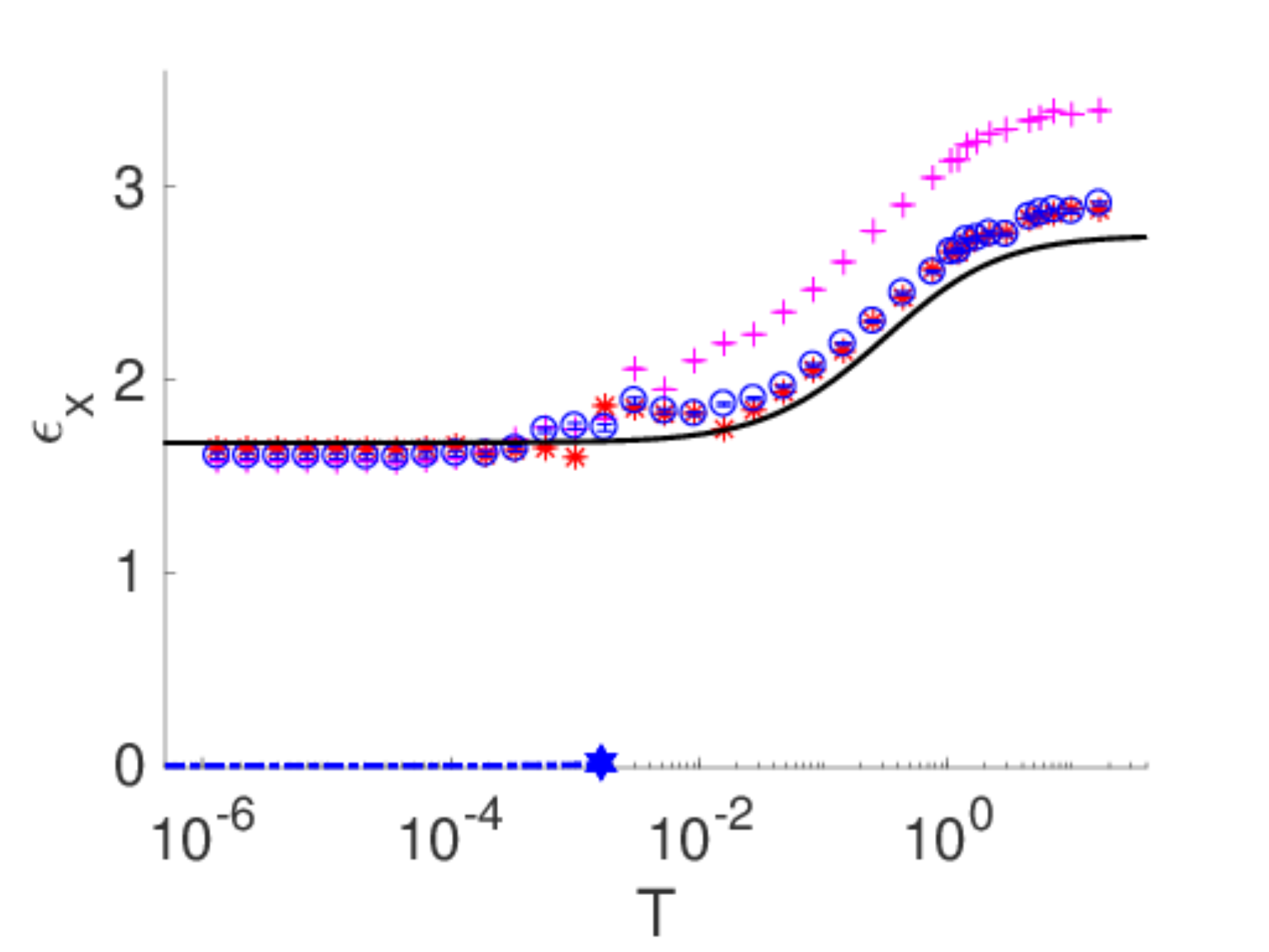}
\caption{
SA performance for the noiseless case at $\alpha=0.5$ and $\rho_0=0.2$. The output MSE $\MSEp$ (upper) and the input MSE $\MSEx$ (lower) are plotted against temperature $T$. The values of $\rho$ are $0.02$ (left), $0.3$ (middle), and $0.45$ (right), respectively. The MCS is fixed at $\tau=100$; the number of averages are $800,~200$ and $50$ for $N=100,~200,$ and $400$, respectively. For visibility, $\MSEp$ is plotted in the double logarithm scale while $\MSEx$ is in the semi-logarithmic one. The black solid and blue dashed dotted lines show the RS analytical solutions. 
}
\Lfig{SA-noiseless}
\end{center}
\end{figure}
The assumed values of $\rho$ are $\rho=0.02$, $\rho=0.3$, and $\rho=0.45$ for the left, middle, and right panels, respectively. The numerical results agree well with the black solid line representing the RS solution connected to the high temperature limit in all cases. The middle panels show the successful region for finding the PR solution, $\rho_{0} < \rho < \rho_{\rm SP}$, and the vanishing $\MSEx$ means that we actually find the PR solution. The right panels are in $\rho_{\rm SP} < \rho $, meaning that the search is trapped in the metastable state connected to high temperatures. The SA result follows the high-temperature branch and cannot reach the low-temperature one denoted by the blue dashed dotted line. Overall, the SA experiments demonstrate that our theoretical predictions are very precise, and the presence of phase transitions strongly degrades the SA's performance in finding the minimum-MSE configuration. 

\paragraph{Noisy case}
Next, we show the results of the noisy case. The SA results for the strong noise case $\sigma_{\xi}^2=10$ at $\alpha=0.5$ and $\rho_0=0.2$ are given in \Rfig{SA-strongnoise}. 
\begin{figure}[htbp]
\begin{center}
 \includegraphics[width=0.32\columnwidth]{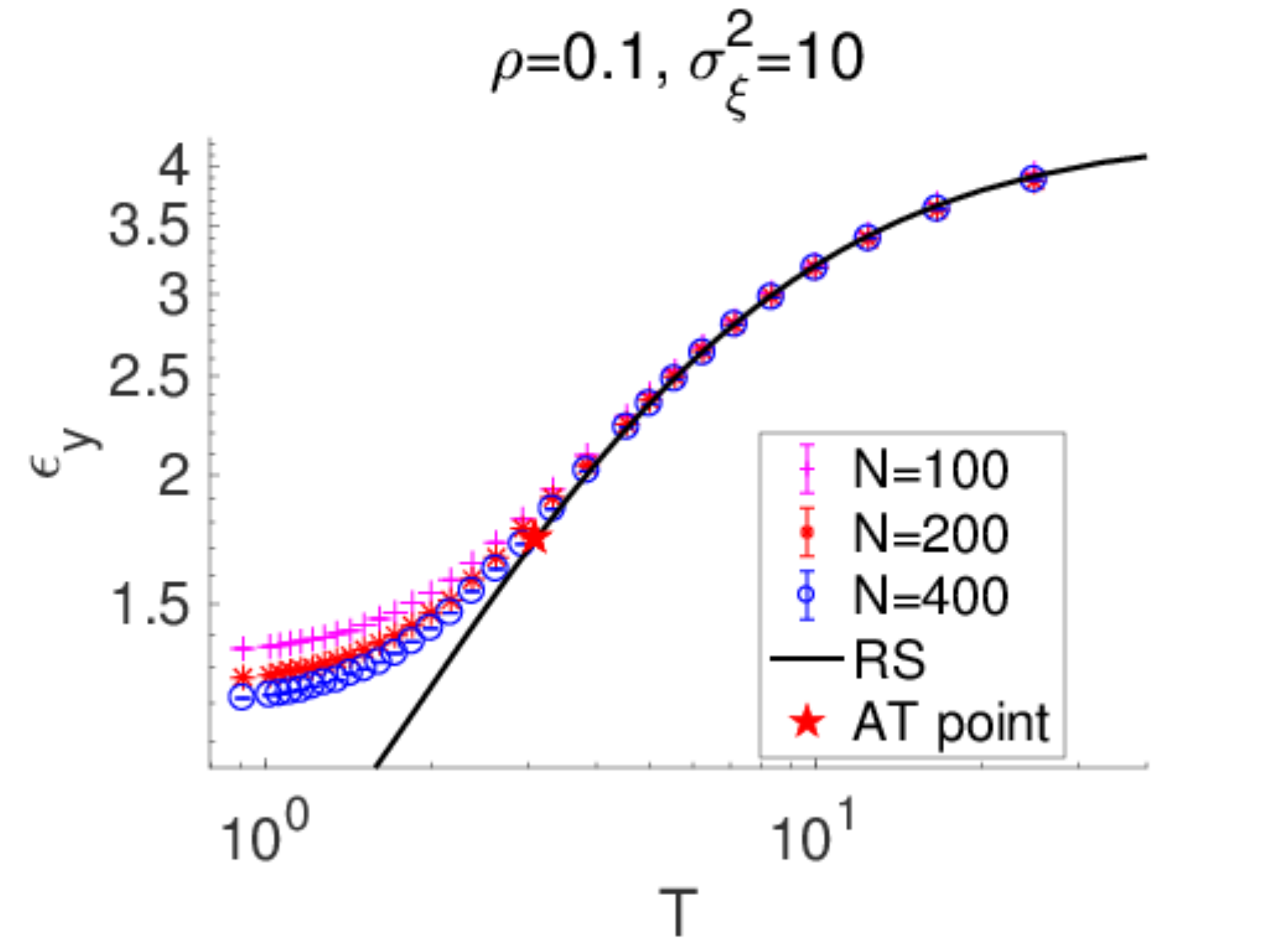}
 \includegraphics[width=0.32\columnwidth]{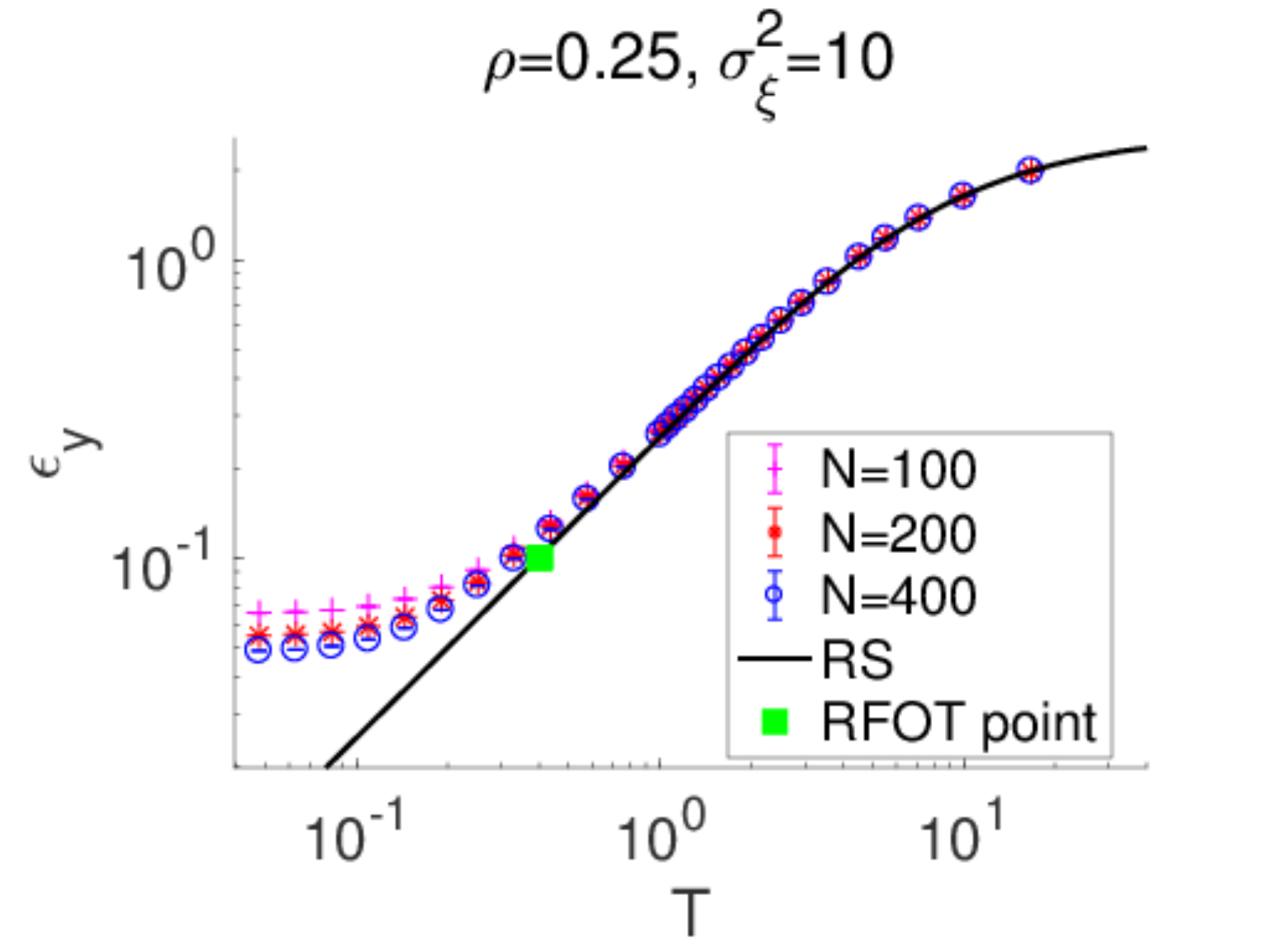}
 \includegraphics[width=0.32\columnwidth]{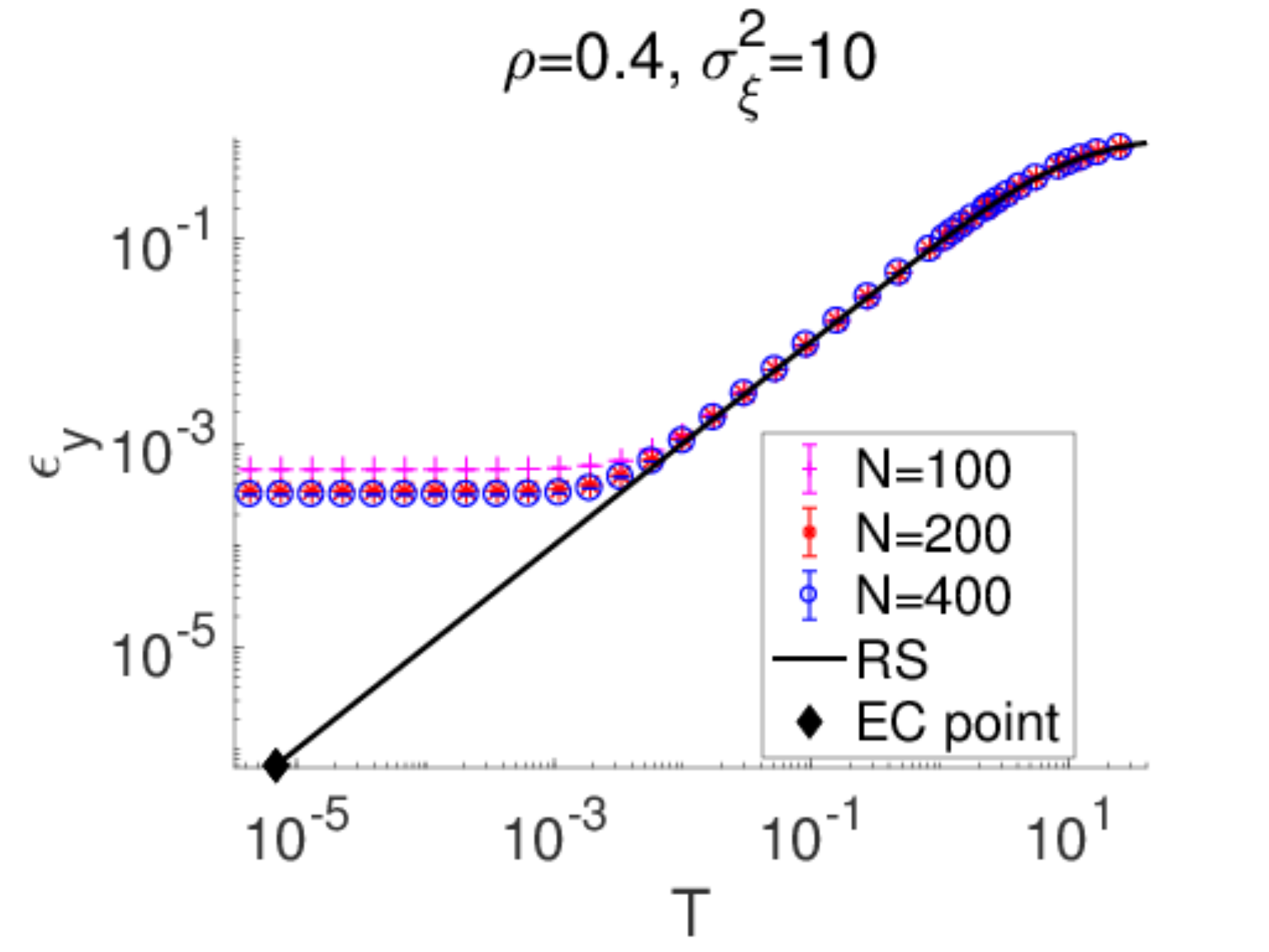}
 \includegraphics[width=0.32\columnwidth]{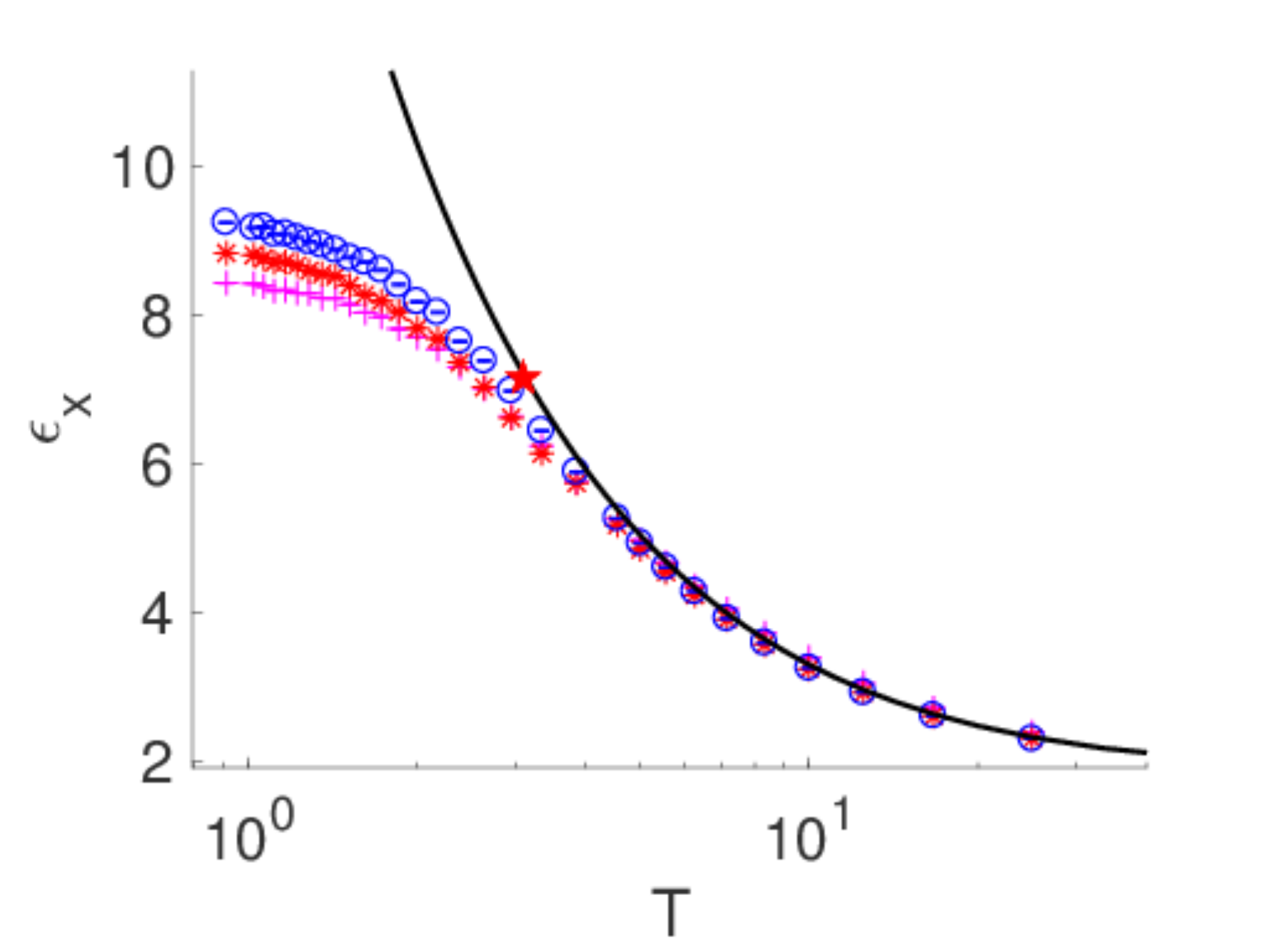}
 \includegraphics[width=0.32\columnwidth]{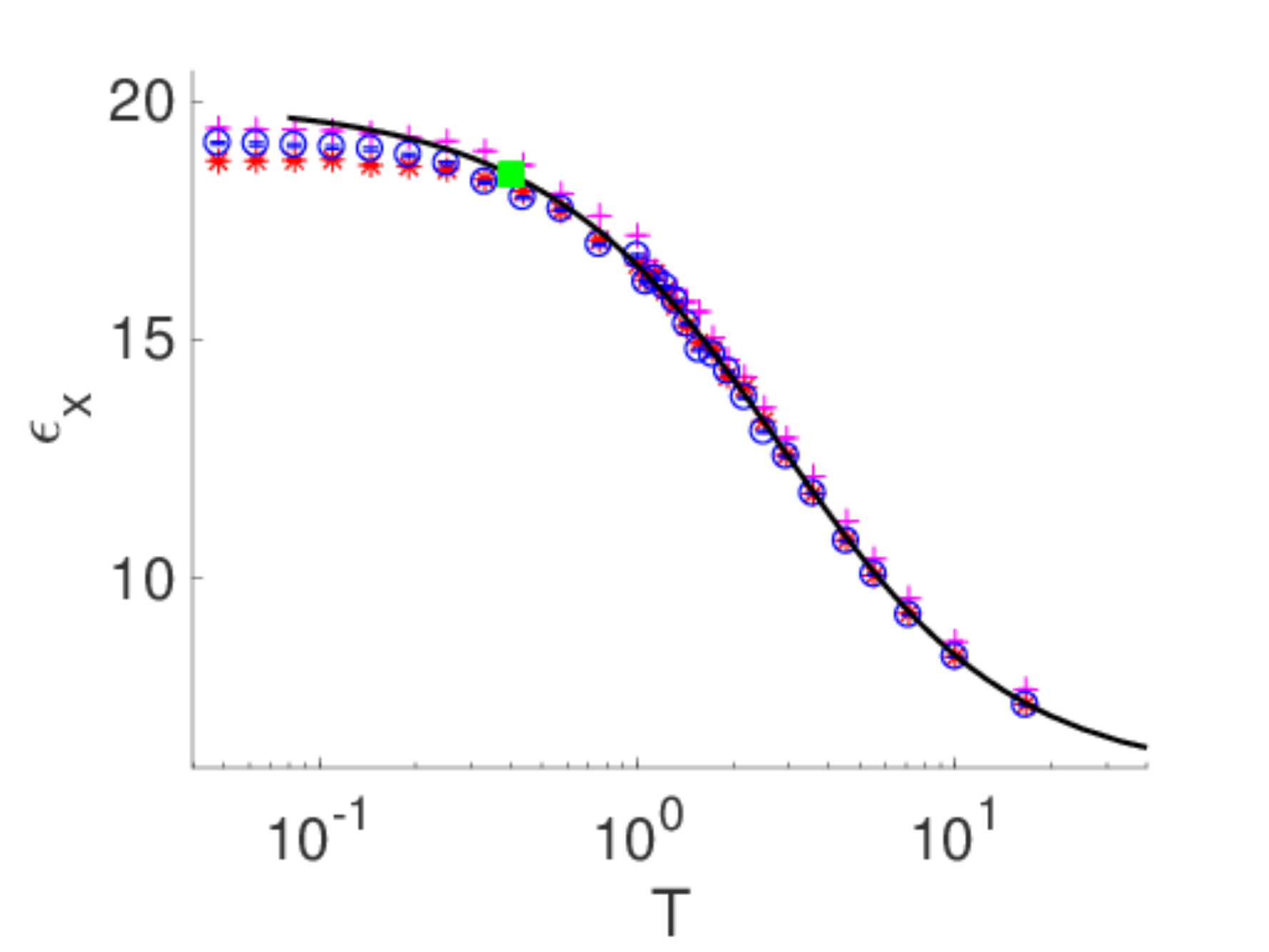}
 \includegraphics[width=0.32\columnwidth]{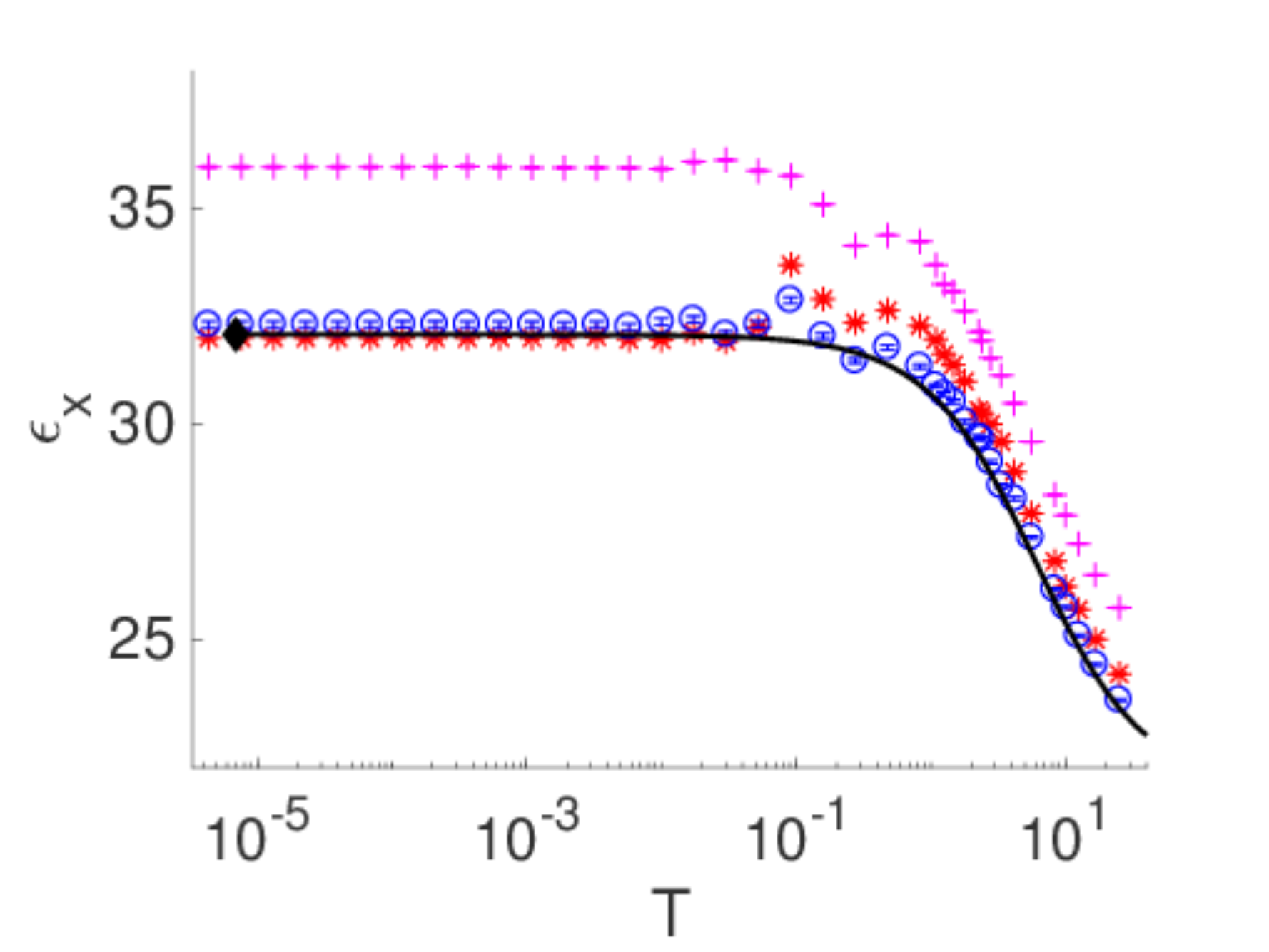}
\caption{
SA performance for the strong noise case $\sigma_{\xi}=10$ at $\alpha=0.5$ and $\rho_0=0.2$. The output MSE $\MSEp$ (upper) and the input MSE $\MSEx$ (lower) are plotted against temperature $T$. The values of $\rho$ are $0.1$ (left), $0.25$ (middle), and $0.4$ (right). The MCS is fixed at $\tau=100$; the number of averages are $800,~200$, and $100$ for $N=100, ~200,$ and $400$, respectively. }
\Lfig{SA-strongnoise}
\end{center}
\end{figure}
As seen from the figure, the MSEs show good agreement to the analytical curve (black solid line) up to the transition points for the left and middle panels. An exceptional deviation is observed at low temperatures in the upper right panel, but this is considered to be a finite-size effect because the range of $\MSEp$ in this region is very small and supposedly unreachable by $N\approx 100$ systems. Hence, these behaviours are very consistent with the analytical predictions that the system's dynamic behaviour is affected by the RSB transitions and ceases to follow the equilibrium state.

The effect of the RSB on the reconstruction performance becomes much clearer by examining the achievable-limit values of the MSEs for a moderate noise case. \Rfig{SA-rhoVSeps} shows the plots against $\rho$ of the limit values obtained at very low temperatures by the rapid SA with $\tau=5$ for $\sigma_{\xi}^2=0.1$. 
\begin{figure}[htbp]
\begin{center}
 \includegraphics[width=0.48\columnwidth]{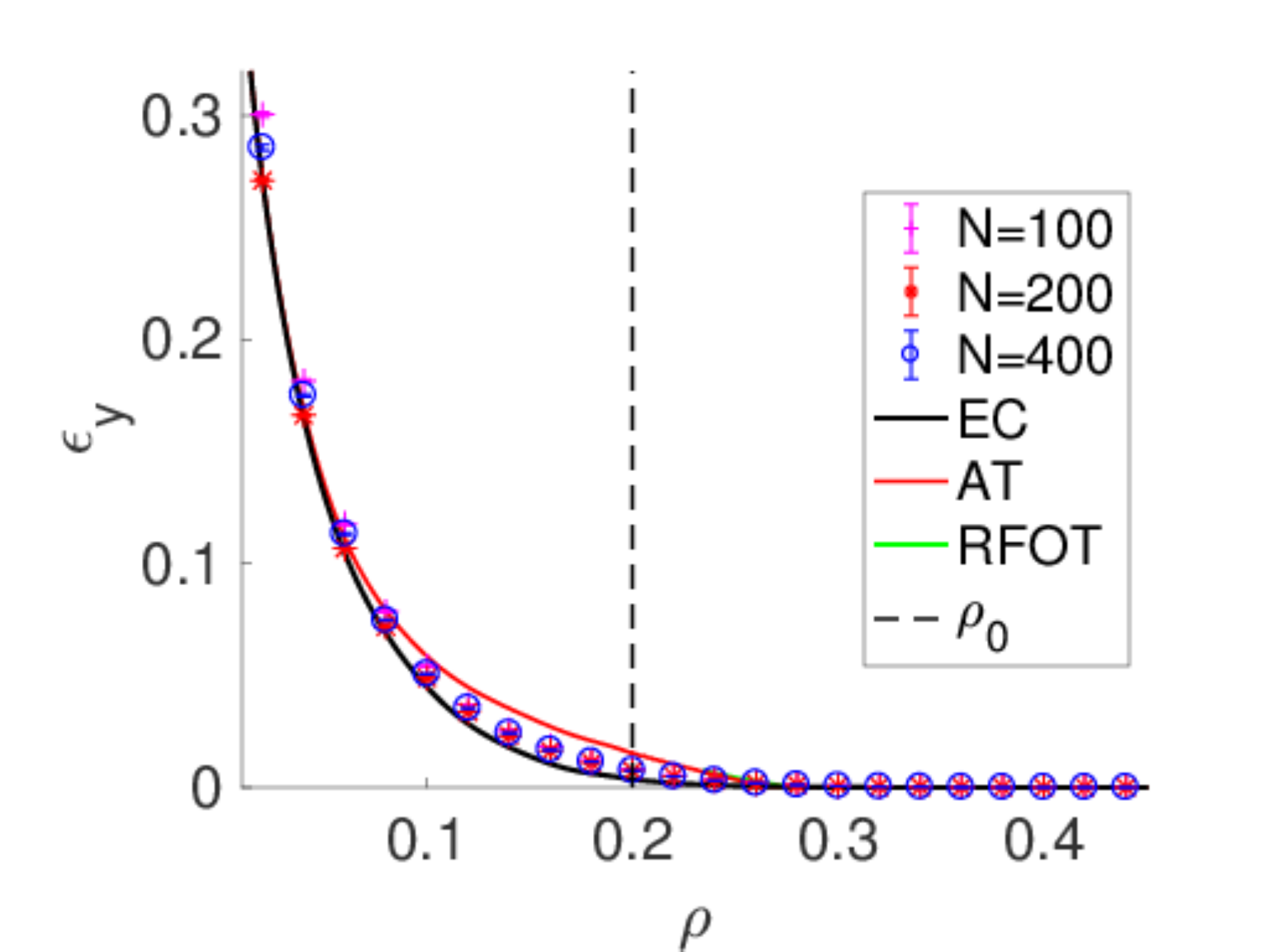}
 \includegraphics[width=0.48\columnwidth]{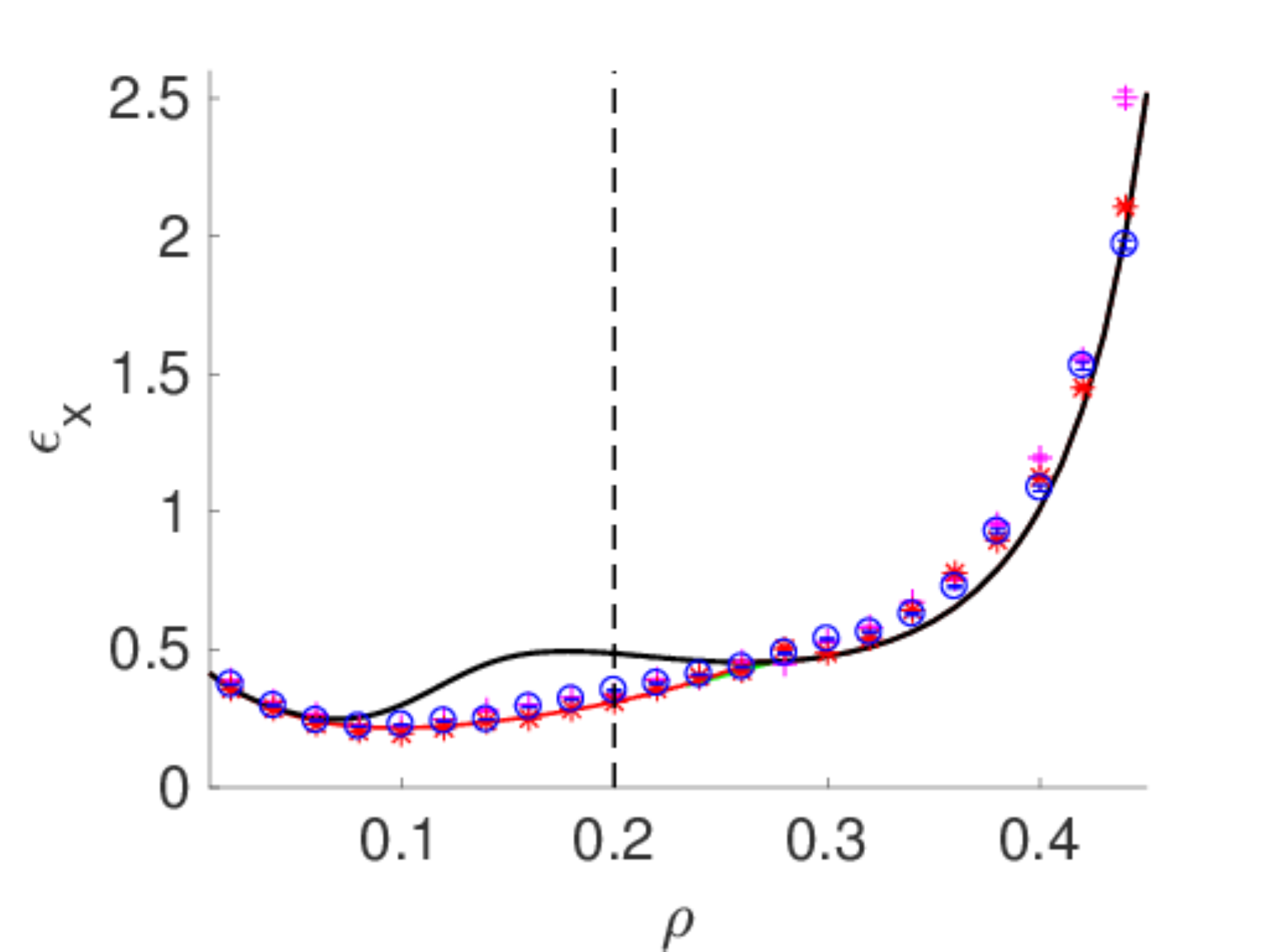}
\caption{
Plots versus $\rho$ of the limiting values of the MSEs $\MSEp$ (left) and $\MSEx$ (right) obtained at very low temperatures for $\alpha=0.5$, $\rho_0=0.2$ and $\sigma_{\xi}^2=0.1$. Three analytical values, where the RS entropy crisis, the AT instability, and the RFOT transition occur, are denoted by black, red, and green solid lines, respectively, although the green line only exists in an extremely small region and is difficult to observe. The simulation results (markers) obtained by the SA with a very rapid schedule $\tau=5$ agree well with the analytical curves. The number of averages are $100,~100$ and $40$ for $N=100,~200,$ and $400$, respectively. }
\Lfig{SA-rhoVSeps}
\end{center}
\end{figure}
The $\MSEx$ values at middle $\rho$ values are clearly dominated by the ones at the AT transition points rather than the ones at the EC points, implying that the system's search is trapped by local minima emerging at the transition points. An interesting outcome of this phenomenon is a better reconstruction of the planted signal $\V{x}_0$. As seen from \Rfig{SA-rhoVSeps}, the $\MSEx$ values at the AT points are lower than the ones at the RS EC points, implying that the reconstruction performance of the local minima induced by the RSB is better than that of the minimum-$\MSEp$ configuration $\hat{\V{c}}$
by solving \Req{L0optimization}. This means that the generalisation capability of the rapid SA is no worse than exactly solving \Req{L0optimization} in this case because the input MSE $\MSEx$ is proportional to the generalisation error when each component of the design matrix and the noise is i.i.d. from the zero-mean Gaussian. This encourages the use of the presented formulation and algorithm for practical purposes, as reported in~\cite{Obuchi:16-2}.

\section{Conclusion}\Lsec{Conclusion}
In this study, we have analytically provided an algorithmic limit of an $\ell_0$-based formulation of compressed sensing through evaluation of the entropy using statistical mechanical techniques. The results are mainly characterised by the ratio of number of observations $\alpha$, the nonzero-components density of the inference and generative models $\rho$ and $\rho_0$, and the strengths of the signal and noise $\sigma_{x}$ and $\sigma_{\xi}$. The entropy curves and the associated phase diagrams have been provided, and their implications to local search algorithms have been discussed. Quantitative analysis of the noisy cases has also been conducted. To validate the analytical computation, we have performed a careful MC simulation using the exchange MC method and the multi-histogram method, the results of which have exhibited fairly good agreement with the analytical results. To test the theoretical predictions on local search algorithms, we have also performed numerical experiments using the SA-based algorithm. The performance of the SA is well understood through the phase diagrams. Over a wide parameter region in the noiseless case, we have actually located the PR solution with reasonable computational cost of $O(N^3)$ although, in other hard regions, the RSB and the RS first-order transitions prevent rapid convergence of the SA to the PR solution. 

To overcome the problems caused by phase transitions, it may be interesting to tailor new algorithms based on MC methods. The idea of extended ensembles can be useful: It will be a promising future work to invent an algorithm relaxing the hard constraint on the nonzero components density, by introducing a ``chemical potential'' softly controlling the density. The idea of multi-canonical sampling such as the Wang--Landau algorithm~\cite{Wang:01} may also be an interesting direction. By extending the work in~\cite{Antenucci:18}, it is also worth trying to take into account glassy natures in algorithms in certain ways. 

Relaxing the i.i.d. random-matrix assumption on $A$ is also an interesting problem. This is even practical because in an ideal situation of compressed sensing, a new observation should be conducted along with maximising the resultant information, implying that all rows of $A$ should be orthogonal. Considering such orthogonal ensembles in the presented framework is a high-priority issue which should be done in near future.

\section*{Acknowledgement}
This work was supported by JSPS KAKENHI Grant Numbers 26870185 and 18K11463 (TO), 17K12749 (YN-O), 25120009 (MO), and 25120013 and 17H00764 (YK). TO is also supported by a Grant for Basic Science Research Projects from the Sumitomo Foundation. YN-O is also supported by JST PRESTO Grant Number JPMJPR1773.

\appendix
\section{Calculations of $g(\mu)$} \Lsec{g-calculation}
Assuming $n$ and $\nu$ are positive integers, we can rewrite
\be
&&
\Phi(n,\beta,\nu) \equiv
  \lsb
   \lbb
    \Tr{\V{c}} 
      \lb 
         \Tr{\V{x}|\V{c}} \mathrm{e}^{-\frac{1}{2}\frac{\mu}{\nu}||\V{y}-\V{A}(\V{c}\circ \V{x}) ||_2^2   }
      \rb^{\nu}
   \rbb^n
  \rsb_{\V{x}_0,\V{\xi},\V{A}},
\no \\ &&
=\Tr{ \{ \V{c} \} } \Tr{ \{ \V{x} \} |\{\V{c} \}}
\lsb
e^{
-\frac{1}{2}\beta \sum_{a=1}^{n}\sum_{\alpha=1}^{\nu}\sum_{\mu=1}^{M}
\lb
\sum_{i=1}^{N}a_{\mu i}\lb x_{0i}-c^{a}_i x^{a\alpha}_{i}\rb 
+\xi_{\mu}
\rb^2
}
\rsb_{\V{x}_0,\V{\xi},A }.
\Leq{Phi}
\ee
These summations over $\V{c}$ and $\V{x}$ are now calculated over all the replicated variables $\{\V{c}^{a}\}_{a=1}^{n}$ and  $\{\V{x}^{a\alpha}\}_{a=1,\cdots,n,\alpha=1,\cdots,\nu}$. Let us set
\be
d^{a\alpha}_{\mu}\equiv \sum_{i=1}^{N}a_{\mu i}\lb  x_{0i}-c^{a}_i x^{a\alpha}_{i}\rb.
\ee
The variable $d$ is an extensive sum of random variables and can be expressed by an appropriate sum of Gaussian variables with a certain covariance. The covariance is expressed by
\be
\lsb d^{a\alpha}_{\mu}d^{b\beta}_{\nu} \rsb_{A}
=\delta_{\mu\nu}\lb 
\frac{1}{N}\sum_{i}x^2_{0i}
-\frac{1}{N}\sum_{i}x_{0i}c^a_i x^{a\alpha}_{i}
-\frac{1}{N}\sum_{i}x_{0i}c^b_i x^{b\beta}_{i}
+\frac{1}{N}\sum_{i}c^a_i c^{b}_i x^{a\alpha}_{i}x^{b\beta}_{i}
\rb.
\Leq{covariance}
\ee
Evaluating this full description is difficult in general. Instead, we consider more amenable subspaces which are described by the RS or RSB ansatz. 

\subsection{RS computation} \Lsec{The RS}
In the RS ansatz, the dominant contribution is assumed to come from the following subspace: 
\be
&&
R=\frac{1}{N}\sum_{i}c^a_i \lb x^{a\alpha}_{i} \rb^2,
\\
&&
Q=\frac{1}{N}\sum_{i}c^a_i x^{a\alpha}_{i} x^{a\beta}_{i}, \,\, (\alpha \neq \beta),
\\
&&
q=\frac{1}{N}\sum_{i}c^a_ic^b_i x^{a\alpha}_{i} x^{b\beta}_{i},\,\, (a \neq b),
\\
&&
m=\frac{1}{N}\sum_{i}x_{0i}c^a_i x^{a\alpha}_{i},
\\
&&
\rho_0\POWx=\frac{1}{N}\sum_{i}x^2_{0i},
\ee
and hence the covariance matrix is described by four order parameters and one external parameter $\rho_0\POWx$. The corresponding useful description of $d^{a\alpha}_{\mu}$ is 
\be
d^{a\alpha}_{\mu}=\sqrt{R-Q}u^{a\alpha}_{\mu}+\sqrt{Q-q}v^{a}_{\mu}+\sqrt{\rho_0\sigma_{x}^2-2m+q}z_{\mu},
\ee
where $u,v$  and $z$ are i. i. d. from $\mathcal{N}(0,1)$. The average with respect to $A$ thus can be replaced by the average over $u,v$ and $z$, yielding
\be
\Phi=\int dR dQ dq dm~\mathcal{I} \times \mathcal{L},
\ee
where $\mathcal{I}$ is the subshell (the state density) characterised by the above four order parameters such that
\be
&&
\mathcal{I}=\Tr{ \{ \V{c} \} } \Tr{ \{ \V{x} \} |\{\V{c} \}}
\prod_{a,\alpha}\delta \lb NR-\sum_{i}c_i^{a}\lb x_i^{a\alpha}\rb^2 \rb 
\prod_{a,\alpha<\beta}\delta \lb N Q-\sum_{i}c_i^{a} x_i^{a\alpha}x_i^{a\beta} \rb 
\no \\
&&
\times
\prod_{a<b,\alpha,\beta}\delta \lb N q-\sum_{i}c_i^{a}c_i^{b} x_i^{a\alpha}x_i^{b\beta} \rb 
\lsb
\prod_{a,\alpha}\delta \lb N m-\sum_{i} x_{0i}c_i^{a} x_i^{a\alpha} \rb,
\rsb_{\V{x}_0}
\ee
and $\mathcal{L}$ describes 
\be
&&
\mathcal{L}=
\prod_{\mu=1}^{M}\lbb 
\int Dz_\mu \int d \xi_\mu \frac{e^{-\frac{1}{2\sigma_{\xi}^2}\xi_{\mu}^2 }}{\sqrt{2\pi \sigma_{\xi}^2}} \int \prod_{a}Dv^{a}_{\mu}\int \prod_{a,\alpha}Du^{a\alpha}_{\mu}
\prod_{a,\alpha}e^{-\frac{1}{2}\beta (d^{a\alpha}_{\mu}+\xi_{\mu})^2}
\rbb
\no \\
&&
=\lbb 
\int Dz \lb \int Dv \lb \int Du~e^{-\frac{1}{2}\beta h^2(u,v,z)} \rb^{\nu} \rb^{n}
\rbb^{M}\equiv L^M,
\Leq{L}
\ee
where we merge two Gaussian variables $(z,\xi)$ into $z$ and
\be
h(u,v,z)=\sqrt{R-Q}u+\sqrt{Q-q}v+\sqrt{\rho_0\sigma_{x}^2+\sigma_{\xi}^2-2m+q}z.
\ee
Direct integrations yield
\be
&&
\log \int Dz \lb  \int Dv \lb \int Du~e^{-\frac{1}{2} \beta h^2}\rb^{\nu} \rb^n
\approx n\int Dz \log \int Dv \lb \int Du e^{-\frac{1}{2} \beta h^2}\rb^{\nu}
\no \\ &&
=-n\lbb
\frac{\nu}{2}\log (1+\beta(R-Q))+\frac{1}{2}\log \frac{1+\beta(R-Q)}{1+\beta(R-Q)+\beta \nu\Delta_{\mathrm{RS}}}
-\frac{1}{2}\frac{\beta \nu (\rho_0\sigma_{x}^2+\sigma_{\xi}^2-2m+q)}{1+\beta(R-Q)+\beta \nu \Delta_{\mathrm{RS}}}
\rbb
\no \\ &&
\to
\frac{n}{2}\lbb \log \frac{1+\chi}{1+\chi+\mu(Q-q)}-\frac{\mu(\rho_0 \POWx+\sigma_{\xi}^2-2m+q)}{1+\chi+\mu(Q-q)}
\rbb,
\Leq{L content}
\ee
where we keep only the linear term with respect to $n$ at the first line and take the limit $\nu\to 0$ while keeping $\beta \nu=\mu$ at the last line and applying $\beta (R-Q)=\chi$ according to \Req{limit-chi}. Summarising \Reqs{L}{L content}, we obtain
\be
\mathcal{L}\approx 
e^{
N n
\frac{\alpha}{2}\lbb \log \frac{1+\chi}{1+\chi+\mu(Q-q)}-\frac{\mu(\rho_0 \POWx+\sigma_{\xi}^2-2m+q)}{1+\chi+\mu(Q-q)}
\rbb
}.
\Leq{L final}
\ee

Evaluation of $\mathcal{I}$ requires additional algebra. We break the delta functions by using the Fourier transform as follows
\subbe
\Leq{subshell}
\be
&&
\delta \lb \sum_{i}c_i^{a}-N\rho \rb
=\int d\tilde{\rho}~e^{N\tilde{\rho}\rho-\tilde{\rho}\sum_{i}c_i^a},
\Leq{rho-subshell}
\\ &&
\delta \lb NR-\sum_{i}c_i^{a}\lb x_i^{a\alpha}\rb^2 \rb
=\int d\tilde{R}~e^{
\frac{1}{2}N\tilde{R}R-\frac{1}{2}\tilde{R}\sum_{i}c_i^{a}( x_i^{a \alpha } )^2
}
\Leq{R-subshell}
\\ &&
\delta \lb NQ-\sum_{i}c_i^{a}x_i^{a\alpha}x_i^{a\beta} \rb
=\int d\tilde{Q}~e^{
-N\tilde{Q}Q+\tilde{Q}\sum_{i}c_i^{a} x_i^{a \alpha }x_i^{a \beta}
}
\Leq{Q-subshell}
\\ &&
\delta \lb Nq-\sum_{i}c_i^{a}c_i^{b}x_i^{a\alpha}x_i^{b\beta} \rb
=\int d\tilde{q}~e^{
-N\tilde{q}q+\tilde{q}\sum_{i}c_i^{a}c_i^{b} x_i^{a \alpha }x_i^{b \beta}
}
\Leq{q-subshell}
\\ &&
\delta \lb Nm-\sum_{i} c_i^{a}x_{0i}x_i^{a\alpha}\rb
=\int d\tilde{m}~e^{
-N\tilde{m}m+\tilde{m}\sum_{i} x_{0i}c_i^{a}x_i^{a\alpha}
}.
\Leq{m-subshell}
\ee
\subee
Then, 
\be
&&
\mathcal{I}=
\int d\tilde{\rho}d\tilde{R}d\tilde{Q}d\tilde{q}d\tilde{m}
~e^{N\lb n\tilde{\rho}\rho+\frac{1}{2}n\nu \tilde{R}R-\frac{1}{2}n\nu(\nu-1)\tilde{Q}Q-\frac{1}{2}n(n-1)\nu^2\tilde{q}q-n\nu \tilde{m}m\rb}
\no \\ &&
\times 
\lsb
\prod_{i=1}^{N}\lbb 
\sum_{\{c_i\}}\Tr{\{x_i\}|\{c_i\}}
e^{f_i\lb \{x_i\},\{c_i\} \rb}
\rbb
\rsb_{\V{x}_{0}}
,
\Leq{I}
\ee
where
\be
&&
f_i\lb \{x\},\{c\} \rb
=
-\tilde{\rho}\sum_{a}c_a
-\frac{1}{2}\tilde{R}\sum_{a}\sum_{\alpha}c_{a}x_{a\alpha}^2
\no \\ &&
+\tilde{Q}\sum_a\sum_{\alpha<\beta}c_ax_{a\alpha}x_{a\beta}
+\tilde{q}\sum_{a<b}\sum_{\alpha,\beta}c_ac_bx_{a\alpha}x_{b\beta}
+\tilde{m}x_{0i}\sum_{a}\sum_{\alpha}c_ax_{a\alpha}.
\ee
The replica indices have been superscripts so far but we rewrite them as subscripts for visibility. Using the Gaussian integrals, we break the sum $\sum_{a<b}$ into a single replica sum as
\be
&&
e^{\tilde{q}\sum_{a<b}\sum_{\alpha,\beta}c_ac_bx_{a\alpha}x_{b\beta}}
=
e^{
\frac{\tilde{q}}{2}\lbb  \lb\sum_{a}\sum_{\alpha}c_ax_{a\alpha}\rb^2
-\sum_{a}c_a\lb \sum_{\alpha}x_{a\alpha}\rb^2 \rbb
}
\no \\ &&
=
\int Dz
e^{
\sqrt{\tilde{q}}z \sum_{a}\sum_{\alpha}c_ax_{a\alpha} 
-\frac{\tilde{q}}{2}\sum_{a}c_a\lb \sum_{\alpha}x_{a\alpha}\rb^2 
}.
\ee
Hence we can take $\sum_{c_a=0,1}$ for each $a$ independently
\be
&&
\sum_{\{c\}} \Tr{\{x\}|\{c\}}
e^{f_i\lb \{x\},\{c\} \rb}
=\int Dz   
\lb 
1+
\int
\prod_{\alpha=1}^{\nu} dx_{\alpha}~
e^{r_{i}(\{x\})}
\rb^n,
\ee
where 
\be
r_{i}(\{x\})
=
-\tilde{\rho}
-\frac{1}{2}\tilde{R}\sum_{\alpha}x_{\alpha}^2
+\tilde{Q}\sum_{\alpha<\beta}x_{\alpha}x_{\beta}
+\sqrt{\tilde{q}}z\sum_{\alpha}x_{\alpha}
-\frac{1}{2}\tilde{q}\lb \sum_{\alpha}x_{\alpha} \rb^2
+\tilde{m}x_{0i}\sum_{\alpha}x_{\alpha}.
\ee
The sums $\sum_{\alpha<\beta}$ and $(\sum_{\alpha}x_{\alpha})^2$ can also be broken
\be
e^{\tilde{Q}\sum_{\alpha<\beta}x_{\alpha}x_{\beta}-\frac{\tilde{q}}{2}\lb \sum_{\alpha}x_{\alpha} \rb^2}
=e^{-\frac{1}{2}\tilde{Q}\sum_{\alpha}x_{\alpha}^2}\int Dy~e^{\sqrt{\tilde{Q}-\tilde{q}}y\sum_{\alpha}x_{\alpha}}
\ee
Hence,
\be
&&
\int \prod_{\alpha=1}^{\nu} dx_{\alpha}~e^{r_{i}(\{x\})}
=
e^{-\tilde{\rho}}
\int Dy
\prod_{\alpha}\lb \int dx_{\alpha} e^{-\frac{1}{2}(\tilde{R}+\tilde{Q})x_{\alpha}^2+\lb \sqrt{\tilde{Q}-\tilde{q}}y+\sqrt{\tilde{q}}z+ \tilde{m}x_{0i}  \rb x_{\alpha}} \rb
\no \\ &&
=
\frac{e^{-\tilde{\rho}}  \sqrt{2\pi}^{\nu} }{\sqrt{\tilde{R}+\tilde{Q}}^{\nu} }
\int Dy 
e^{
\frac{1}{2}\nu \frac{ \lb \sqrt{\tilde{Q}-\tilde{q}}y+\sqrt{\tilde{q}}z+ \tilde{m}x_{0i}  \rb^2 }{\tilde{R}+\tilde{Q}} 
}.  
\ee
Summarising the result yields
\be
\sum_{\{c_i\}}\Tr{\{x_i\}|\{c_i\}}
e^{f_i\lb \{x_i\},\{c_i\} \rb}
=
\int Dz \lb 1+\frac{e^{-\tilde{\rho}} \sqrt{2\pi}^{\nu} }{\sqrt{\tilde{R}+\tilde{Q}}^{\nu} }
\int Dy 
e^{
\frac{1}{2}\nu \frac{ \lb \sqrt{\tilde{Q}-\tilde{q}}y+\sqrt{\tilde{q}}z+\tilde{m}x_{0i} \rb^2}{\tilde{R}+\tilde{Q}} 
}  
\rb^n
\equiv F(x_{0i}),
\ee
and the law of large number implies
\be
&&
\frac{1}{N}\log \lb \prod_{i=1}^{N}\lbb 
\sum_{\{c_i\}}\Tr{\{x_i\}|\{c_i\}}
e^{f_i\lb \{x_i\},\{c_i\} \rb}
\rbb
\rb
=\frac{1}{N}\sum_{i}\log F(x_{0i})
=
\lsb \log F(x_0) \rsb_{x_0}
\no \\ &&
=  
\rho_0 \int dx_0 P_0(x_0) \log \int Dz (1+X_{\tilde{m}})^n 
+
(1-\rho_0)\log \int Dz (1+X_{0})^n.
\ee
where
\be
&&
X_{\tilde{m}}=\frac{e^{-\tilde{\rho}} \sqrt{2\pi}^{\nu} }{\sqrt{\tilde{R}+\tilde{Q}}^{\nu}}
\sqrt{
\frac{\tilde{R}+\tilde{Q}}{\tilde{R}+\tilde{Q}-\nu(\tilde{Q}-\tilde{q})}
}
e^{\frac{1}{2} \nu \frac{(\sqrt{\tilde{q}}z+\tilde{m}\tilde{x})^2 }{\tilde{R}+\tilde{Q}-\nu(\tilde{Q}-\tilde{q})}},
\ee
and $X_{0}$ is obtained by inserting $\tilde{m}=0$ in $X_{\tilde{m}}$. This relation means that the average over $\V{x}_0$ in \Req{I} is actually not needed, which comes from the absence of correlations among components of the design matrix $A$. This validates the use of the factorised prior \NReq{factorised}. 

To take the limit $\nu \to 0$, we rescale the tilde variables as follows
\subbe
\Leq{rescaling_RS}
\be
&&
\nu (\tilde{R}+\tilde{Q})\to \tilde{\chi}+\tilde{Q},
\\ &&
\nu^2 \tilde{R}\to -\tilde{\chi},
\\ &&
\nu^2 \tilde{Q}\to \tilde{\chi},
\\ &&
\nu^2 \tilde{q}\to \tilde{q}, 
\\ &&
\nu \tilde{m}\to \tilde{m}.
\ee
\subee
Applying this rescaling and keeping only the linear term with respect to $n$, we get
\be
&&
\mathcal{I} \approx
\int d\tilde{\rho}d\tilde{R}d\tilde{Q}d\tilde{q}d\tilde{m}~
\exp Nn \Big\{
\tilde{\rho} \rho+\frac{1}{2}\tilde{Q}Q-\frac{\tilde{\chi}\chi}{2\mu}+\frac{1}{2}\tilde{q}q-\tilde{m}m
\no \\ &&
+\rho_0\int dx_0 P_0(x_0) \int Dz  \log  \lb 1+Y^{\mathrm{RS}}_{\tilde{m}}\rb
+(1-\rho_0)\int Dz  \log  \lb 1+Y^{\mathrm{RS}}_{0}\rb
 \Big\}.
 \Leq{I final}
\ee
Combining \Reqs{L final}{I final} and using the saddle-point method yield \Req{g_RS}.

\subsection{1RSB computation} \Lsec{The 1RSB}
In the 1RSB level, the overlap between $c^a$ and $c^b$ with different $a,b=1,\cdots,n$ should be treated in a more involved manner. In the standard 1RSB construction, the $n$ replicas are separated into $n/\tau$ blocks of the size $\tau$. We may label the blocks by $B=1,\cdots,n/\tau$, and the replica index $a$ is represented by two indices as $a=(B_a,I_a)$ where $I_a$ denotes the component index inside the block. For simplicity, we identify the index set of those components with the block label $B_a$, allowing us to represent them as $I_a \in B_a$. The overlap $q_{b\beta}^{a\alpha}=(1/N)\sum_{i}c^a_i c^{b}_i x^{a\alpha}_{i}x^{b\beta}_{i}$ is assumed to take two discriminative values depending on whether the replica indices $a,b$ are in the same block or not. This can be written as
\be
q_{b\beta}^{a\alpha}
=
\left\{
\begin{array}{cc}
q_1  &   (a\neq b~\&~B_a=B_b)    \\
q_0  &   (a\neq b~\&~B_a\neq B_b)
\end{array}
\right..
\ee
Correspondingly, $d^{a\alpha}_{\mu}$ can be expressed as
\be
d^{a\alpha}_{\mu}=d^{B I \alpha}_{\mu}=\sqrt{R-Q}u^{B I \alpha}_{\mu}
+\sqrt{Q-q_1}v^{B I}_{\mu}
+\sqrt{q_1-q_0}w^{B }_{\mu}
+\sqrt{\rho_0\sigma_{x}^2-2m+q_0}z_{\mu},
\ee
where $u,v,w,$ and $z$ are i.i.d. from the normal distribution. Then, \Req{Phi} is rewritten as
\be
\Phi=\int dR dQ dq_1 dq_0 dm~\mathcal{I}_{\mathrm{1RSB}} \times \mathcal{L}_{\mathrm{1RSB}},
\Leq{Phi_1RSB}
\ee
where 
\be
&&
\mathcal{I}_{\mathrm{1RSB}}=\Tr{ \{ \V{c} \} } \Tr{ \{ \V{x} \} |\{\V{c} \}}
\prod_{a,\alpha}\delta \lb NR-\sum_{i}c_i^{a}\lb x_i^{a\alpha}\rb^2 \rb 
\prod_{a,\alpha<\beta}\delta \lb N Q-\sum_{i}c_i^{a} x_i^{a\alpha}x_i^{a\beta} \rb 
\no \\ &&
\times
\lsb
\prod_{a,\alpha}\delta \lb N m-\sum_{i} x_{0i}c_i^{a} x_i^{a\alpha} \rb,
\rsb_{\V{x}_0}
\prod_{B}
\prod_{I<J}
\prod_{\alpha,\beta}\delta \lb N q_1-\sum_{i}c_i^{B I}c_i^{B J} x_i^{BI \alpha}x_i^{B J\beta} \rb 
\no \\ &&
\times
\prod_{B_a<B_b}
\prod_{I \in B_a}\prod_{J \in B_b}
\prod_{\alpha,\beta}\delta \lb N q_0-\sum_{i}c_i^{B_a I}c_i^{B_b J} x_i^{B_a I \alpha}x_i^{B_b J\beta} \rb, 
\ee
and 
\be
&&
\mathcal{L}_{\mathrm{1RSB}}=
\prod_{\mu=1}^{M}\Blbb
\int Dz_\mu \int d \xi_\mu \frac{e^{-\frac{1}{2\sigma_{\xi}^2}\xi_{\mu}^2 }}{\sqrt{2\pi \sigma_{\xi}^2}} 
\no \\ &&
\times
\int \prod_{B}Dw^{B}_{\mu}
\int \prod_{B}\prod_{I\in B}Dv^{BI}_{\mu}
\int \prod_{B}\prod_{I\in B}\prod_{\alpha}Du^{BI\alpha}_{\mu}
\prod_{B}\prod_{I\in B}\prod_{\alpha}e^{-\frac{1}{2}\beta (d^{BI\alpha}_{\mu}+\xi_{\mu})^2}
\Brbb
\no \\
&&
=\lbb 
\int Dz \lb \int Dw \lb \int Dv \lb \int Du~e^{-\frac{1}{2}\beta h^2_2(u,v,w,z)} \rb^{\nu}\rb^{\tau}  \rb^{n/\tau}
\rbb^{M},
\ee
where
\be
h_2(u,v,w,z)=\sqrt{R-Q}u+\sqrt{Q-q_1}v+\sqrt{q_1-q_0}w+\sqrt{\rho_0\sigma_{x}^2+\sigma_{\xi}^2-2m+q_0}z.
\ee

$\mathcal{L}_{\mathrm{1RSB}}$ can be computed by recurring Gaussian integrations as in the RS case, and the result is
\be
\mathcal{L}_{\mathrm{1RSB}}\approx
e^{Nn\frac{\alpha}{2}
\lbb 
\log\frac{1+\chi}{D_1}
+\frac{1}{\tau}\log\frac{D_1}{D_0}
-\frac{\mu (V+q_0)}{D_0}
\rbb
}.
\Leq{L_1RSB final}
\ee

The computation of $\mathcal{I}_{\mathrm{1RSB}}$ can also be performed in parallel with the RS case. The delta functions of common variables with the RS case are broken in the same manner as \Req{subshell}, and the ones of $q_1$ and $q_0$ are broken similarly to \Req{q-subshell}. These transforms yield 
\be
\mathcal{I}_{\mathrm{1RSB}}=
\int d\tilde{\rho}d\tilde{R}d\tilde{Q}d\tilde{q}_1d\tilde{q}_0d\tilde{m}~e^{Nf_{\times}}
 \lsb  
 \prod_i \Tr{ \{ c_i \} }\Tr{ \{ x_i \} | \{ c_i \} }e^{ f_{2i}(\{x_i \},\{c_i \}) } 
 \rsb_{\V{x}_0},
\ee
where
\be
&&
\hspace{-10mm}
f_{\times}=n\tilde{\rho}\rho+\frac{1}{2}n\nu \tilde{R}R-\frac{1}{2}n\nu(\nu-1) \tilde{Q}Q
-\frac{1}{2}n(\tau-1)\nu^2 \tilde{q}_1 q_1-\frac{1}{2}n(n-\tau)\nu^2 \tilde{q}_0 q_0
-n\nu \tilde{m}m,
\\ &&
\hspace{-10mm}
f_{2i}\lb \{x\},\{c\} \rb
=
-\tilde{\rho}\sum_{a}c_a
-\frac{1}{2}\tilde{R}\sum_{a}\sum_{\alpha}c_{a}x_{a\alpha}^2
+\tilde{Q}\sum_a\sum_{\alpha<\beta}c_ax_{a\alpha}x_{a\beta}
+\tilde{m}x_{0i}\sum_{a}\sum_{\alpha}c_ax_{a\alpha}
\no \\ &&
+\tilde{q_1}\sum_{B}\sum_{I<J} \sum_{\alpha,\beta} c_{BI}c_{BJ}x_{BI\alpha}x_{BJ\beta}
+\tilde{q_0}\sum_{B_a<B_b}\sum_{I\in B_a}\sum_{J\in {B_b}} 
\sum_{\alpha,\beta} c_{B_a I}c_{B_b J}x_{B_a I\alpha}x_{B_b J\beta}.
\Leq{f_2i}
\ee
To derive a factorised form with respect to the replica index, we again use the trick of the Gaussian integrations as the RS case. For example, the last two terms in \Req{f_2i} are factorised as
\be
&&
\hspace{-10mm}
e^{\tilde{q_1}\sum_{B}\sum_{I<J} \sum_{\alpha,\beta} c_{BI}c_{BJ}x_{BI\alpha}x_{BJ\beta}
+\tilde{q_0}\sum_{B_a<B_b}\sum_{I\in B_a}\sum_{J\in {B_b}} 
\sum_{\alpha,\beta} c_{B_a I}c_{B_b J}x_{B_a I\alpha}x_{B_b J\beta}}
\no \\ &&
\hspace{-10mm}
=
\int Dz_0 \int \prod_{B}^{n/\tau}Dz_B~
e^{\sqrt{\tilde{q}_0}z_0\sum_{B}\sum_{I\in B}c_{BI}X_{BI}
+\sqrt{ \tilde{q}_1 - \tilde{q}_0 }\sum_{B} z_B \sum_{I\in B} c_{BI}X_{BI}
-\frac{1}{2}\hat{q}_1\sum_B \sum_{I\in B}c_{BI}X_{BI}^2
 },
\ee
where we insert $X_{BI}= \sum_{\alpha}x_{BI\alpha}$. Repeating similar computations, in the leading order of $n$ we finally get
\be
&&
\mathcal{I}_{\mathrm{1RSB}} \approx
\int d\tilde{\rho}d\tilde{R}d\tilde{Q}d\tilde{q}_1d\tilde{q}_0d\tilde{m}~
\exp Nn \Big\{
\tilde{\rho} \rho+\frac{1}{2}\tilde{Q}Q-\frac{\tilde{\chi}\chi}{2\mu}
-\frac{1}{2}(\tau-1)\tilde{q}_1q_1+\frac{1}{2}\tau\tilde{q}_0q_0-\tilde{m}m
\no \\ &&
+\frac{\rho_0}{\tau}\int dx_0P_0(x_0) \int Dz_0 \log \int Dz_1 \lb 1+Y^{\mathrm{1RSB}}_{\tilde{m}}\rb^{\tau}
\no \\ &&
+\frac{1-\rho_0}{\tau}\int Dz_0 \log  \int Dz_1 \lb 1+Y^{\mathrm{1RSB}}_{0}\rb^{\tau}
 \Big\}.
\Leq{I_1RSB final}
\ee
To derive this, when taking the limit $\nu\to 0$, we have rescaled $\tilde{q}_1$ and $\tilde{q}_0$ as
\subbe
\be
&&
\nu^2 \tilde{q}_1 \to \tilde{q}_1,
\\ &&
\nu^2 \tilde{q}_0 \to \tilde{q}_0.
\ee
\subee
Other tilde variables are rescaled in the same manner as \Req{rescaling_RS}. 

Inserting \Reqs{L_1RSB final}{I_1RSB final} into \Req{Phi_1RSB} and using the saddle-point method, we obtain \Req{g_1RSB}. The EOS are obtained by taking the extremisation condition and the result is
\subbe
\Leq{EOS_1RSB}
\be
&&
\hspace{-10mm}
\tilde{\chi}=\alpha \lbb 
\frac{\mu^2\Delta_1 }{(1+\chi)D_1}
+\frac{\mu^2\Delta_0 }{D_1 D_0}
+\frac{ \mu^2(V+q_0) }{D_0^2}\rbb,
\Leq{chitilde_1RSB}
\\ &&
\hspace{-10mm}
\tilde{Q}=\alpha \lbb 
\frac{\mu}{D_1}
-\frac{\mu^2{\Delta_0} }{D_1 D_0}
-\frac{ \mu^2(V+q_0) }{D_0^2}\rbb,
\Leq{Qtilde_1RSB}
\\ &&
\hspace{-10mm}
\tilde{q}_1=\alpha \lbb 
\frac{\mu^2{\Delta_0} }{D_1 D_0}
+\frac{ \mu^2(V+q_0) }{D_0^2}\rbb,
\Leq{q1tilde_1RSB}
\\ &&
\hspace{-10mm}
\tilde{q}_0=\alpha 
\frac{ \mu^2(V+q_0) }{D_0^2}
,
\Leq{q0tilde_1RSB}
\\ &&
\hspace{-10mm}
\tilde{m}= \frac{\alpha \mu}{D_0},
\Leq{mtilde_1RSB}
\\ &&
\hspace{-10mm}
\rho=\rho_0\int dx_0 P_0(x_0) \int Dz_0 
\Ave{\frac{Y^{\mathrm{1RSB}}_{\tilde{m}} }{1+Y^{\mathrm{1RSB}}_{\tilde{m}} }}^{\mathrm{1RSB}}_{\tilde{m}}
+(1-\rho_0)\int Dz_0 
\Ave{
\frac{Y^{\mathrm{1RSB}}_{0}}{1+Y^{\mathrm{1RSB}}_{0}}
}^{\mathrm{1RSB}}_{0},
\Leq{rho_1RSB}
\\ &&
\hspace{-10mm}
\chi=\frac{\mu}{\tilde{\chi}+\tilde{Q}}
\Blbb
\rho_0\int dx_0 P_0(x_0) \int Dz_0 
\Ave{\frac{Y^{\mathrm{1RSB}}_{\tilde{m}} }{1+Y^{\mathrm{1RSB}}_{\tilde{m}} }}^{\mathrm{1RSB}}_{\tilde{m}}
\no \\ &&
+(1-\rho_0)\int Dz_0 
\Ave{
\frac{Y^{\mathrm{1RSB}}_{0}}{1+Y^{\mathrm{1RSB}}_{0}}
}^{\mathrm{1RSB}}_{0}
\Brbb,
\Leq{chi_1RSB}
\\ &&
\hspace{-10mm}
Q=\frac{\tilde{\chi}-\tilde{q}_1}{(\tilde{\chi}+\tilde{Q})(\tilde{Q}+\tilde{q}_1)}
\Blbb
\rho_0\int dx_0 P_0(x_0) \int Dz_0 
\Ave{\frac{Y^{\mathrm{1RSB}}_{\tilde{m}} }{1+Y^{\mathrm{1RSB}}_{\tilde{m}} }}^{\mathrm{1RSB}}_{\tilde{m}}
\no \\ &&
\hspace{5mm}
+(1-\rho_0)\int Dz_0 
\Ave{
\frac{Y^{\mathrm{1RSB}}_{0}}{1+Y^{\mathrm{1RSB}}_{0}}
}^{\mathrm{1RSB}}_{0}
\Brbb
\no \\ &&
+\frac{1}{(\tilde{Q}+\tilde{q}_1)^2}\Blbb
\rho_0
\int dx_0 P_0(x_0)\int Dz_0~
\Ave{
\frac{ \lb h^{\mathrm{1RSB} }_{\tilde{m}} \rb^2 Y^{\mathrm{1RSB} }_{\tilde{m}} }{1+Y^{\mathrm{1RSB}}_{\tilde{m}}}
}^{\mathrm{1RSB} }_{\tilde{m}} 
\no \\ &&
\hspace{5mm}
+
(1-\rho_0)\int Dz_0~
\Ave{
\frac{\lb h^{\mathrm{1RSB} }_{0} \rb^2  Y^{\mathrm{1RSB}}_{0} }{1+Y^{\mathrm{1RSB}}_{0}}
}^{\mathrm{1RSB} }_{0} 
\Brbb,
\Leq{Q_1RSB}
\\ && \hspace{-10mm}
q_1=
\frac{1}{(\tilde{Q}+\tilde{q}_1)^2}
\Blbb
\rho_0\int dx_0 P_0(x_0)
\int Dz_0~  
   \Ave{ 
      \lb \frac{ h^{\mathrm{1RSB} }_{\tilde{m}}  Y^{\mathrm{1RSB}}_{\tilde{m}}}{1+Y^{\mathrm{1RSB}}_{\tilde{m}}} 
      \rb^2
   }^{\mathrm{1RSB} }_{\tilde{m}} 
\no \\ &&
+(1-\rho_0)\int Dz_0~
 \Ave{ 
   \lb \frac{h^{\mathrm{1RSB} }_{0}  Y^{\mathrm{1RSB}}_{0}}{1+Y^{\mathrm{1RSB}}_{0}} \rb^2
  }^{\mathrm{1RSB} }_{0} 
\Brbb
,
\Leq{q1_1RSB}
\\ && \hspace{-10mm}
q_0=
\frac{1}{(\tilde{Q}+\tilde{q}_1)^2}
\Blbb
\rho_0\int dx_0 P_0(x_0)
\int Dz_0~ 
\lb 
   \Ave{ \frac{ h^{\mathrm{1RSB} }_{\tilde{m}}  Y^{\mathrm{1RSB}}_{\tilde{m}}}{1+Y^{\mathrm{1RSB}}_{\tilde{m}}} 
   }^{\mathrm{1RSB} }_{\tilde{m}} 
\rb^2
\no \\ &&
+(1-\rho_0)\int Dz_0~
\lb 
 \Ave{ 
   \frac{h^{\mathrm{1RSB} }_{0}  Y^{\mathrm{1RSB}}_{0}}{1+Y^{\mathrm{1RSB}}_{0}} 
  }^{\mathrm{1RSB} }_{0} 
\rb^2
\Brbb
,
\Leq{q0_1RSB}
\\ &&
\hspace{-10mm}
m=
\frac{1}{\tilde{Q}+\tilde{q}_1}\lbb
\rho_0\int dx_0 P_0(x_0) \int Dz_0 
\Ave{
 \frac{x_0 h^{\mathrm{1RSB}}_{\tilde{m}} Y^{\mathrm{1RSB}}_{\tilde{m}}}{1+Y^{\mathrm{1RSB}}_{\tilde{m}}}
 }^{\mathrm{1RSB} }_{\tilde{m}} 
\rbb
.
\Leq{m_1RSB}
\ee
\subee
where
\be
\Ave{\lb \cdots \rb}^{\mathrm{1RSB}}_{\tilde{m}}
=\frac{
\int Dz_1 (\cdots)  \lb 1+Y^{\mathrm{1RSB}}_{\tilde{m}} \rb^{\tau}
}{\int Dz_1  \lb 1+Y^{\mathrm{1RSB}}_{\tilde{m}} \rb^{\tau}}
,~
\Ave{\lb \cdots \rb}^{\mathrm{1RSB}}_{0}
=\frac{
\int Dz_1 (\cdots)  \lb 1+Y^{\mathrm{1RSB}}_{0} \rb^{\tau}
}{\int Dz_1  \lb 1+Y^{\mathrm{1RSB}}_{0} \rb^{\tau}}.
\ee

\end{document}